\documentclass[12pt,preprint]{aastex}

\usepackage{morefloats}
\usepackage{color}
\usepackage{amsmath}

\shorttitle{A Statistical Consideration of the Habitable Zone}
\shortauthors{Truitt et al.}

\begin{document}

\title{A Flexible Bayesian Framework for Assessing Habitability with Joint Observational and Model Constraints}

\author{Amanda R. Truitt\altaffilmark{1}, Patrick A. Young\altaffilmark{2}, Sara I. Walker\altaffilmark{2}, Alexander Spacek\altaffilmark{1}}
\altaffiltext{1}{Los Alamos National Laboratory, Los Alamos, New Mexico 87545}
\altaffiltext{2}{School of Earth and Space Exploration, Arizona State University, Tempe, AZ 85287}

\newcommand{\Msol}{\mbox{$\rm{M_{\odot}\ }$}}
\newcommand{\msol}{\mbox{$\rm{M_{\odot}\ }$}}
\newcommand{\Rsol}{\mbox{$\rm{R_{\odot}\ }$}}
\newcommand{\rsol}{\mbox{$\rm{R_{\odot}\ }$}}
\newcommand{\sol}{\mbox{$\rm{_{\odot}\ }$}}
\newcommand{\Ni}{\mbox{$^{56}$Ni}}

\begin{abstract}

The catalog of stellar evolution tracks discussed in our previous work is meant to help characterize exoplanet host-stars of interest for follow-up observations with future missions like JWST. However, the utility of the catalog has been predicated on the assumption that we would precisely know the age of the particular host-star in question; in reality, it is unlikely that we will be able to accurately estimate the age of a given system. Stellar age is relatively straightforward to calculate for stellar clusters, but it is difficult to accurately measure the age of an individual star to high precision. Unfortunately, this is the kind of information we should consider as we attempt to constrain the long-term habitability potential of a given planetary system of interest. This is ultimately why we must rely on predictions of accurate stellar evolution models, as well a consideration of what we can observably measure (stellar mass, composition, orbital radius of an exoplanet) in order to create a statistical framework wherein we can identify the best candidate systems for follow-up characterization. In this paper we discuss a statistical approach to constrain long-term planetary habitability by evaluating the likelihood that at a given time of observation, a star would have a planet in the 2 Gy continuously habitable zone (CHZ$_2$). Additionally, we will discuss how we can use existing observational data (i.e. data assembled in the {\it Hypatia} catalog and the {\it Kepler} exoplanet host star database) for a robust comparison to the catalog of theoretical stellar models.
\end{abstract}

\keywords{(stars:) evolution, astrobiology, (stars:) planetary systems, exoplanets, catalogs}

\section{Introduction}

The search for a habitable\footnote{In this paper, by habitable we generally mean a rocky planet with liquid water and the potential to host life as we know it.} Earth-like planet is one of the largest current undertakings in the fields of astronomy and astrobiology. We now have over 3700 confirmed exoplanets, with over 2200 additional planetary candidates. There are several planned space telescope missions such as CHEOPS \citep{broeg14}, PLATO \citep{rauer13}, TESS \citep{rick14}, WFIRST \citep{sper15}, LUVOIR \citep{bolcar16}, JWST (e.g. \citet{bar16,greene16}), OST \citep{battersby18}, and HabEx \citep{gaudi18} that are intended to help discover or further characterize systems \citep{borucki10,traub12,bat14,liss14}, and to narrow down the search for habitable worlds. There is also the future possibility of probing the atmospheres of Earth-like exoplanets using ground-based telescopes like the ELT (e.g. \citet{hawk2019}). We now know that exoplanets are relatively abundant in the habitable zones of Sun-like and M-type stars (e.g. \citet{tar07,bat11,pet2013,dc2015}), including small, rocky Earth-like planets in the habitable zones of their host stars (e.g. \citet{ditt2017,gill2017}). The question of how we would detect life on another planet has long been considered, including the idea of estimating the likelihood of life existing elsewhere. 

Given the sheer number of planetary systems that have been found and are still being discovered, detailed characterization of every system is impractical if not impossible, so it is necessary to prioritize targets for followup. It makes sense to approach the search for life on another planet from a statistical perspective, which allows us to evaluate the likelihood of detectability progressively as more information is obtained. The most famous example of this type of probabilistic consideration would be the Drake Equation, but other more recent work has also outlined the necessity for a statistical approach to calculating the potential for planetary habitability (e.g. \citet{bean17}). 

The advantage of approaching the complex problem of planetary habitability within a statistical framework is that we can assign probability distributions, regardless of the level at which the important stellar or planetary parameters have been measured. Better observations would mean a narrower and more accurate distribution, but even a poorly characterized system can be broadly constrained by limits imposed by observational data. We can rely on statistical values from large catalogs of measured values for similar systems in order to assess the habitability potential of a particular planetary system of interest for more in-depth, follow-up characterization. For example, \citet{desch18} have outlined a specific observational strategy to move toward the prioritization of exoplanet observations. They first focus on observational data that is relatively more easily obtained, and rank planetary systems for more difficult follow-up observations and characterization.

There is a plethora of statistical data available already considered, including exoplanet orbital properties, characteristics, and distributions of planet types/sizes from radial velocity and transit surveys \citep{kane12,fm14,kane16,kopp18}, and the frequency of planets around stars of specific spectral types (effective temperatures and mass) and/or metallicity values (e.g. \citet{all99,san03,ida05,bean06,adib12,buc12,jj12,berg14}). Additionally, recent work by \citet{bean17} discusses a survey of planetary atmospheric compositions for terrestrial planets, for a ``comparative planetology" approach to habitability that utilizes measurable planetary characteristics and a statistical analysis to answer large-scale questions using only a small sample of objects. This method was developed for calculating estimates for the frequency of habitable environments and the existence of life in terms of the exoplanet catalog as a whole, rather than considering planetary systems individually. 

In this paper, we consider the inverse problem. It is important to evaluate a planet's habitability history in a probabilistic sense, especially if important characteristics of the system are not known precisely, or at all. From the stellar side of the problem this most prominently includes a host star's mass, composition, and age. Composition may or may not be measured for a particular star. Age is rarely determined to better than a gigayear for low mass field stars of ages comparable to the sun, as techniques for determining more precise ages are far more effective for stars younger than 1 Gy (e.g. \citet{sod10,silva15,af2018}). Given constraints for the properties of a planetary system from observations (if available) or population statistics (if not), and appropriate theoretical models, can we derive a statistical likelihood that a planet will have particular properties? From an astrobiological perspective, we are interested particularly in whether a planet is likely to have detectable life. It is desirable to develop a flexible formalism that can incorporate various constraints as they are contributed by the community. 
 
We implement a Bayesian statistical approach to considering the long-term habitability potential for exoplanetary systems based on joint constraints from modeling and observation. We present an example focused on host star properties. This uses a theoretical model grid of stellar and habitable zone evolutionary tracks and associated observable distributions for mass and composition of exoplanet host stars and planetary orbital distance. Since Bayesian statistics use mathematical rules of posterior probability to combine observable data and prior information, we produce an inference more precise than gathered from either source alone. It also accommodates disparate sources of information. The utility of this approach is demonstrated by previous work on a Bayesian analysis to understand the probabilities related to emergence of life \citep{spi12}, an inherently interdisciplinary enterprise.

High among the problems of considering detectability is the issue of long-term habitability. We need to be able to assess the likelihood that life might not only be present on a particular exoplanet, but also that we would eventually be able do observe it. It is assumed that the ability of life to establish itself and flourish on a planet's surface remains heavily impacted by how long the environment remains stable to even slightly habitable conditions (e.g. \citet{mckay14,dong18}), thus instantaneous habitability is an insufficient consideration. Life must be present for long enough to measurably alter the environment. We choose a time of 2 Gy as the minimum time for a planet to be in the HZ in order for life to possibly be detectable following \citet{truitt15} and \citet{truitt17} and motivated by, e.g., \citet{brocks99}, \citet{kopp05}, and \citet{crowe13}. This limit can also easily be adjusted without compromising the framework presented here. In this case study we use the host star as a metric for predicting the long-term habitability potential of a given system; this is an important first step, but an incomplete consideration. We recognize that in order to fully assess the habitability potential of a given planetary system, we must consider not only how the characteristics of a parent star will influence the evolution of the habitable zone, but also the nature of the planets themselves (orbital properties, material, atmosphere, etc.) and the conditions of the circumstellar environment \citep{rush13,walt17}. In this framework, constraints can be relatively easily added to refine predictions as they are developed.

We acknowledge that the relationship between a ``habitable zone'' and habitability itself is a complicated concept, and that the conditions for ``habitability'' and ``detectability" are difficult to define. Indeed, each might be characterized differently based on whether the associated calculations focus on planetary atmosphere contributions (perhaps related to the search for biosignatures) and the incoming stellar flux, or if  approached from a perspective more related to the thermal evolution and the chemical compositions of the planets themselves \citep{lam09,gud14}. One could consider the effect on habitability depending on whether the planet is a ``water world'' or if it is more of an Earth-like planet with continents and plate tectonics \citep{kite18}. The relevant data, though potentially possible to obtain, are challenging to confidently interpret outside of a statistical framework \citep{sea10,rein14,cat17,dem17,schwi17,walker17}, and there are simply a large number of possibly habitable exoplanet candidates.

We are also aware of the likely complications to habitability when we consider stellar activity of low-mass (M-type) stars, and that planets orbiting in the habitable zones of these stars are almost necessarily tidally locked \citep{joshi1997}. These complications are actually drivers of the statistical approach. Even very approximate habitable zone predictions can place broad constraints on follow-up prioritization, and the habitable zone model prior can be treated as a ``module" that can be swapped for predictions from an improved model. 

In section~\ref{method} we discuss the methodology and the viability of a direct comparison between the data assembled in the {\it Hypatia} Catalog and the {\it Kepler} database. We also present a Bayesian framework for habitable zones with four specific cases for different combinations of mass and metallicity priors. In section~\ref{results} we discuss the interpretation of the posterior probability distributions and present several example planetary systems for consideration. In section~\ref{conclusions} we discuss our conclusions and the implications for this approach.

\section{Methodology \label{method}}

Our overarching goal is to evaluate the relative promise of known planets for hosting detectable life as a way of prioritizing candidates for followup observations. The proximate goal of this paper is to address one component by determining the likelihood that a given planet has been in the host star's habitable zone for at least 2 Gy (denoted CHZ$_2$), based on measured stellar properties jointly with stellar models. To be clear, CHZ$_2$ simply means the distances from a star that are in the star's HZ for at least 2 Gy during the star's lifetime, limited to 12 Gy. We use Bayes' theorem with priors derived from observations and and outputs from the Tycho code to evaluate the posterior probability that an exoplanet with certain orbital parameters is in the CHZ$_2$.  

The statistical consideration for the distribution of stellar compositions is based on the data assembled in {\it Hypatia} \citep{hinkburg17}, a catalog comprised of high-resolution stellar abundance measurements. We also implement data from NASA's Exoplanet Archive \citep{ea} to understand whether the {\it Kepler} exoplanet host stars are well-represented by the abundance distribution derived from the {\it Hypatia} catalog. Though the measurements included in {\it Hypatia} may be inherently biased due to the requirements for obtaining high-resolution abundance measurements, these are stars in the solar neighborhood that give us an idea of the representative distribution of elemental compositions.

We used the stellar evolution code Tycho \citep{ya05} to create the catalog of evolutionary tracks discussed in detail in \citet{truitt15} and \citet{truitt17}, which are the basis for the model priors discussed here. The database currently contains models between 0.5-1.2 solar masses, with metallicities that fall between $0.1-1.5$ of solar $Z$-value, so this is the parameter space we analyze here. (Models are also available for non-solar elemental abundance ratios, but we consider only mass and metallicity here.) Tycho outputs information on stellar surface quantities for each time-step of the evolution. For this proof of concept, we use those values to define the inner and outer boundaries of the habitable zone as a function of stellar age, utilizing equations from \citet{kopp13,kopp14}, following from \citet{sel07} and \citet{kwr93}. These prescriptions parameterize the habitable zone as a function of the host star luminosity and effective temperature, from which we can calculate the associated time-dependent habitable zone distance for each stellar evolution track. For a given orbital distance from any star we can predict how long and at what stellar age a planet would remain habitable. Thus, for a perfectly characterized star (mass, metallicity, and age known exactly), we can say with certainty whether a planet has been continuously habitable for $\geq$ 2 Gy for a given habitable zone model. 

HZ boundaries are very uncertain and model dependent. This is especially true given the use of both 1D and 3D models (e.g. Kopparapu et al. 2013, Leconte et al. 2013). We use the HZ formulation from Kopparapu et al. (2013, 2014) here because they represent combined 1D and 3D model results that present a relatively simple, conservative HZ estimate. For example, the inner HZ may actually be farther in according to some 3D models that require a higher stellar flux for the runaway greenhouse limit (Leconte et al. 2013). Similarly, the outer HZ may be at a larger distance if additional greenhouse gasses like H$_2$, CH$_4$, and NH$_3$ are present (Pierrehumbert \& Gaidos 2011). Since we are trying to constrain potentially habitable planets, the conservative HZ estimates we use are appropriate since we would rather err on throwing out potentially habitable planets than further polluting our results with planets that aren't really in the HZ.

Real habitability is also very complicated and depends on much more than just the star characteristics and the planet's distance from the star, including the planet's rotation, size, mass, composition, and atmosphere. In this work we are trying to use as simple of a framework as reasonably possible to show how a statistical, Bayesian analysis can be made regarding potentially habitable planets. We emphasize that more sophisticated HZ prescriptions can be substituted in the framework presented here, but the main goal of this work is as a proof of concept.

The models are created for fixed stellar masses and compositions, and observations must be used to associate a predicted track with a given star. The observed values will not be precise values. There will be greater or lesser uncertainty. This may reflect unavoidable observational errors for a precise measurement technique all the way to a property that is not measured at all for a given star. It is very difficult to ascertain the ages of low mass field stars \citep{sod10}, so we must rely on predictive theoretical models and statistical analyses (e.g. \citet{silva15,af2018}). For a star with a given metallicity and mass and unknown age, we can directly derive a posterior probability that a given orbital distance is within CHZ$_2$, which is what we do in this case. 

To obtain the likelihood P(CHZ$_2$) that a planet has spent 2 Gy in the habitable zone, we convolve the probability that a star has a specified value of mass and metallicity, based upon observational constraints, with the theoretical probability of a planet being in the CHZ$_2$ for the range of M and Z represented by the observational constraints in a Bayesian sense. Below we describe this process for the general case and the following priors: mass M with a Gaussian uncertainty $\sigma_M$, metallicity Z with a Gaussian uncertainty $\sigma_Z$, and unmeasured metallicity with no assumption about the shape of the metallicity distribution P(Z), and unmeasured metallicity with P(Z) based on the metallicity distribution derived from the {\it {\it Hypatia}} abundance catalog of nearby stars, as well as combinations of these priors.

\subsection{Comparing to Observations: {\it Hypatia} and {\it Kepler}}

We started with data from the {\it Hypatia} catalog for 3,861 stars with iron abundance measurements within the defined limits of our catalog (0.1 $\leq$ Fe $\leq$ 1.5 Z\sol). {\it Hypatia} includes measurements for almost any particular element of interest, and it would be important to consider specific elemental abundance ratios (particularly for oxygen); unfortunately, the {\it Kepler} host star dataset only provides measurements for ``metallicity'' as the iron abundance [Fe/H] and our closest method for comparison is to use the Tycho catalog scaled metallicity $Z$-value range. We were able to extract a total of 382 host stars from the {\it Kepler} exoplanet database that have metallicity measurements between 0.1 and 1.5 Z\sol.

\begin{figure}
\centerline{\includegraphics[height=12cm,width=15cm]{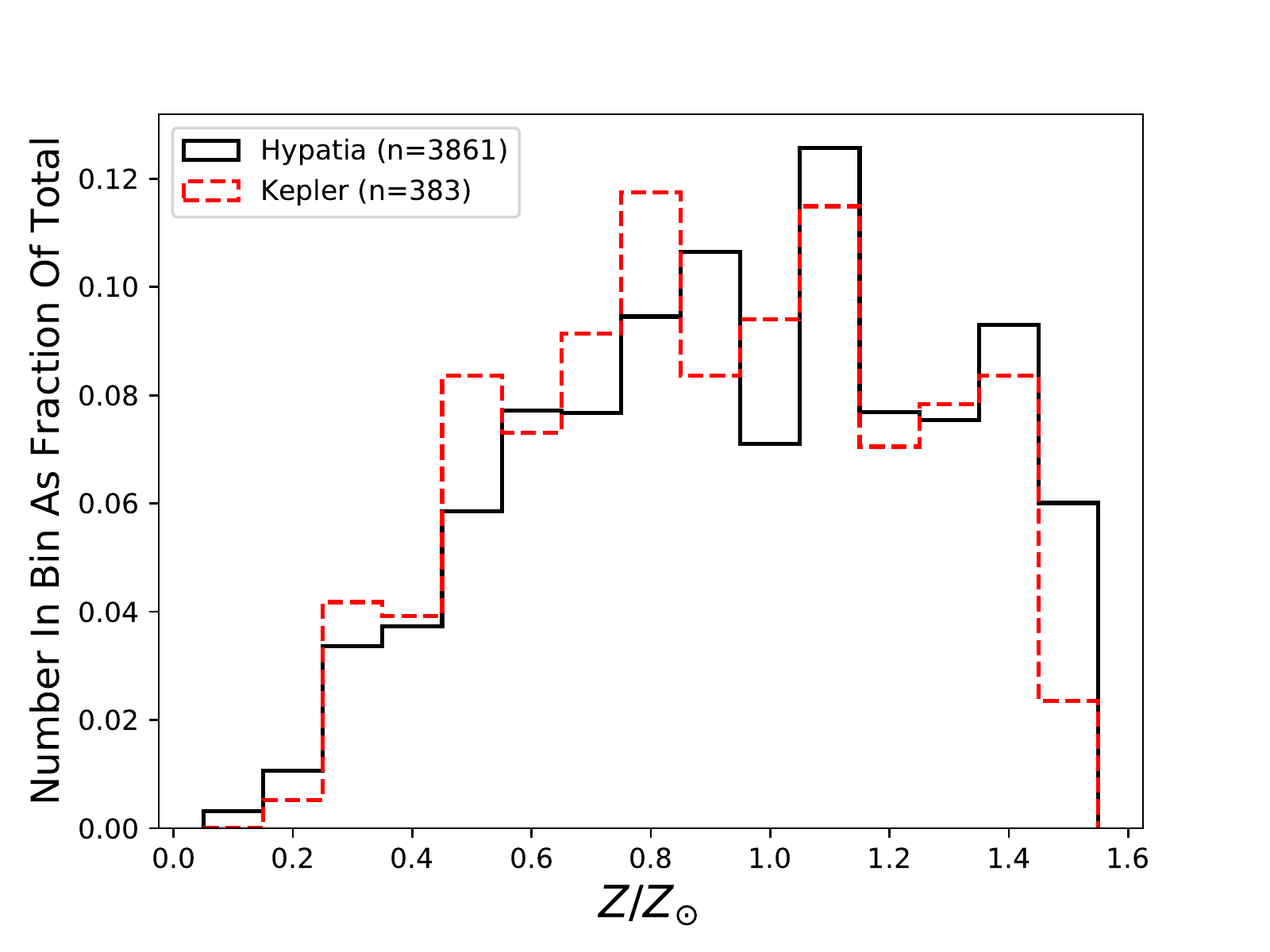}}
\caption{KS-Test comparing a subset of {\it Kepler} host stars with the distribution of stellar chemical abundance values documented in the {\it Hypatia} catalog, calculated in units of $Z/Z\sol$ so each bin is representative of a value in the Tycho grid of stellar evolutionary tracks.\label{fig:kephyp}}
\end{figure}

We used bins from 0.1-1.5, centered on those values, with widths of 0.1. Since the measured abundance values in {\it Hypatia} have low precision, they tend to fall on the values between bins, like 0.15 or 0.25. For each bin division we took all values at that number and put half of them in the lower bin half in the higher bin. We used the IDL routine KSTWO to compute a Kolmogorov-Smirnov (KS) comparison test \citep{mass51}, which outputs the corresponding initial test D-value, $D_{KS}$. The KS test compares the distributions (cumulatively add up the fraction of data points that are less than or equal to a value), with $D_{KS}$ being the greatest vertical distance between the two dataset distributions. The comparison of the differences between the {\it Kepler} and {\it Hypatia} datasets is shown in Figure~\ref{fig:kephyp}.
\\
\begin{table}[h!]
\centering
\resizebox{10cm}{!}{
\begin{tabular}{lcccccccc} 
\hline
& KS-Test (D-values) & AD-Test ($A^{2}$-values)\\
\hline
\hline
Initial values & 0.0656673 & 1.11140\\
10,000 trials, avg & 0.0518119 & 1.15103\\
10,000 trials, sig & 0.0145691 & 0.826723\\
Final distribution & 0.951011$\sigma$ & -0.0479336$\sigma$\\
\hline
\end{tabular}
}
\caption{Comparing the Distributions of the {\it Hypatia} and {\it Kepler} Datasets}
\label{table5.2}
\end{table}

To test the significance of the D-values, we used 3,861 relevant values from {\it Hypatia} and 382 values from {\it Kepler}; we wanted to know what the D-value would be if it was {\it Hypatia} compared with itself (ultimately the test is to see whether two datasets are representative of the same distribution). We performed 10,000 trials, selecting a random subset of 382 {\it Hypatia} values, and did the KS test between the random 382-value subset and the full {\it Hypatia} dataset. This represents our ideal D-values, since we know that the subset comes directly from the full dataset. With these 10,000 D-values, we compute the mean and standard deviation and compare to our original {\it Hypatia}-{\it Kepler} D-values. We calculate $\sigma$ = ($D_{KS}$ - $\left<D_{KS,trial}\right>$)/$\sigma_{D_{KS,trial}}$. Thus, we conclude from these tests that {\it Hypatia} and {\it Kepler} are representative of non-matching distributions at a level of 0.95$\sigma$ (see Table~\ref{table5.2}).

We then applied the Anderson-Darling test \citep{ss87}, given by

\footnotesize
\begin{equation}
A^{2} = \frac{n+m-1}{(n+m)^{2}} (\Sigma_{j} \frac{h_{j}((n+m)F_{j}-nH_{j})^{2}}{H_{j}(n+m-H_{j})-(n+m)h_{j}/4}
+ \frac{1}{m} \Sigma_{j} \frac{h_{j}((n+m)G_{j}-nH_{j})^{2}}{H_{j}(n+m-H_{j})-(n+m)h_{j}/4})
\end{equation}
\normalsize

\noindent where j runs over the Fe values (low-to-high); F is the {\it Hypatia} set; n is the number of values in F; G is the {\it Kepler} set; m is the number of values in G; F$_{j}$ represents the number of {\it Hypatia} values less than j plus 1/2 of those equal to j; G$_{j}$ is the same but with {\it Kepler}; H$_{j}$ is the same but with both datasets combined; and h$_{j}$ is the number of values equal to j from the combined datasets. We compute this value and do 10,000 more trials comparing a {\it Hypatia} subset to itself. $A^{2}$ represents non-matching distributions at a level of -0.0479336$\sigma$.

For the normal standard deviation definition, these values can be translated to mean that the KS-test gives a 34\% chance that these two datasets are representative of the same distribution, and the AD-test gives a 96\% chance. While these two results differ significantly, neither of them can confidently rule out the possibility that the {\it Kepler} and {\it Hypatia} datasets are representative of the same distribution. We conclude from these tests that the {\it Hypatia} catalog is a fairly good representation of the Fe distribution of the {\it Kepler} planet-hosts, and thus we can move forward in comparing these two datasets with our catalog of stellar models.

\subsection{A Bayesian Framework for the Habitable Zone}

We use Bayes' Theorem to create a statistical framework for assessing long-term planetary habitability. Ultimately, we can calculate conditional probabilities based on a mathematical formulation of observables, in order to estimate the habitability potential for a given system. We need two mathematical functions in order to define this framework: a function for the likelihood of a certain outcome, and a distribution of defined priors. The product of these will give us the posterior distribution. Bayes' Theorem is commonly written as

\begin{equation}
P(A|B) = \frac{P(B|A) \times P(A)}{P(B)}
\label{eq:bayes}
\end{equation}

\noindent where $P(A|B)$ is the \textbf{posterior} probability of $A$ occurring given that $B$ is true, $P(B|A)$ is the probability of $B$ occurring given that $A$ is true, $P(A)$ is probability of $A$, and $P(B)$ is the probability of $B$. Given two possible outcomes $A_1$ and $A_2$, the denominator of Eq. (\ref{eq:bayes}) can be rewritten using $P(B) = P(B|A_1) \times P(A_1) + P(B|A_2) \times P(A_2)$. We therefore have

\begin{equation}
P(A_1|B) = \frac{P(B|A_1) \times P(A_1)}{P(B|A_1) \times P(A_1) + P(B|A_2) \times P(A_2)}
\label{eq:bayesX}
\end{equation}

We apply Equation (\ref{eq:bayesX}) to Tycho models and observational distributions of stellar properties to calculate the Bayesian posterior probability that an exoplanet with certain orbital parameters is in the CHZ$_2$. We limit the stellar models to a total age of 12 Gy to take into account the age of the Universe. We perform this analysis for four cases, first using $B = Z$ where $Z$ is the metallicity taken from {\it Hypatia}, then using $B = Z$ where $Z$ is a measured stellar metallicity with an assumed Gaussian error distribution, next using $B = Z, M$ where $Z$ is metallicity from {\it Hypatia}, and mass $M$ is a measured value with an assumed Gaussian error, and finally $B = Z, M$, where both $Z$ and $M$ are measured values with Gaussian errors. 

\subsubsection{Case 1: Using a Generalized Stellar Metallicity Distribution}

Here, we use the metallicity $Z$ as the only prior in our Bayes' equation and assume a fixed mass, so from Equation (\ref{eq:bayesX}) we have $A_1 = \textrm{CHZ}_2$, $A_2 = \textrm{not CHZ}_2$, and $B = Z$. $A_1$ is the possible outcome where a given radius has been in the CHZ for 2 Gy, and $A_2$ is the outcome wherein the radius is not in the CHZ$_2$. Equation (\ref{eq:bayesX}) therefore becomes

\begin{equation}
P(\textrm{CHZ}_2|Z) = \frac{P(Z|\textrm{CHZ}_2)P(\textrm{CHZ}_2)}{P(Z|\textrm{CHZ}_2)P(\textrm{CHZ}_2)+P(Z|\textrm{not CHZ}_2)P(\textrm{not CHZ}_2)}
\label{eq:bayes2}
\end{equation}

To obtain the observational $Z$ probabilities, we use the distribution from the {\it Hypatia} catalog of nearby stars. For $Z$ we use the iron abundance $[Fe/H]$ values, where $Z/Z\sol$ = 10$^{[Fe/H]}$. For each calculation we use a fixed mass $M_j$ where the index $j$ refers to the specific mass model used, and only the Tycho model outputs for that mass, linearly interpolated if necessary. We describe how we compute each component of Equation (\ref{eq:bayes2}) for a desired metallicity $Z_j$ and the fixed mass $M_j$, for any given radius from the star. We compute $P(\textrm{CHZ}_2)$, the probability that a given radius around a star of mass $M_j$ is in the CHZ$_2$, by

\begin{equation}
P(\textrm{CHZ}_2) = \frac{\sum_i t_{\textrm{CHZ}_2,i} \Delta Z_i}{\sum_i t_{\textrm{tot},i} \Delta Z_i}
\label{eq:bayesterm1}
\end{equation}

\noindent where the index $i$ runs through the Tycho model metallicities from $Z$ = $0.1-1.5$ $Z\sol$ , $t_{\textrm{CHZ}_2,i}$ is the total time the radius is in the CHZ$_2$ for a given ($Z_i$, $M_j$) Tycho model, $t_{\textrm{tot},i}$ is total lifetime (with a maximum of 12 Gy) for the given ($Z_i$, $M_j$) Tycho model, and $\Delta Z_i$ is $Z_{i+1}-Z_i$, the distance between $Z$ values used. The value of $P(\textrm{not CHZ}_2)$ is given by

\begin{equation}
P(\textrm{not CHZ}_2) = 1 - P(\textrm{CHZ}_2)
\label{eq:bayesterm2}
\end{equation}

\noindent and the probability that a star has metallicity $Z_j$ if a given radius is in the CHZ$_2$ is 

\begin{equation}
P(Z_j|\textrm{CHZ}_2) = \frac{t_{\textrm{CHZ}_2,j} \times P(Z_j)}{\sum_i (t_{\textrm{CHZ}_2,i} \times P(Z_i))}
\label{eq:bayesterm3}
\end{equation}

\noindent $P(Z)$ is a Gaussian fit to the fraction of stars in {\it Hypatia} with metallicity $Z$, given by

\begin{equation}
P(Z_i) = 0.0937309 \times e^{-((Z_i-0.927451)/0.498708)^2/2}
\label{eq:HypGauss}
\end{equation}

\noindent We similarly compute $P(Z_j|\textrm{not CHZ}_2)$, the probability that a star has metallicity $Z_j$ if a given radius around the star is not in the CHZ$_2$, by

\begin{equation}
P(Z_j|\textrm{not CHZ}_2) = \frac{(t_{\textrm{tot},j}-t_{\textrm{CHZ}_2,j}) \times P(Z_j)}{\sum_i ((t_{\textrm{tot},i}-t_{\textrm{CHZ}_2,i}) \times P(Z_i))}
\label{eq:bayesterm4}
\end{equation}

When calculating Bayesian posterior probability from Eq. (\ref{eq:bayes2}), we combine Equations (\ref{eq:bayesterm1}), (\ref{eq:bayesterm2}), (\ref{eq:bayesterm3}), and (\ref{eq:bayesterm4}) to find $P(\textrm{CHZ}_2|Z_j)$. With a factor of $P(Z_j)$ cancelling out, we find

\small
\begin{equation}
P(\textrm{CHZ}_2|Z_j) = \frac{\frac{t_{\textrm{CHZ}_2,j}}{\sum_i (t_{\textrm{CHZ}_2,i} P(Z_i))} \frac{\sum_i t_{\textrm{CHZ}_2,i}}{\sum_i t_{\textrm{tot},i}} }{ \frac{t_{\textrm{CHZ}_2,j}}{\sum_i (t_{\textrm{CHZ}_2,i} P(Z_i))} \frac{\sum_i t_{\textrm{CHZ}_2,i}}{\sum_i t_{\textrm{tot},i}} + \frac{t_{\textrm{tot},j}-t_{\textrm{CHZ}_2,j}}{\sum_i ((t_{\textrm{tot},i}-t_{\textrm{CHZ}_2,i}) P(Z_i))} \left(1 - \frac{\sum_i t_{\textrm{CHZ}_2,i}}{\sum_i t_{\textrm{tot},i}}\right) }
\label{eq:Pfin}
\end{equation}
\normalsize

\noindent We now have a Bayesian posterior probability at each radius for a given $Z$ and $M$. We can marginalize over $Z$ in $P(\textrm{CHZ}_2|Z)$ for a given mass and make a contour plot of all masses, shown in Figures~\ref{fig:M1_Zall}-\ref{fig:Pcontour}, and additionally over-plot exoplanets with known stellar masses and orbital distances from the {\it Kepler} exoplanet archive, shown in the right panel of Figure~\ref{fig:Pcontour}. As an exercise, we can also apply this process to a few known planets, with the results summarized in Table~\ref{tab:planettab} and shown in Figures~\ref{fig:planets-contour} and \ref{fig:ZPplanets}.

\subsubsection{Case 2: We Have a Stellar Metallicity Measurement}

If we know the $Z$ measurement in the form of Gaussian errors, as the $Z' \pm \sigma_Z$ generally quoted in abundance surveys, we can use this as a more accurate representation of the term $P(Z)$ by using the Gaussian probability
\begin{equation}
P(Z) = \frac{1}{\sqrt{2 \pi}} e^{\frac{1}{2} \frac{\Delta' Z^2}{\sigma_Z^2}}
\label{eq:pz}
\end{equation}

\noindent where $\Delta' Z$ is defined as $Z' - Z$. This $P(Z)$ is used in Equation (\ref{eq:Pfin}). We apply this method to our planets of interest, with results in Table~\ref{tab:planettab} and Figure~\ref{fig:ZPplanetsMeasure}. For the Sun-related calculations that use measurements and corresponding errors for the $M$ and $Z$-values, we have simply assumed very small errors of 0.001 for both $M$ (M/M\sol) and $Z$ (Z/Z\sol).

\subsubsection{Case 3: Mass Measurement But No Metallicity Measurement}

If we know the mass measurement in the form of Gaussian errors, like $M' \pm \sigma_M$, we can, like before, use this as a representation of the term $P(M)$ by using the Gaussian probability

\begin{equation}
P(M) = \frac{1}{\sqrt{2 \pi}} e^{\frac{1}{2} \frac{\Delta' M^2}{\sigma_M^2}}
\label{eq:pm}
\end{equation}

\noindent where $\Delta' M$ is defined as $M' - M$. This $P(M)$ is then used in Equation (\ref{eq:PfinZM}). If we use mass as another known property in our Bayes' equation, and if we also assume that the stellar mass and metallicity values are independent, Equation (\ref{eq:Pfin}) becomes

\footnotesize
\begin{equation}
P(\textrm{CHZ}_2|Z_j,M_j) = \frac{\frac{t_{\textrm{CHZ}_2,j}}{\sum_i (t_{\textrm{CHZ}_2,i} P(Z_i) P(M_i))} \frac{\sum_i t_{\textrm{CHZ}_2,i}}{\sum_i t_{\textrm{tot},i}} }{ \frac{t_{\textrm{CHZ}_2,j}}{\sum_i (t_{\textrm{CHZ}_2,i} P(Z_i) P(M_i))} \frac{\sum_i t_{\textrm{CHZ}_2,i}}{\sum_i t_{\textrm{tot},i}} + \frac{t_{\textrm{tot},j}-t_{\textrm{CHZ}_2,j}}{\sum_i ((t_{\textrm{tot},i}-t_{\textrm{CHZ}_2,i}) P(Z_i) P(M_i))} \left(1 - \frac{\sum_i t_{\textrm{CHZ}_2,i}}{\sum_i t_{\textrm{tot},i}}\right) }
\label{eq:PfinZM}
\end{equation}
\normalsize

\noindent For $P(Z)$ we use the Gaussian fit of the fraction of stars in {\it Hypatia} with the metallicity $Z$ given by Equation (\ref{eq:HypGauss}). We apply this method to the same planets of interest as previously discussed, with the results summarized in Table~\ref{tab:planettab} and shown in Figure~\ref{fig:ZMPplanetsHypatia}.

\subsubsection{Case 4: Mass Measurement and a Metallicity Measurement}

If we know both the mass and metallicity measurements for a star, we can use Equations (\ref{eq:pz}) and (\ref{eq:pm}) along with Equation (\ref{eq:PfinZM}). We can also apply this method to the same planets as discussed previously, with results summarized in Table~\ref{tab:planettab} and in Figure~\ref{fig:ZMPplanetsMeasure}.

\begin{table}[t]
\begin{center}
\resizebox{16.5cm}{!}{
\begin{tabular}{|c|c|c|c|c|c|c|c|c|c|c|} \hline
Planet & Distance & mass & $\sigma_M$ & metallicity & $\sigma_Z$ & radius & $\left<P(\textrm{CHZ}_2|Z)\right>$ & $\left<P(\textrm{CHZ}_2|Z)\right>$ & $\left<P(\textrm{CHZ}_2|Z,M)\right>$ & $\left<P(\textrm{CHZ}_2|Z,M)\right>$ \\ 
 & (ly) & ($M_\odot$) & (+/-) & ($Z/Z\sol$) & (+/-) & (AU) & ({\it Hypatia}-$Z$) & (measure-$Z$) & ({\it Hypatia}-$Z$) & (measure-$Z$) \\ 
 & & & & & & & Case 1 & Case 2 & Case 3 & Case 4 \\ \hline
Venus & - & 1.0 & 0.001 & 1.0 & 0.001 & 0.72 & 0.000 & 0.000 & 0.000 & 0.000  \\
Earth & - & 1.0 & 0.001 & 1.0 & 0.001 & 1.00 & 0.355 & 0.322 & 0.380 & 0.322 \\
Mars & - & 1.0 & 0.001 & 1.0 & 0.001 & 1.52 & 0.802 & 0.791 & 0.810 & 0.791 \\ \hline
Tau Ceti e & 12 & 0.78 & 0.012 & 0.28 & 0.05 & 0.52-0.58 & 0.000 & 0.000 & 0.000 & 0.000 \\
Tau Ceti f & 12 & 0.78 & 0.012 & 0.28 & 0.05 & 1.29-1.35 & 0.064 & 0.054 & 0.194 & 0.370 \\
Kepler-186f & 561 & 0.54 & 0.02 & 0.55 & 0.12 & 0.42-0.44 & 0.837 & 0.835 & 0.831 & 0.835 \\
Kepler-62f & 1200 & 0.69 & 0.02 & 0.43 & 0.04 & 0.71-0.73 & 0.529 & 0.787 & 0.780 & 0.834 \\ \hline
 \end{tabular} }
\end{center}
\caption{Results of Cases 1-4 for several Solar system planets and exoplanets of interest. The radius ranges indicated here are representative of $\pm 1 \sigma$ semi-major axis orbits, and $\left<P(\textrm{CHZ}_2|X)\right>$ is a marginalization of $P(\textrm{CHZ}_2|X)$ over the entire radius span. \vspace{2mm}}
\label{tab:planettab}
\end{table}

\section{Results \label{results}}

\begin{figure}
\centerline{\includegraphics[height=7cm,width=8cm]{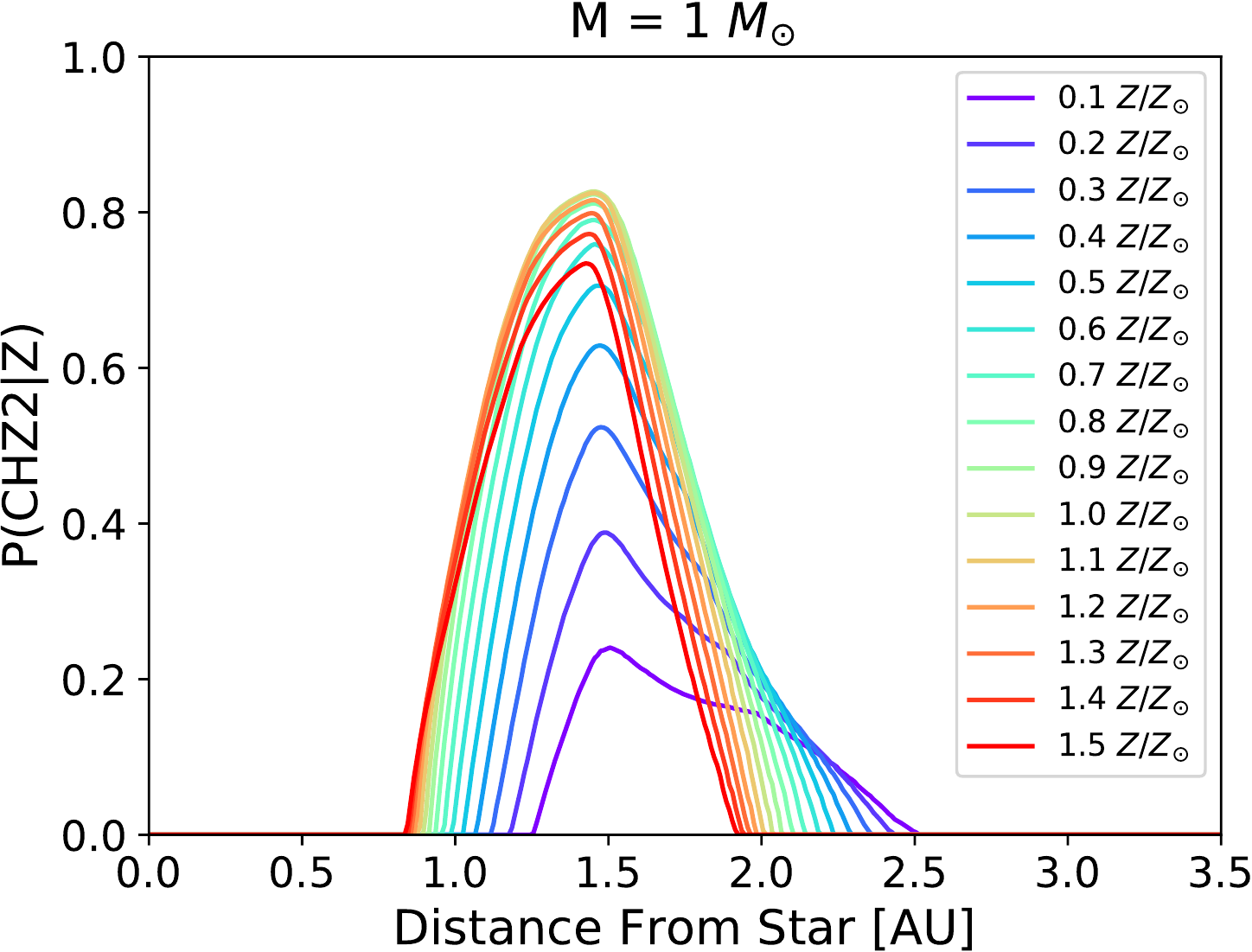}
\includegraphics[height=7cm,width=8cm]{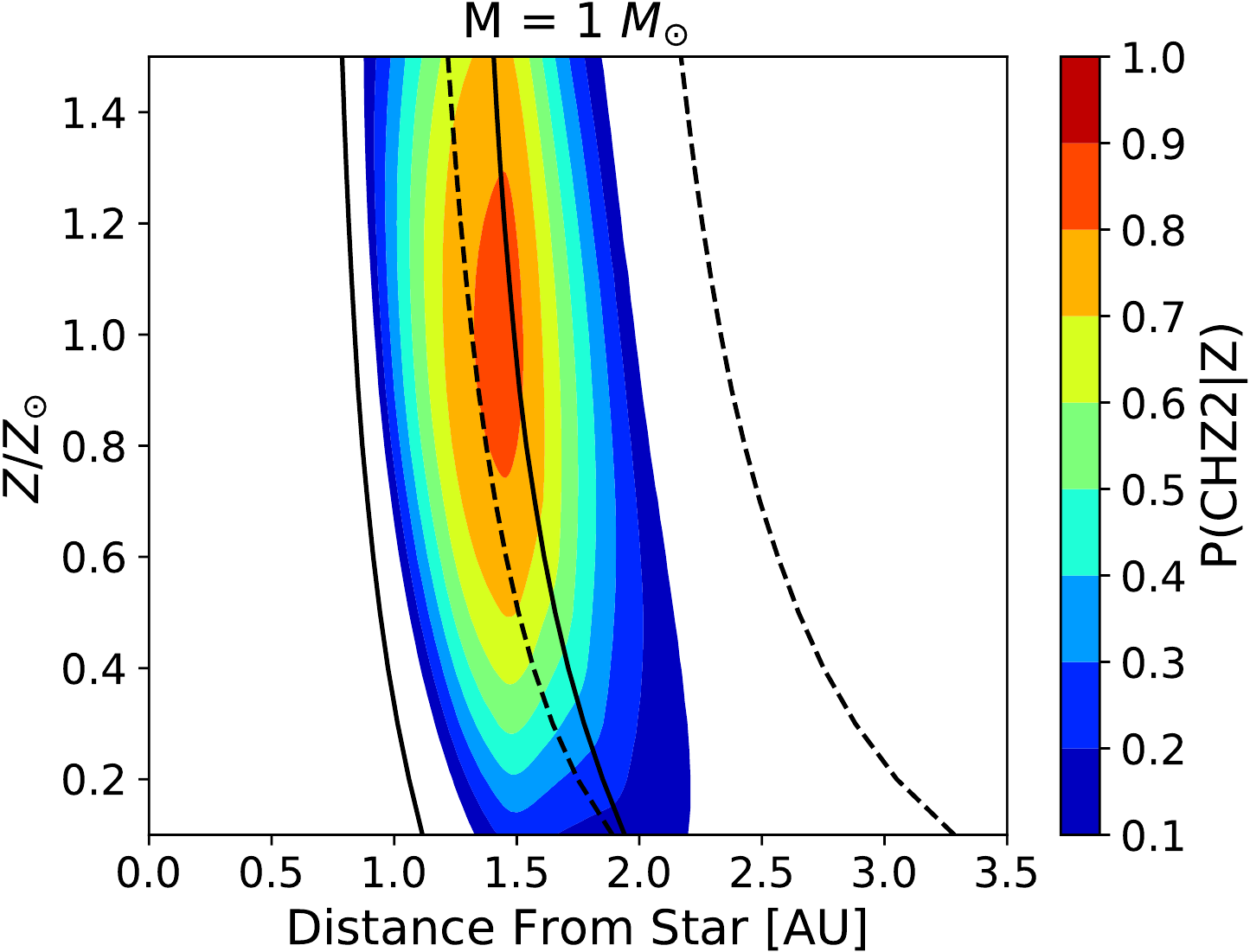}}
\caption{Left: Bayesian posterior probability distributions for each relevant [Fe/H] value in the {\it Hypatia} catalog, converted to $Z$-value ($Z/Z\sol$), assuming a fixed mass of 1 $M\sol$. Red lines indicate a higher metallicity, whereas purple lines indicate lower metallicity (as expected, they show a more distant habitable zone and a lower posterior probability of long-term habitability). Right: the same posterior probability distribution mapped to a contour plot. The posterior probability values at each $Z$-value for any particular orbital distance from the star are given by the color bar at the top. The inner and outer habitable zone limits at ZAMS and TAMS are also included (solid is ZAMS inner/outer, dashed is TAMS inner/outer). Note the overlap, meaning there are a significant subset of orbits that are continuously habitable for the entire MS lifetime. Scalloped edges in the contours noticeable at the outer limits of the habitable zone are an artifact of finite mass resolution in the models. \label{fig:M1_Zall}}
\end{figure}

\begin{figure}[t]
\centerline{\includegraphics[height=2.3cm]{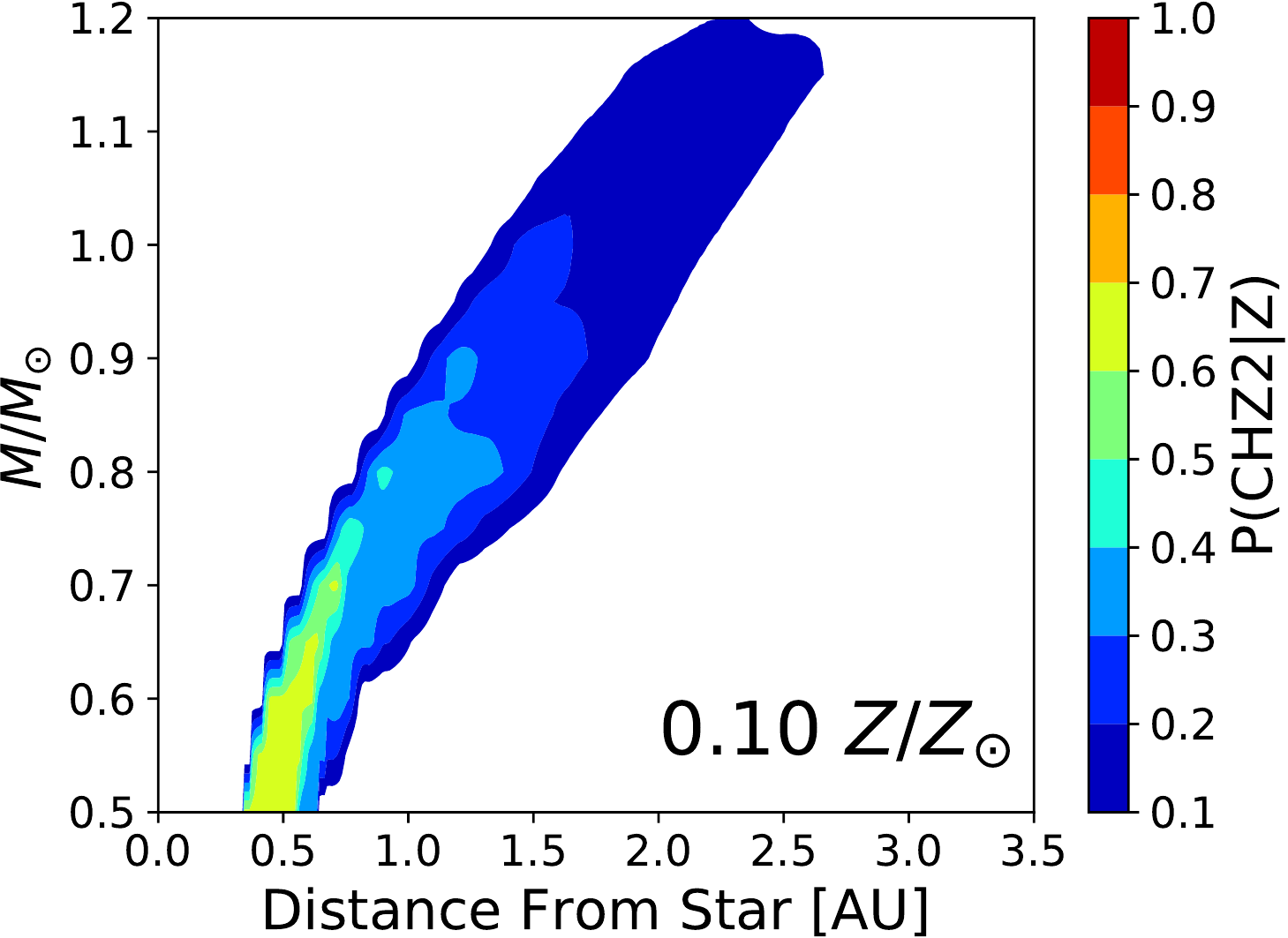}
\includegraphics[height=2.3cm]{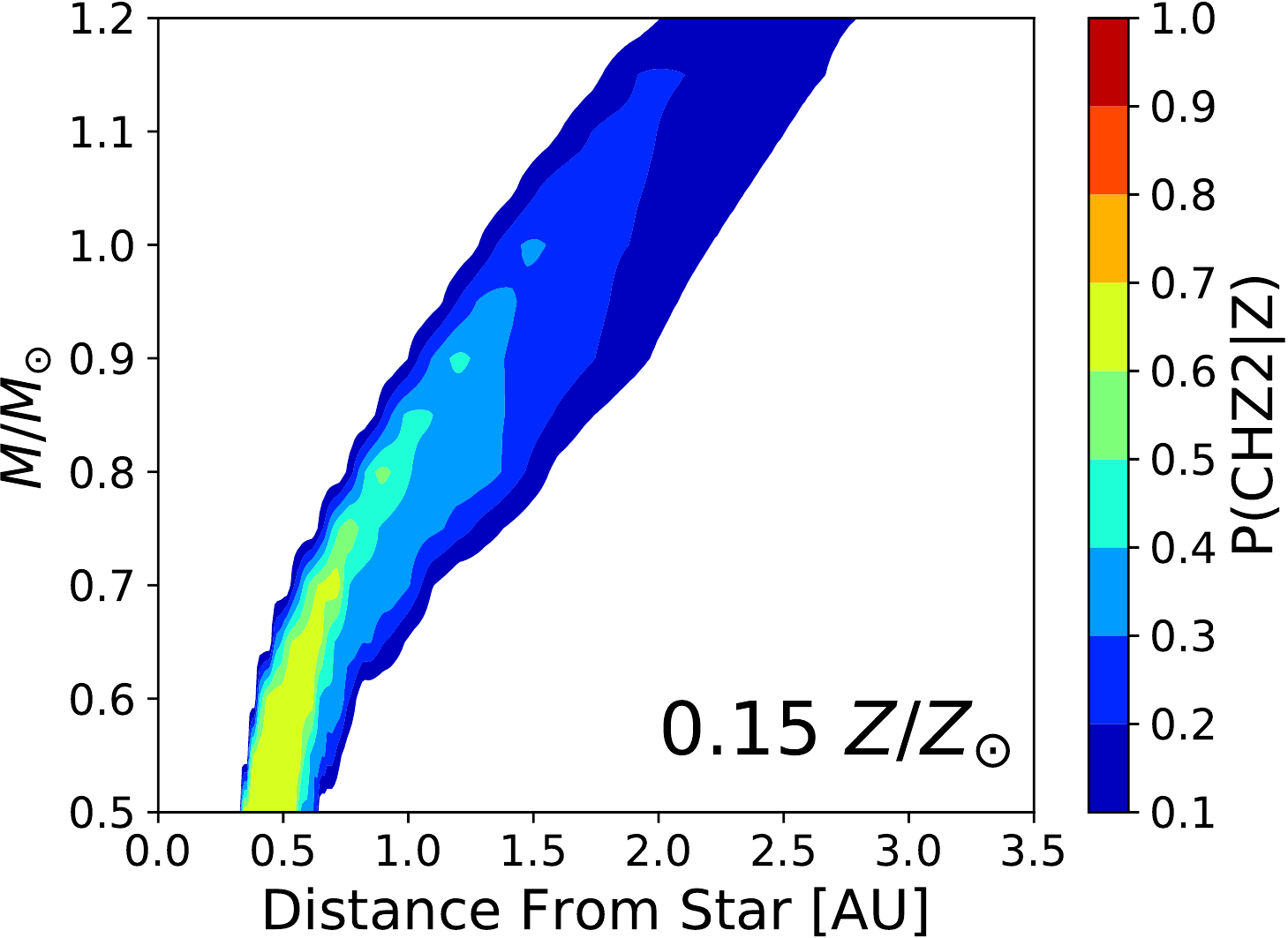}
\includegraphics[height=2.3cm]{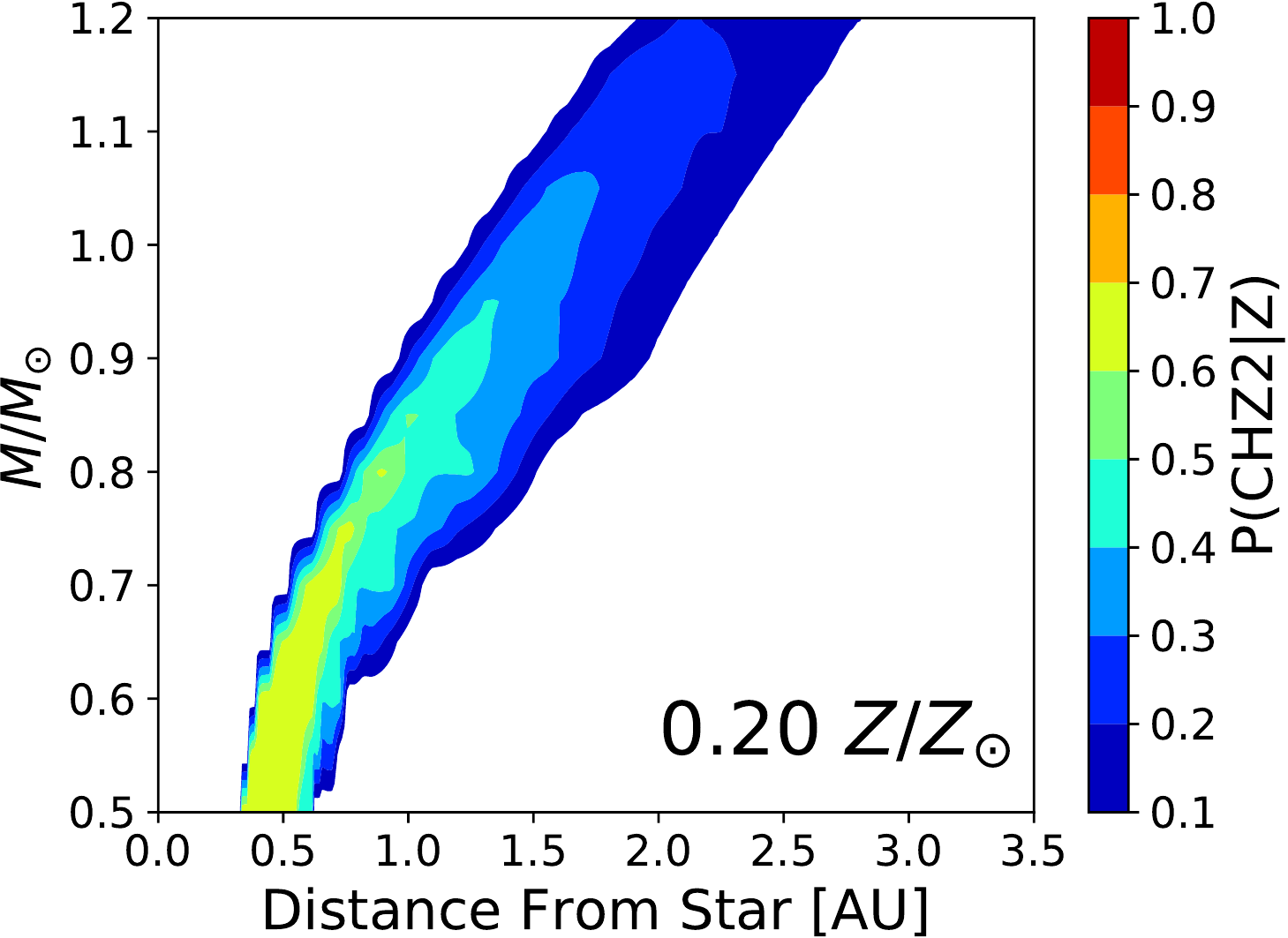}
\includegraphics[height=2.3cm]{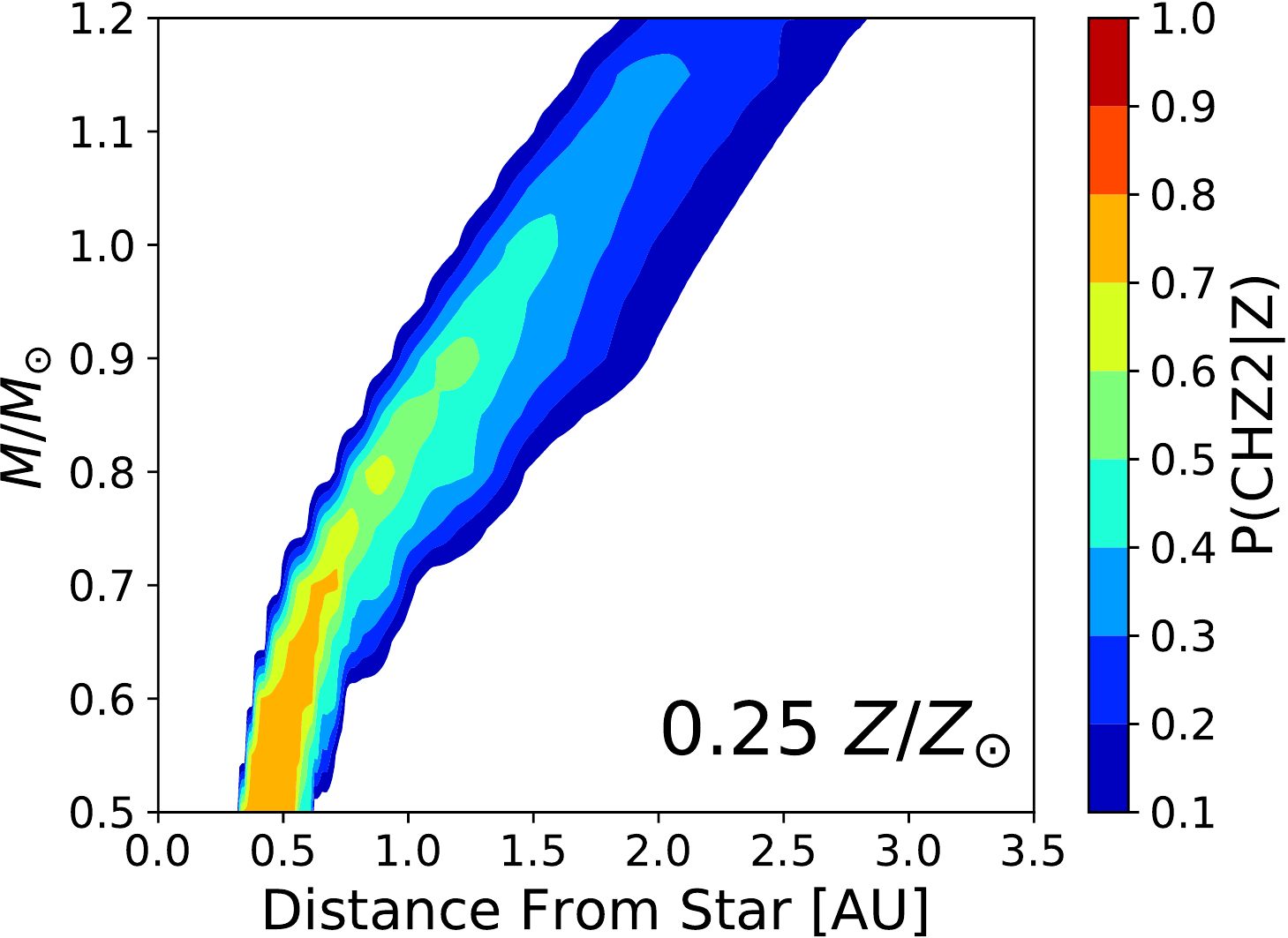}
\includegraphics[height=2.3cm]{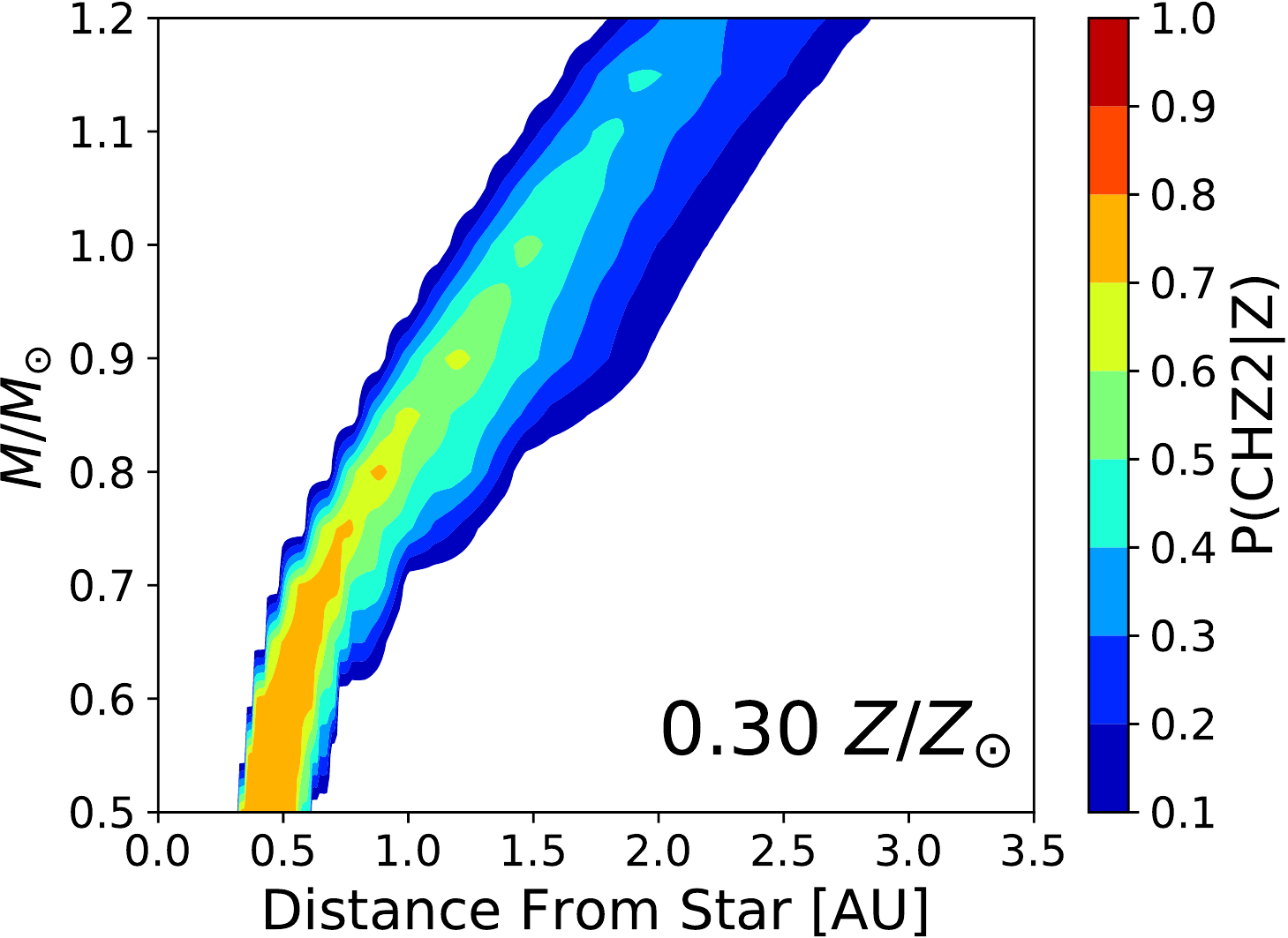}}
\centerline{\includegraphics[height=2.3cm]{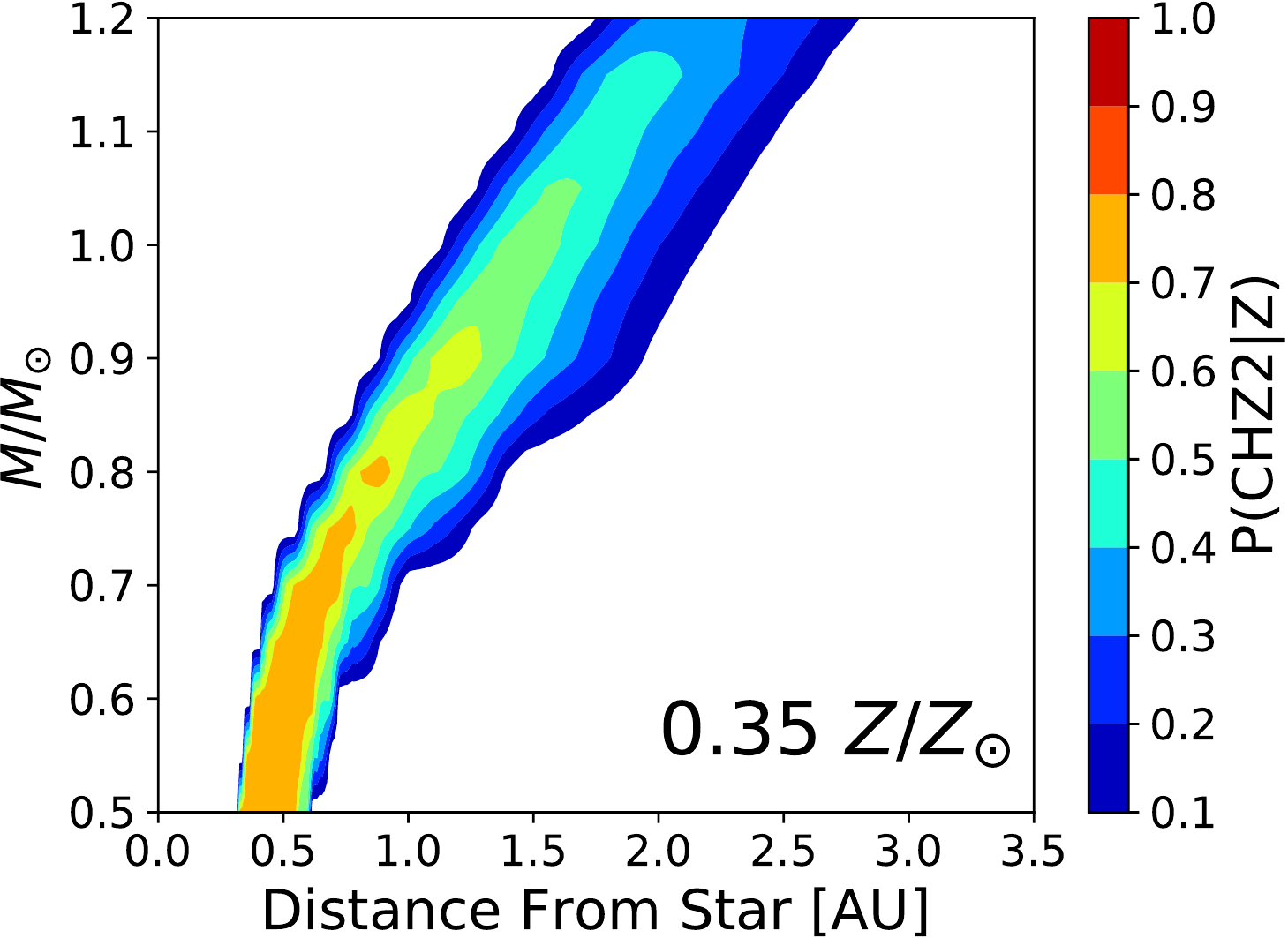}
\includegraphics[height=2.3cm]{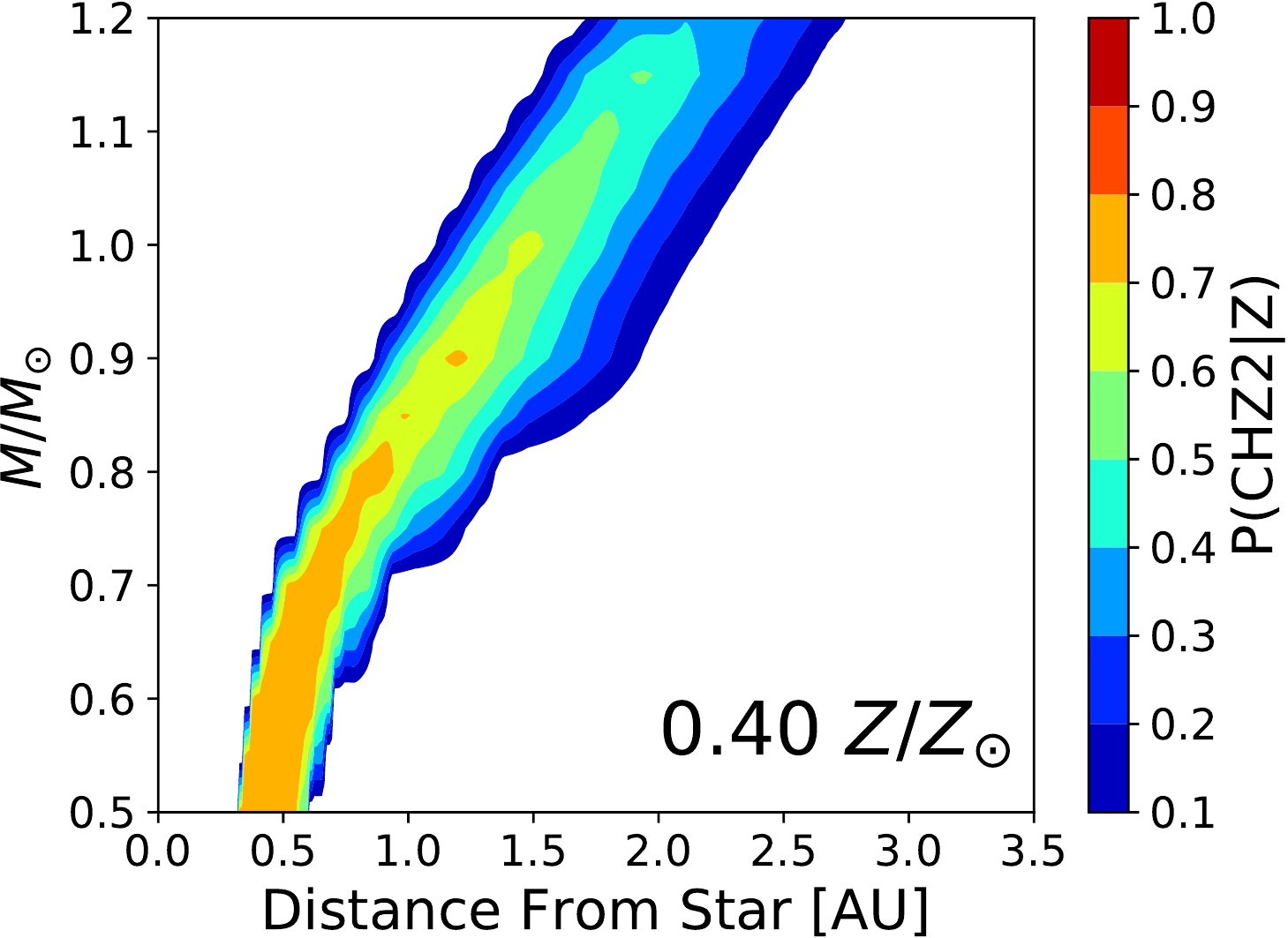}
\includegraphics[height=2.3cm]{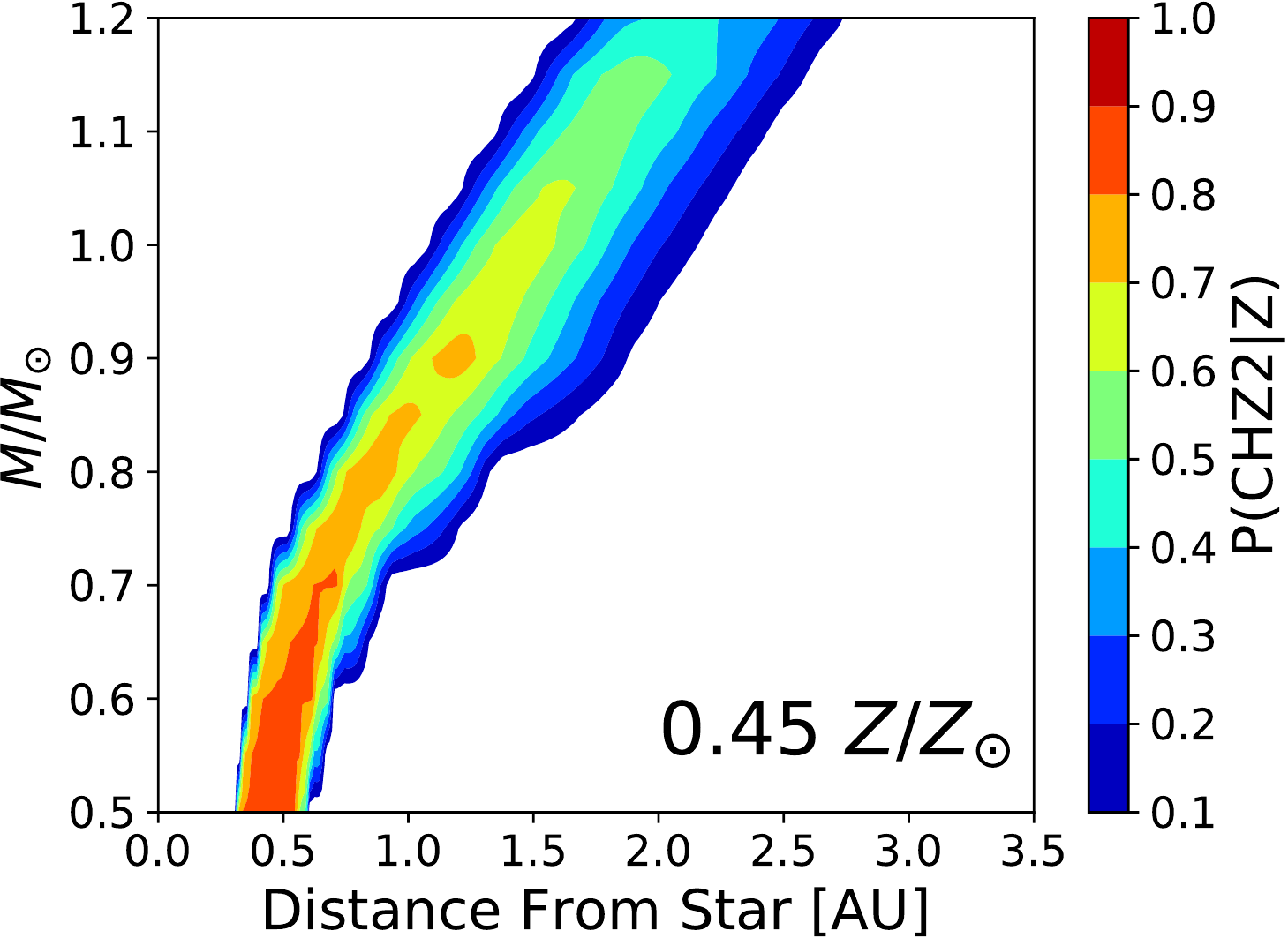}
\includegraphics[height=2.3cm]{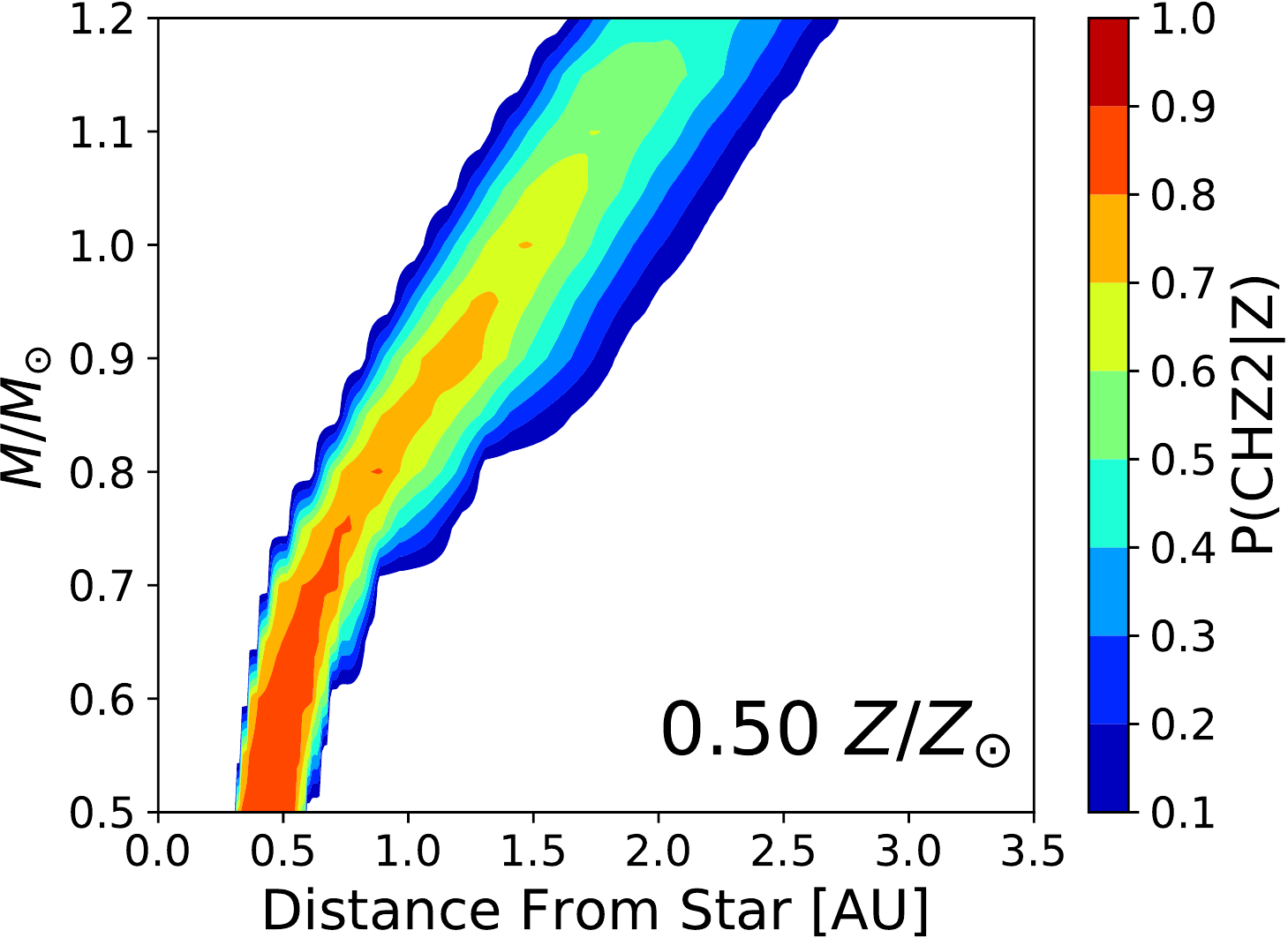}
\includegraphics[height=2.3cm]{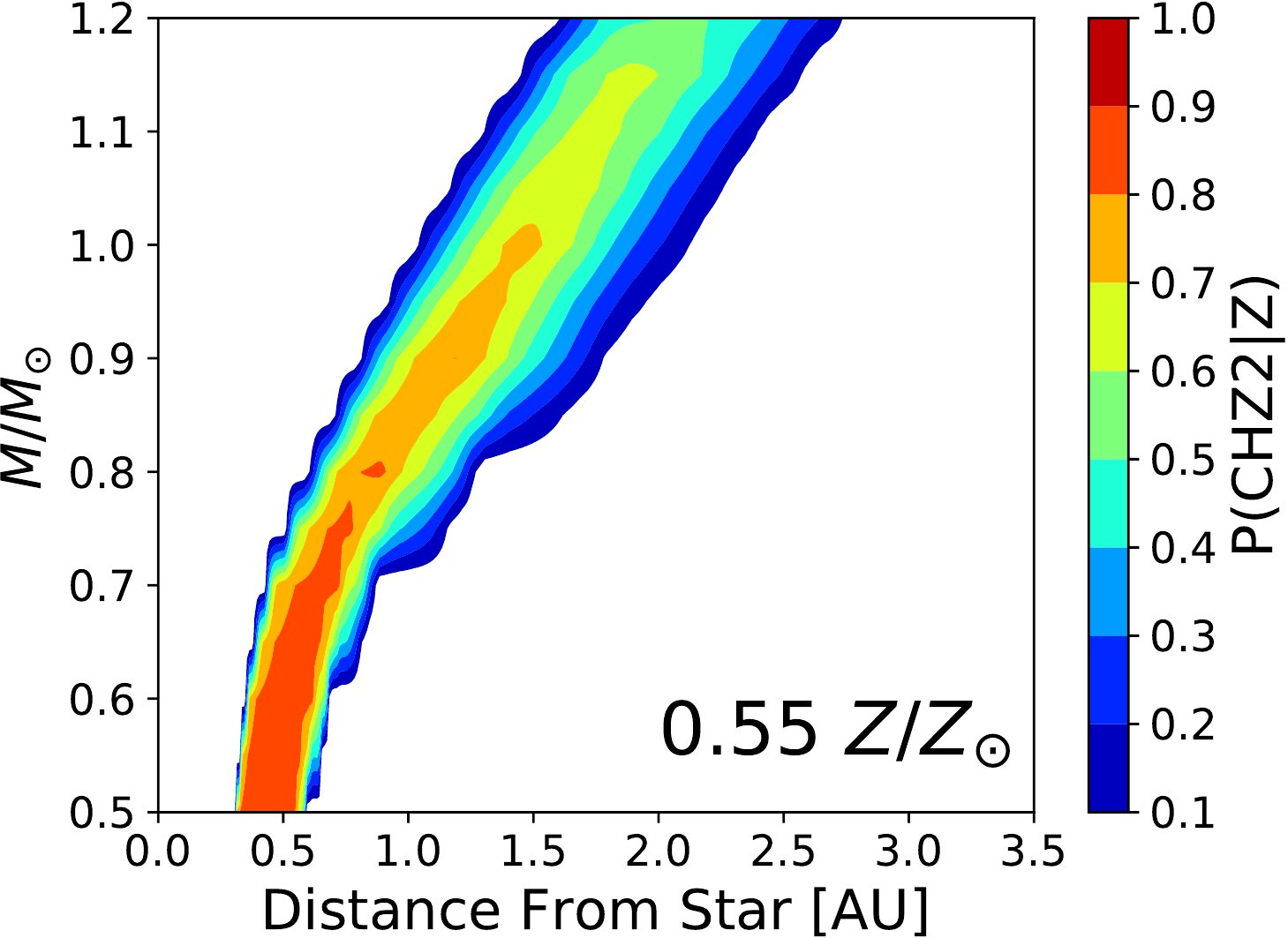}}
\centerline{\includegraphics[height=2.3cm]{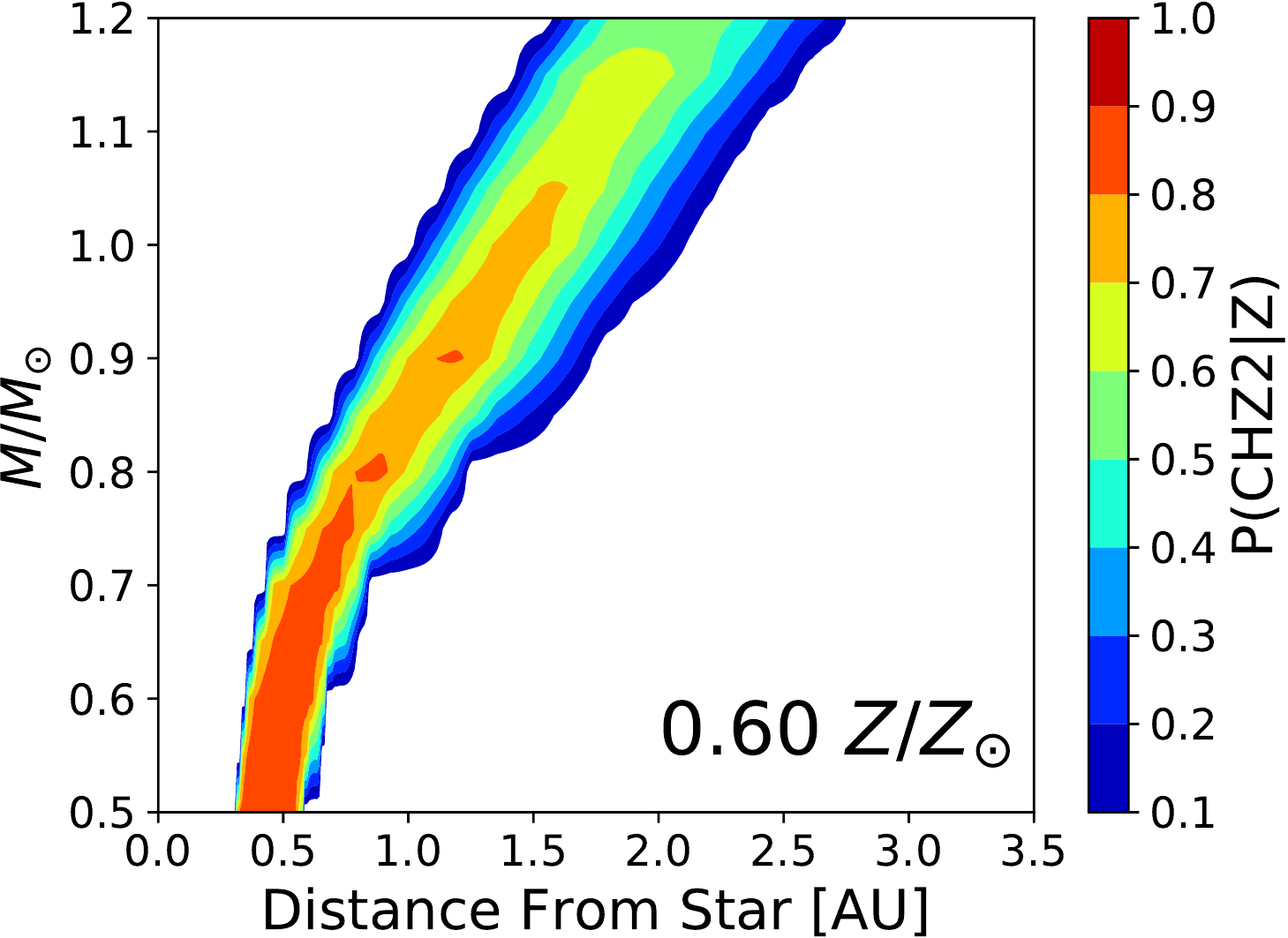}
\includegraphics[height=2.3cm]{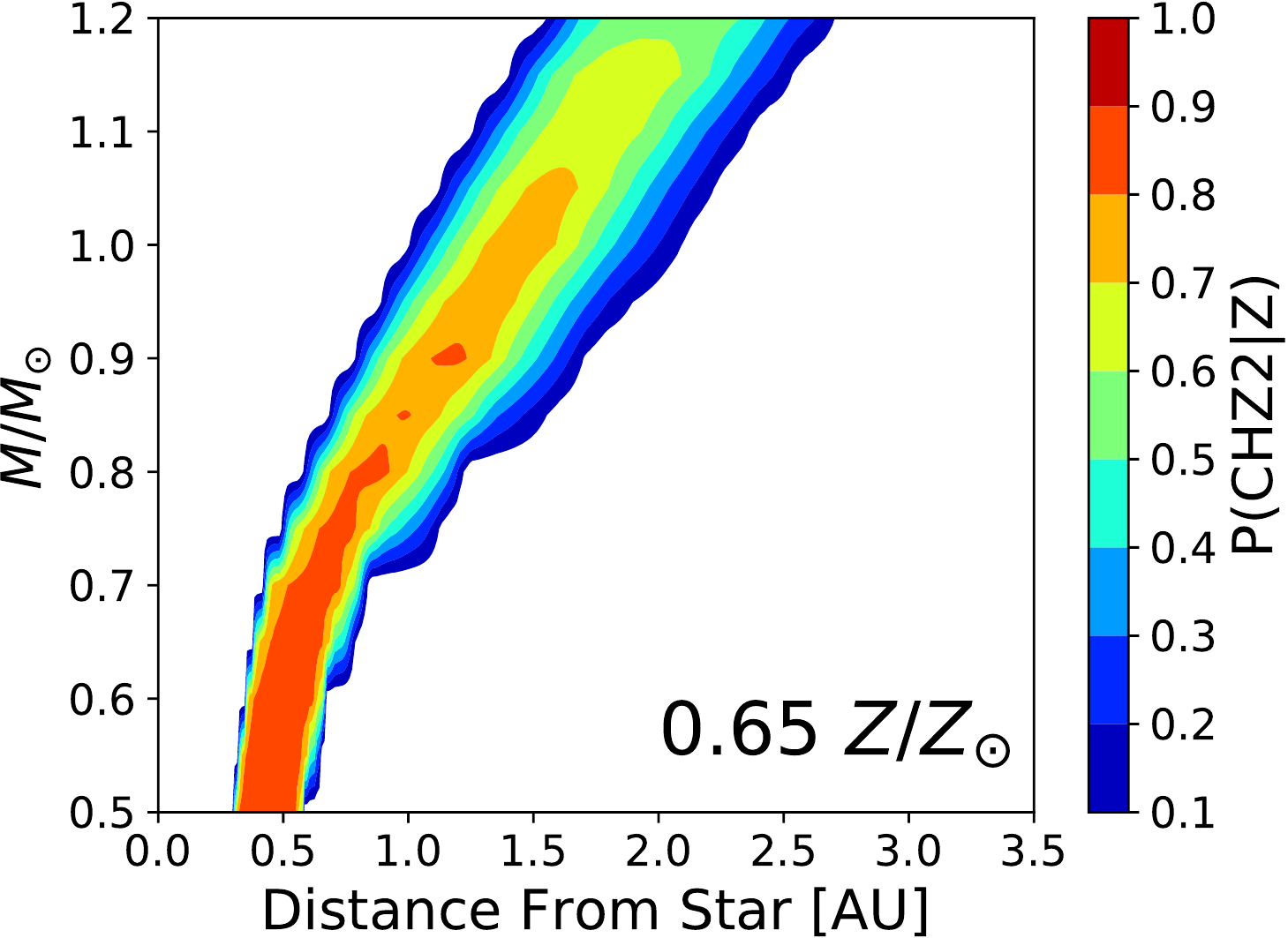}
\includegraphics[height=2.3cm]{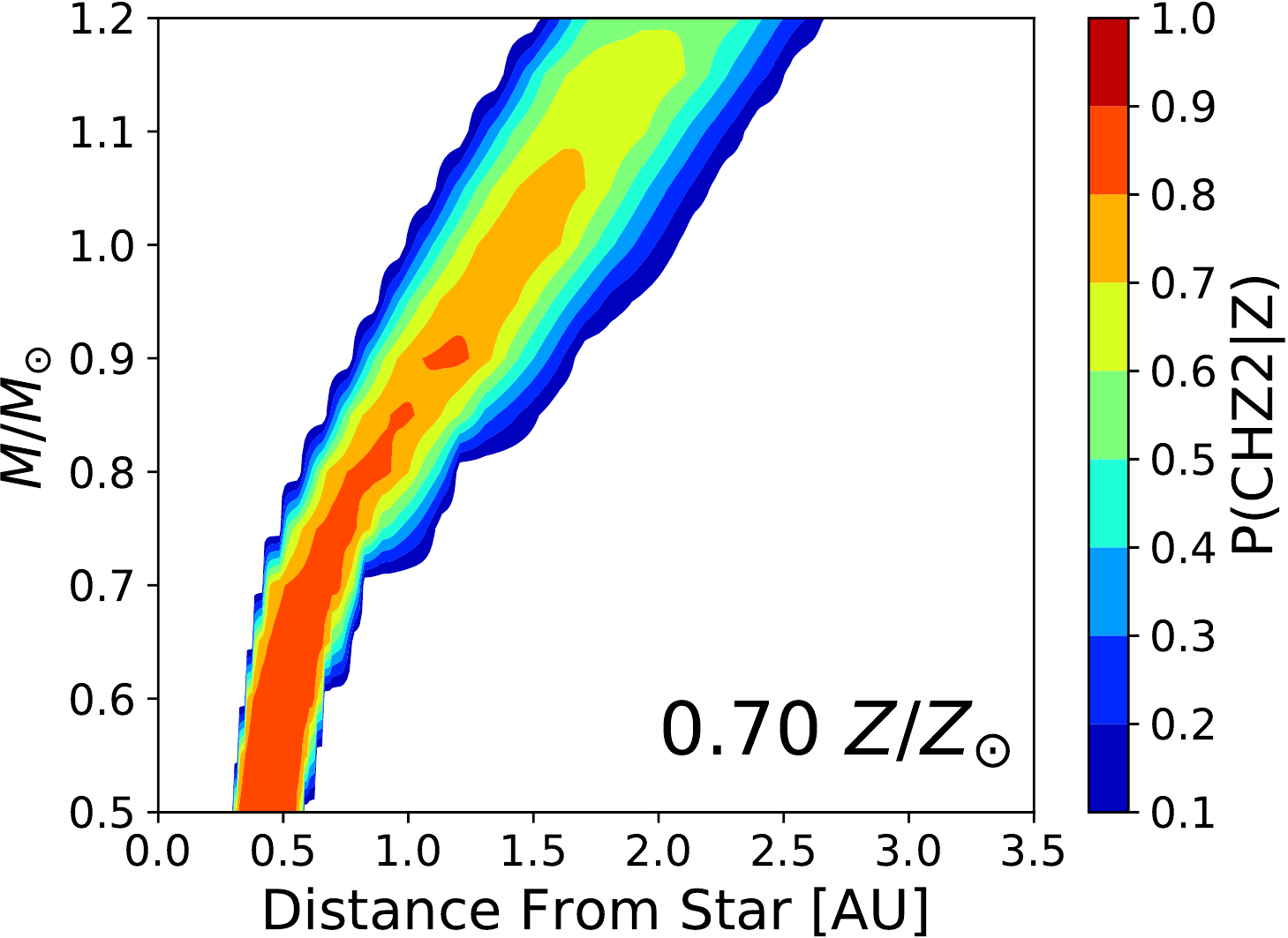}
\includegraphics[height=2.3cm]{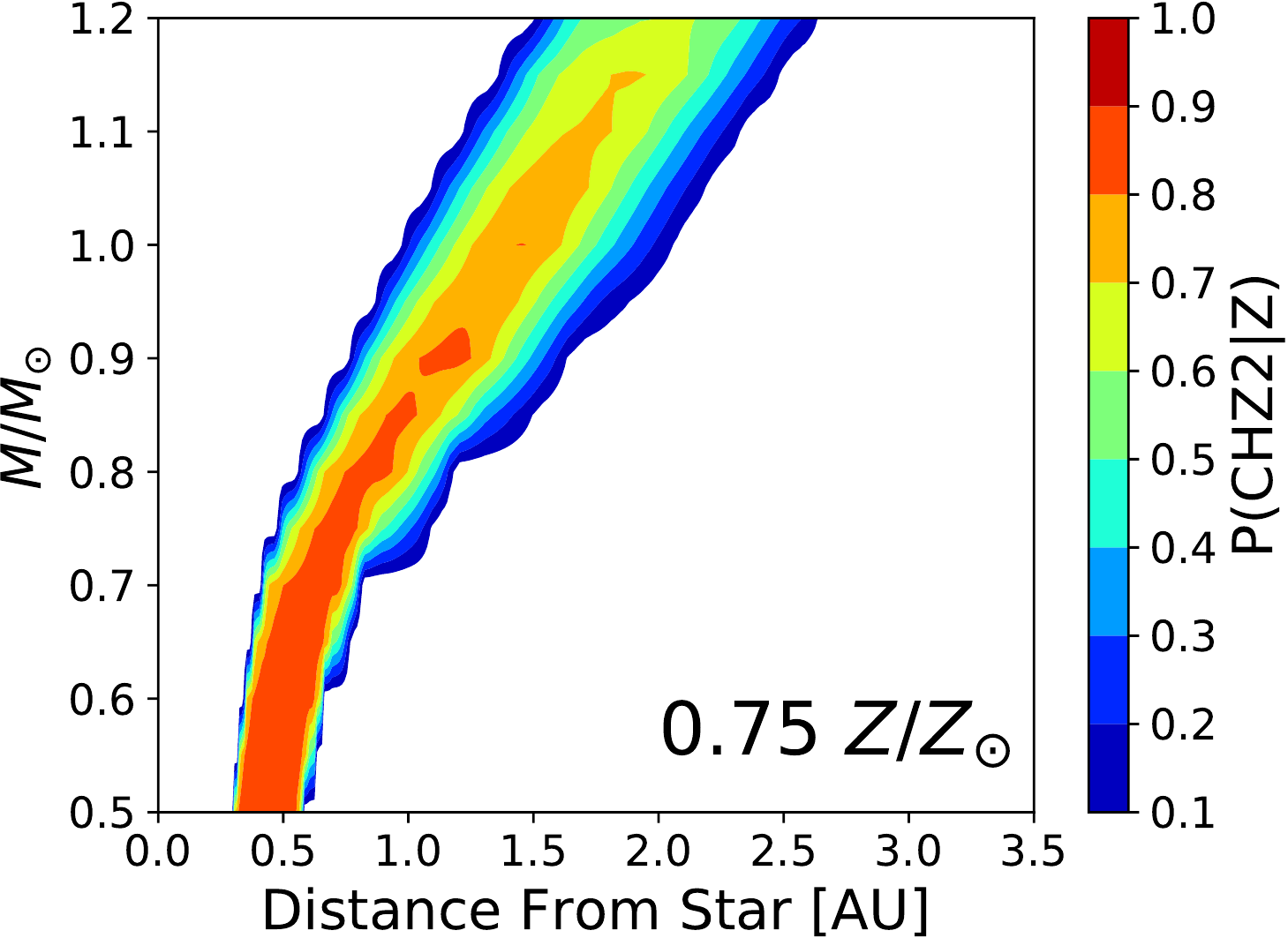}
\includegraphics[height=2.3cm]{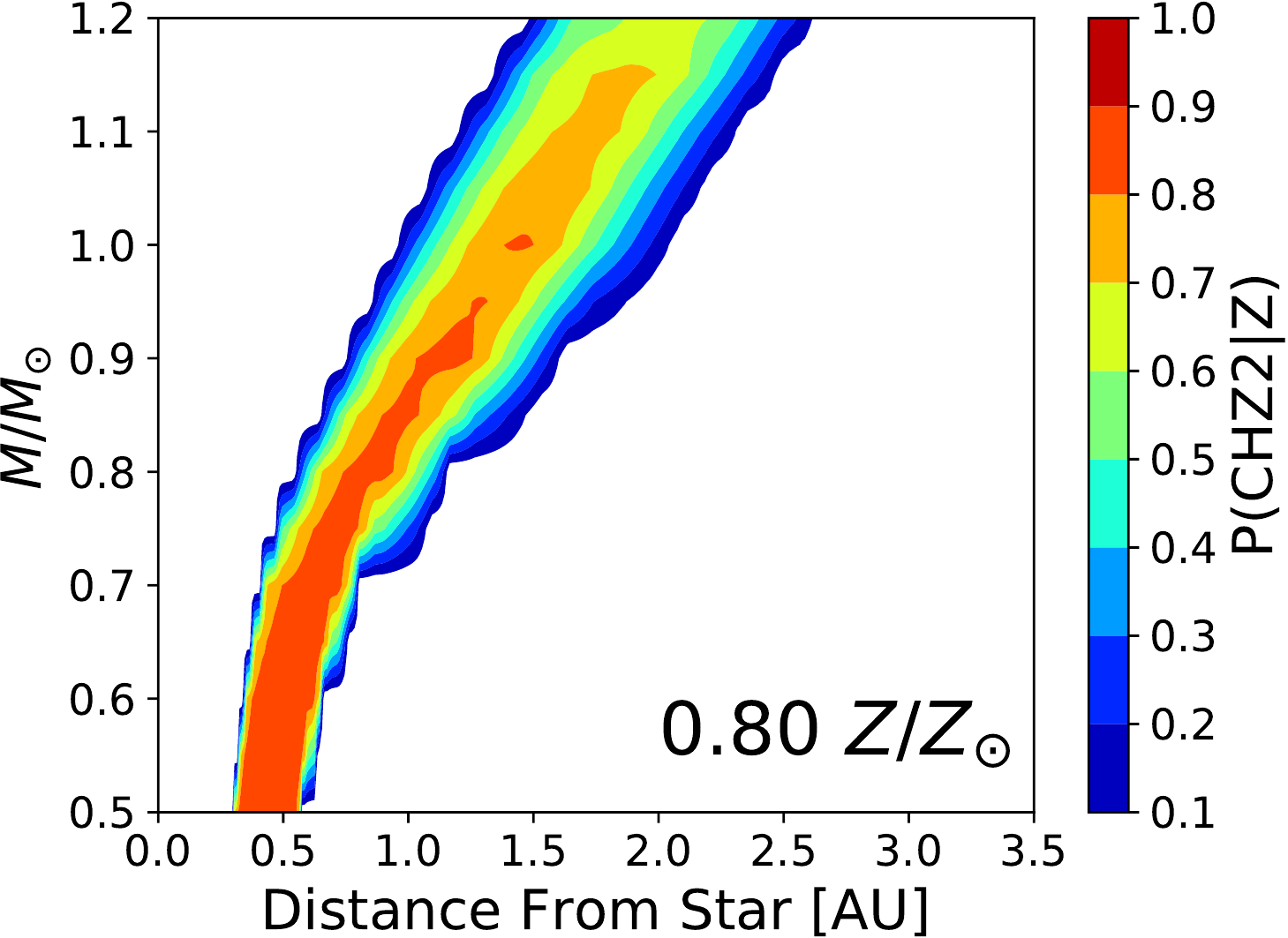}}
\centerline{\includegraphics[height=2.3cm]{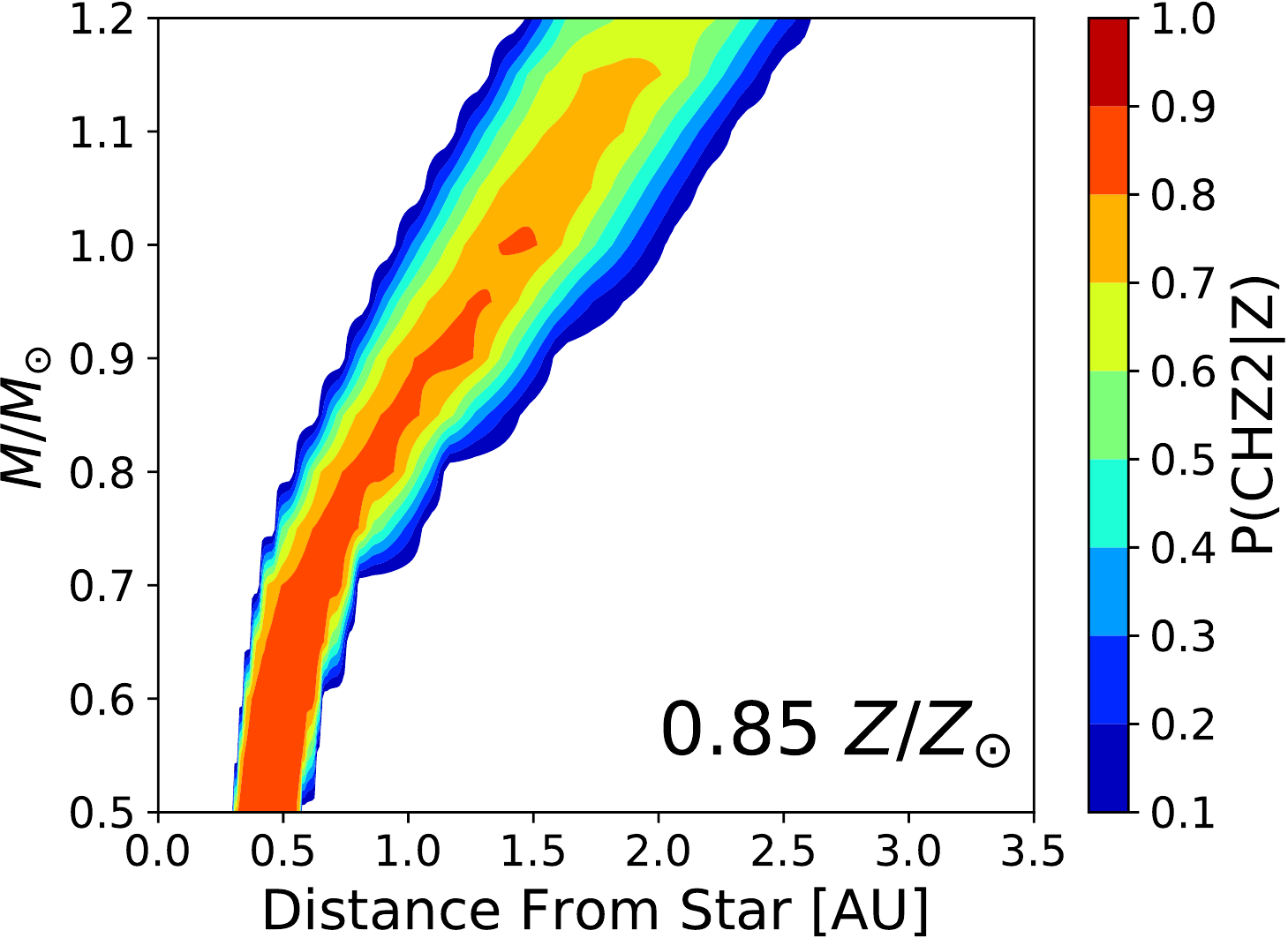}
\includegraphics[height=2.3cm]{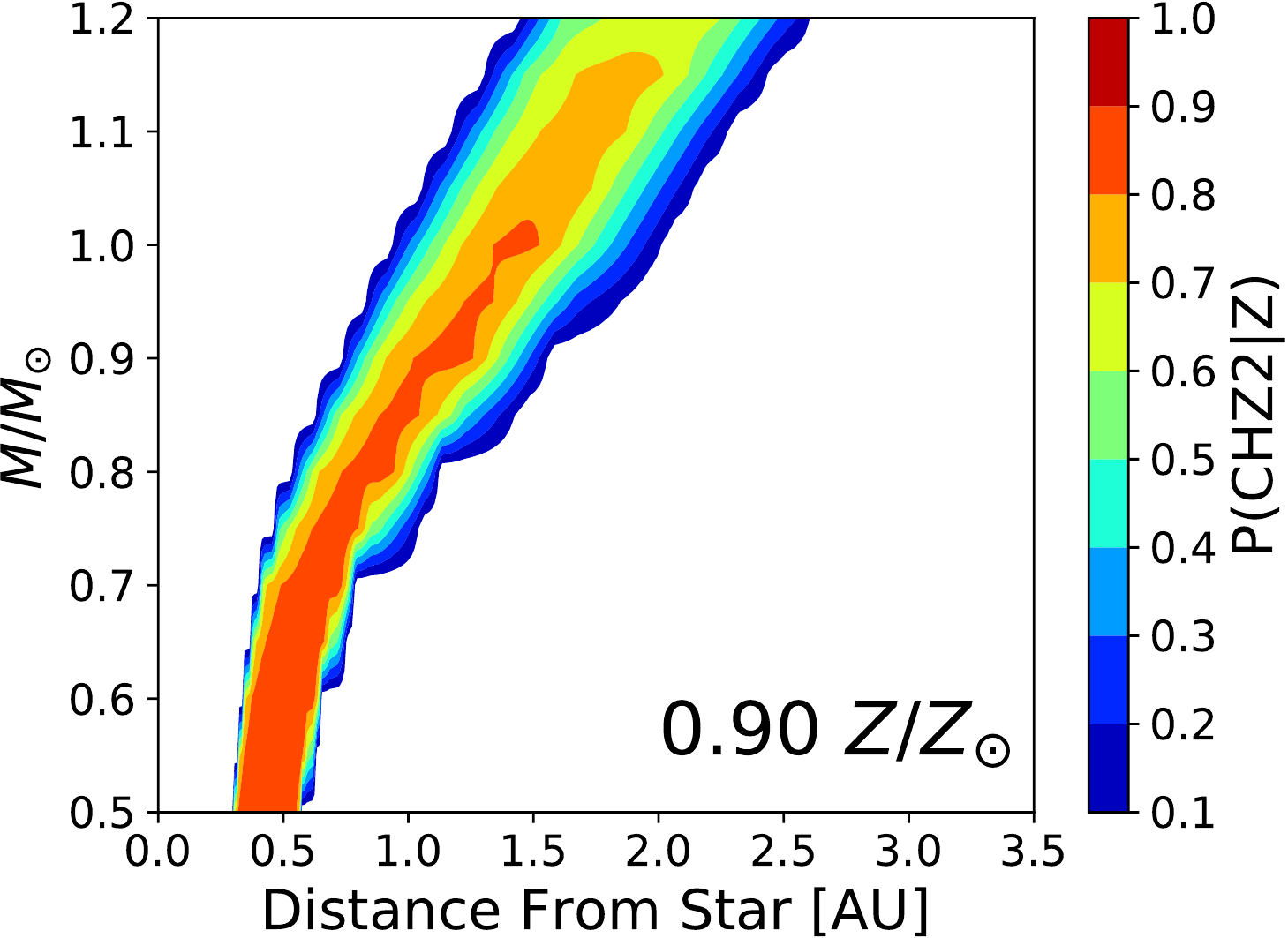}
\includegraphics[height=2.3cm]{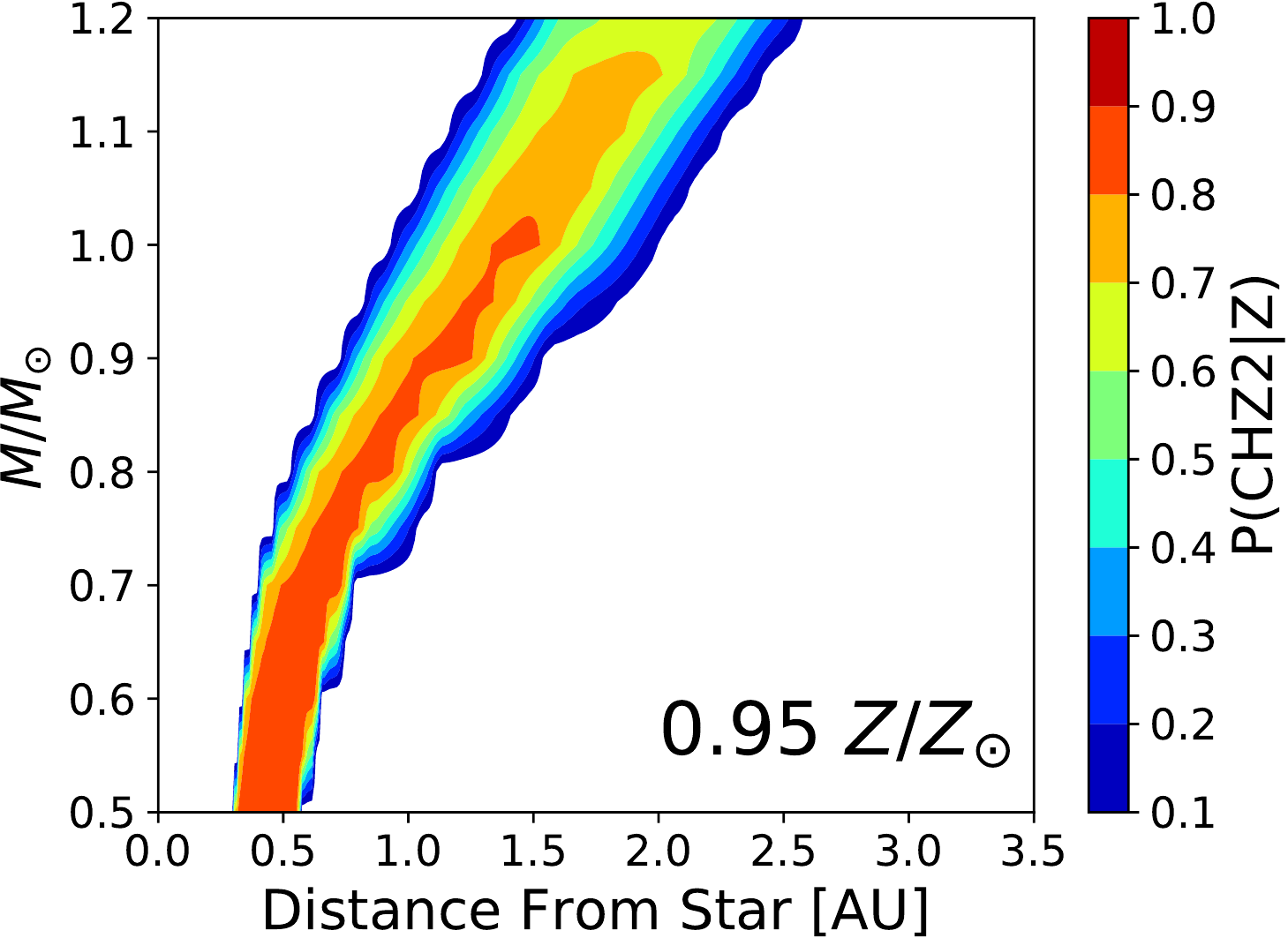}
\includegraphics[height=2.3cm]{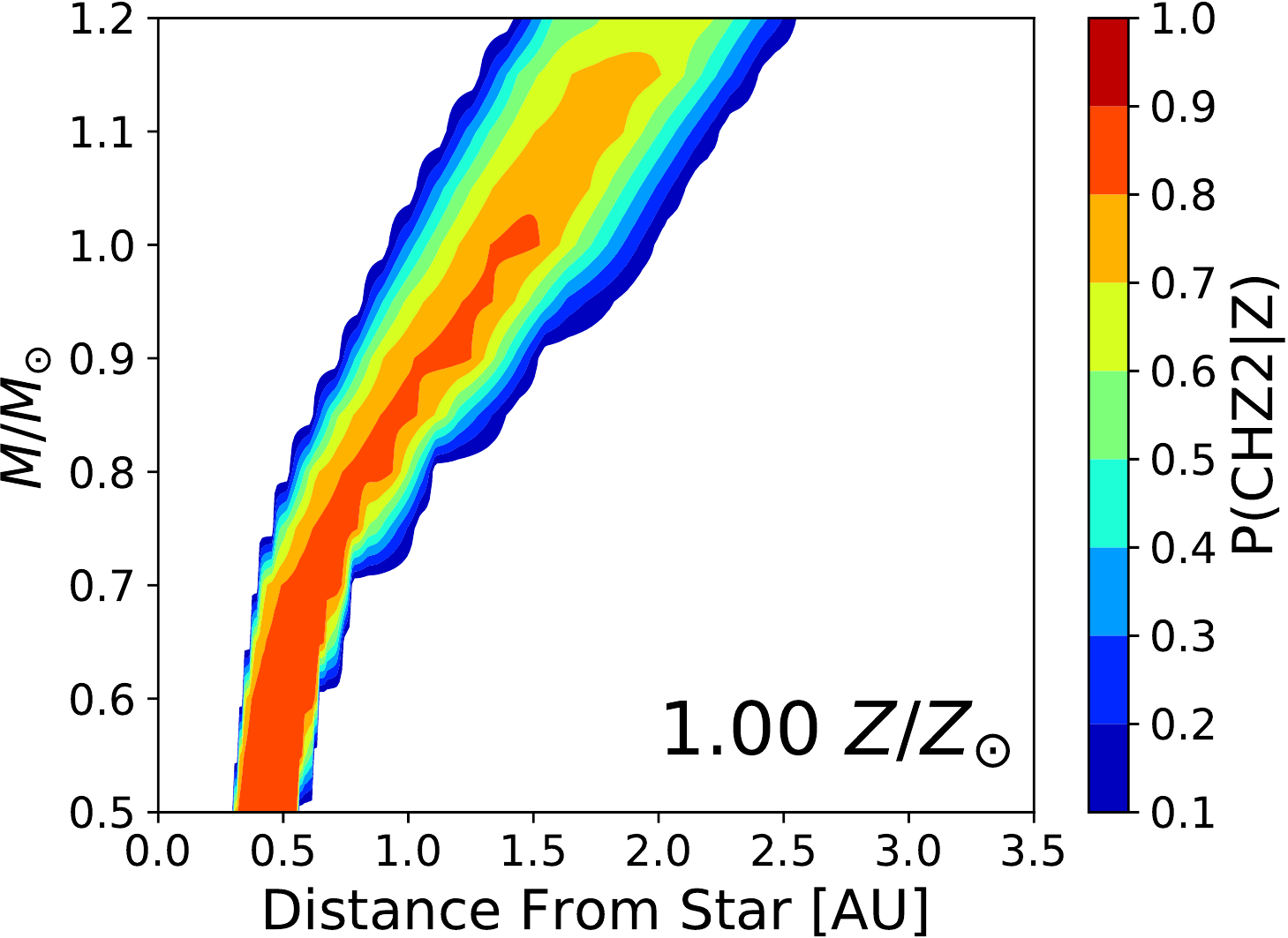}
\includegraphics[height=2.3cm]{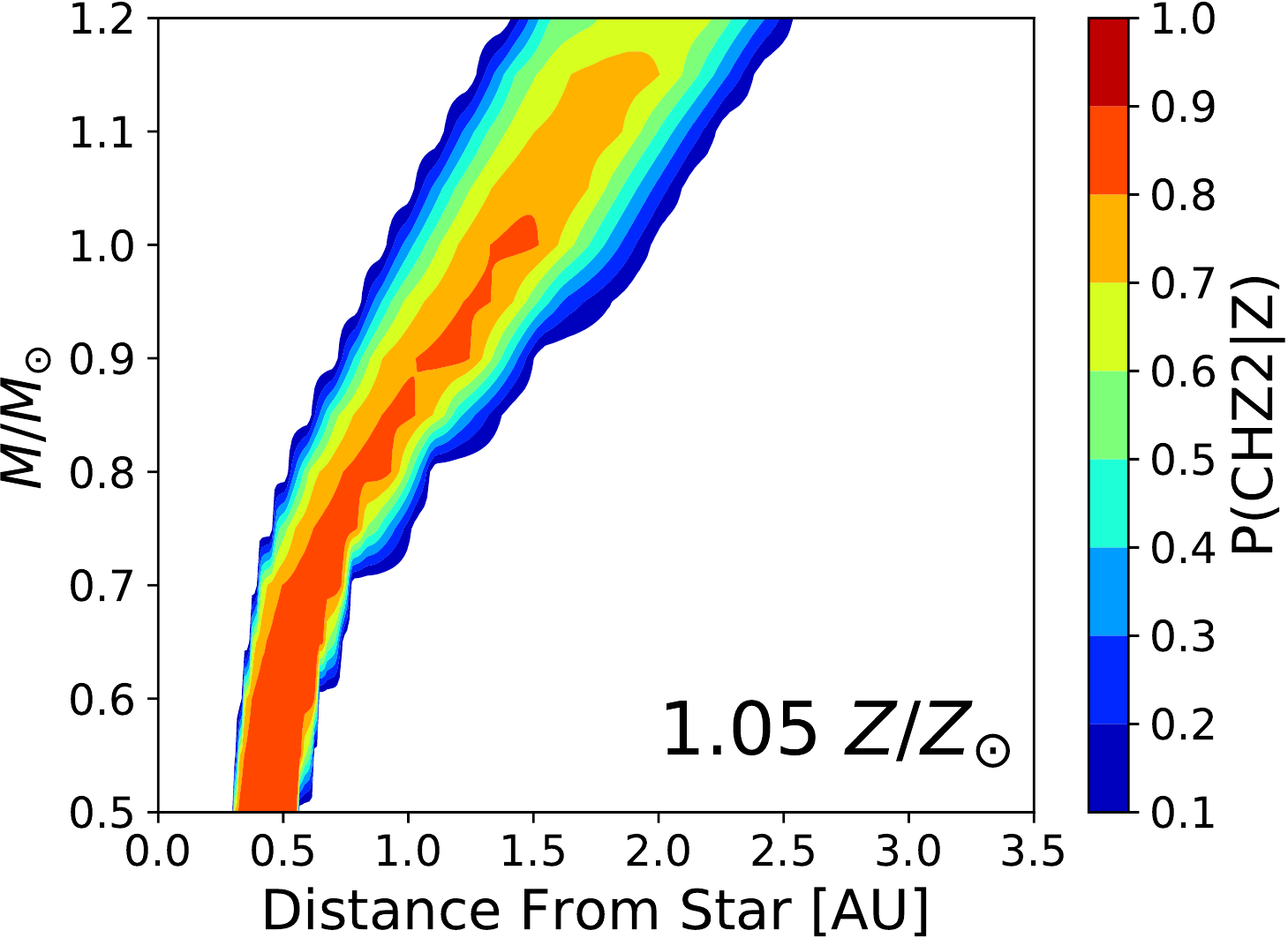}}
\centerline{\includegraphics[height=2.3cm]{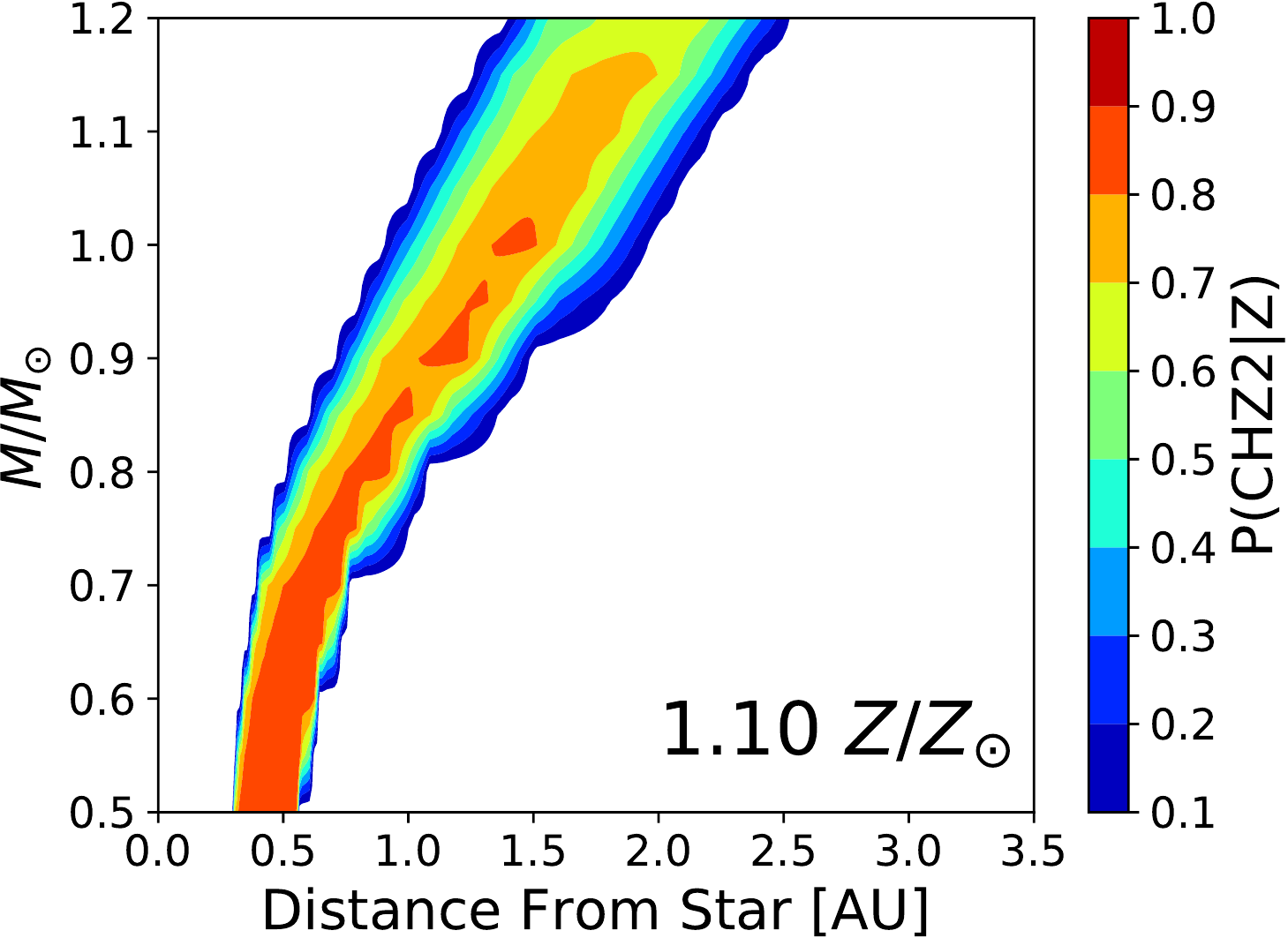}
\includegraphics[height=2.3cm]{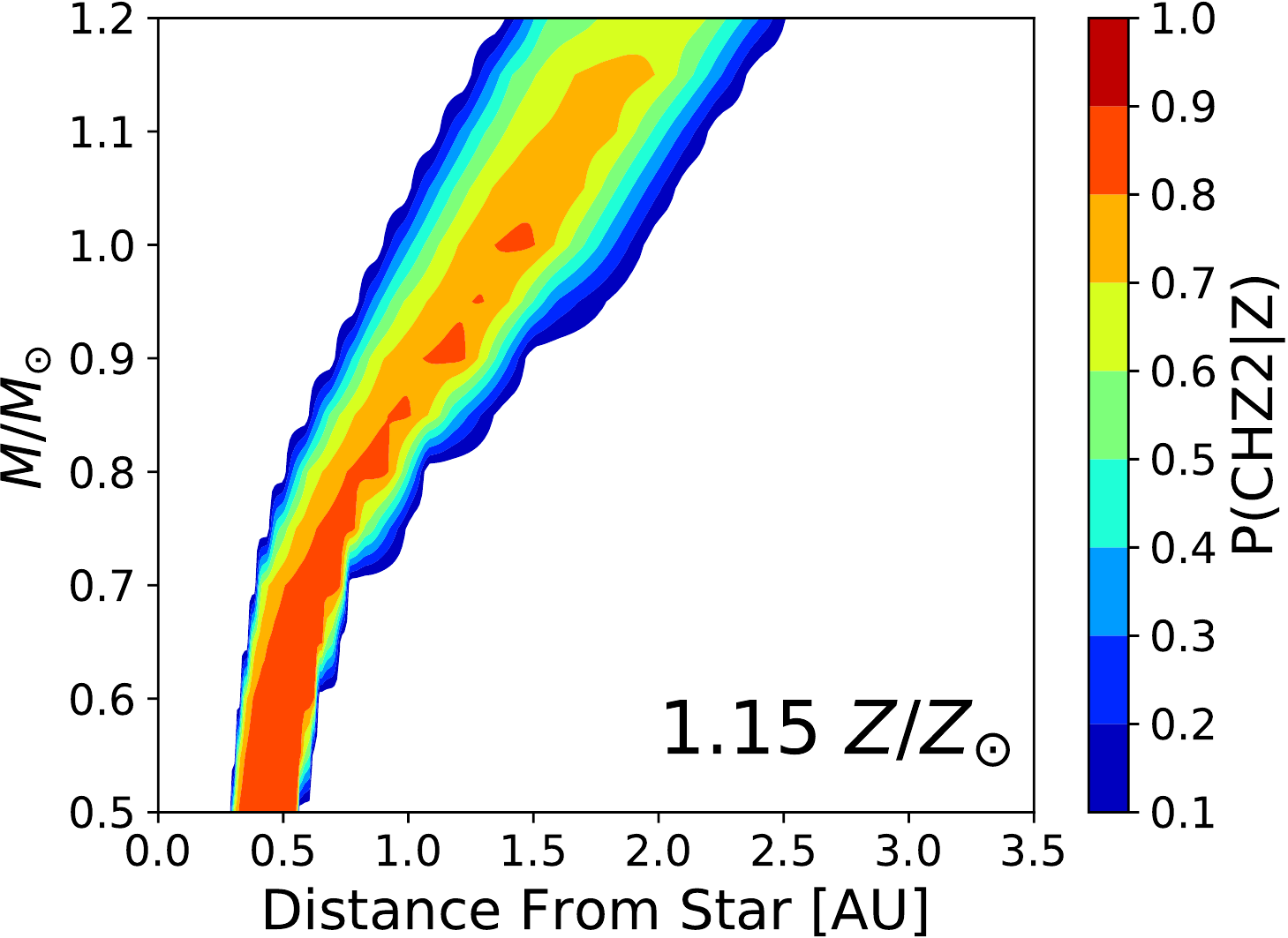}
\includegraphics[height=2.3cm]{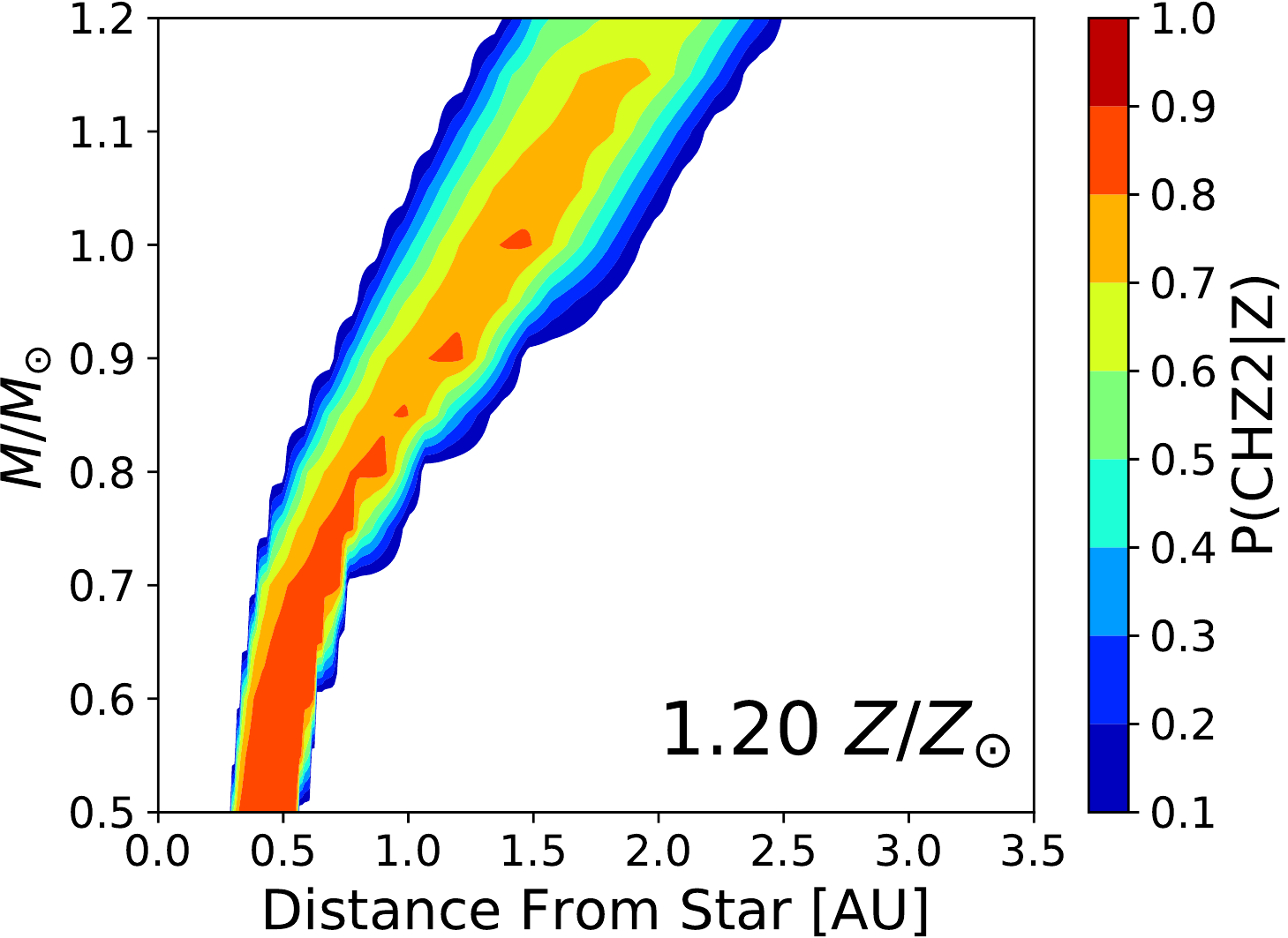}
\includegraphics[height=2.3cm]{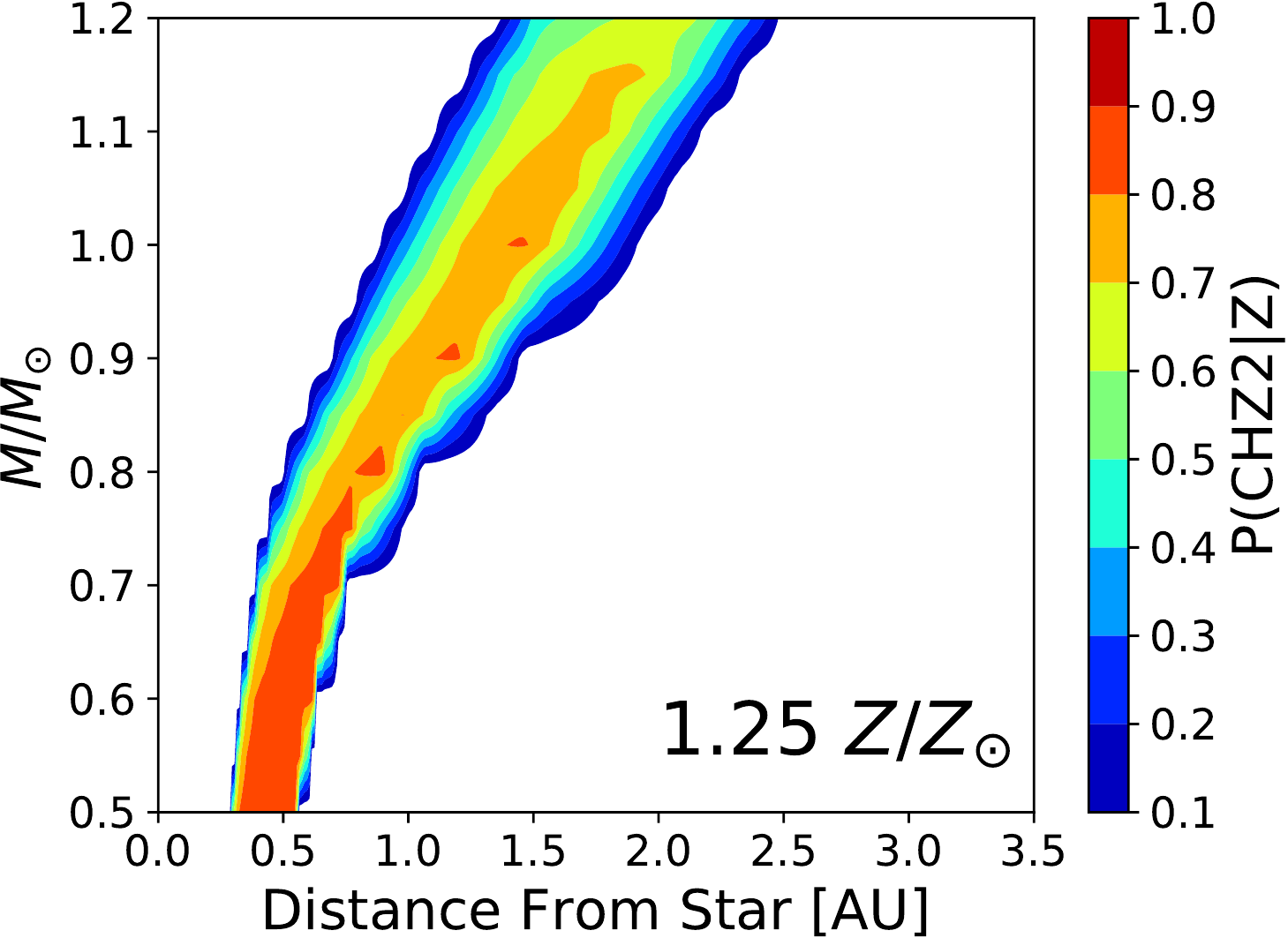}
\includegraphics[height=2.3cm]{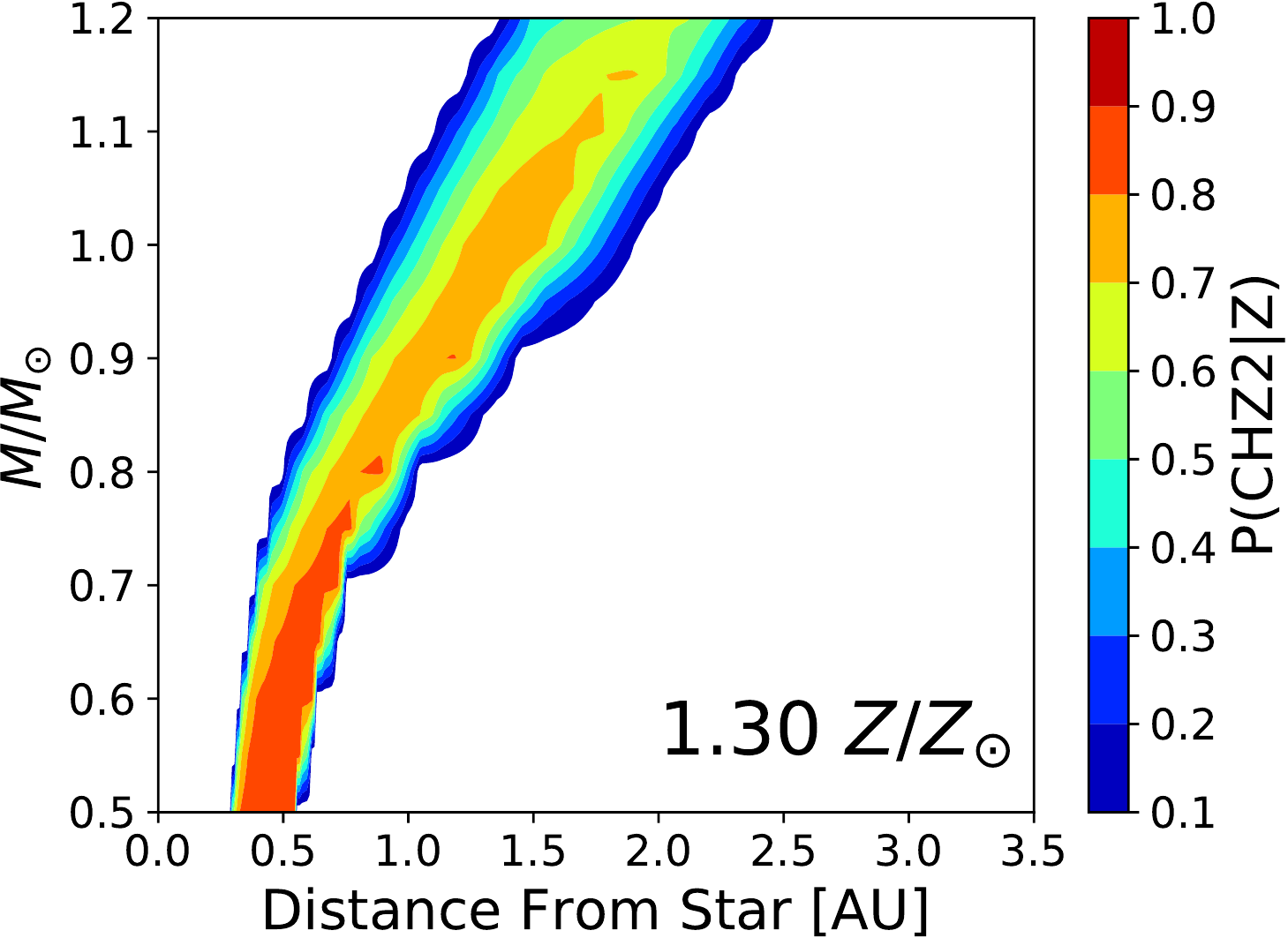}}
\centerline{\includegraphics[height=2.3cm]{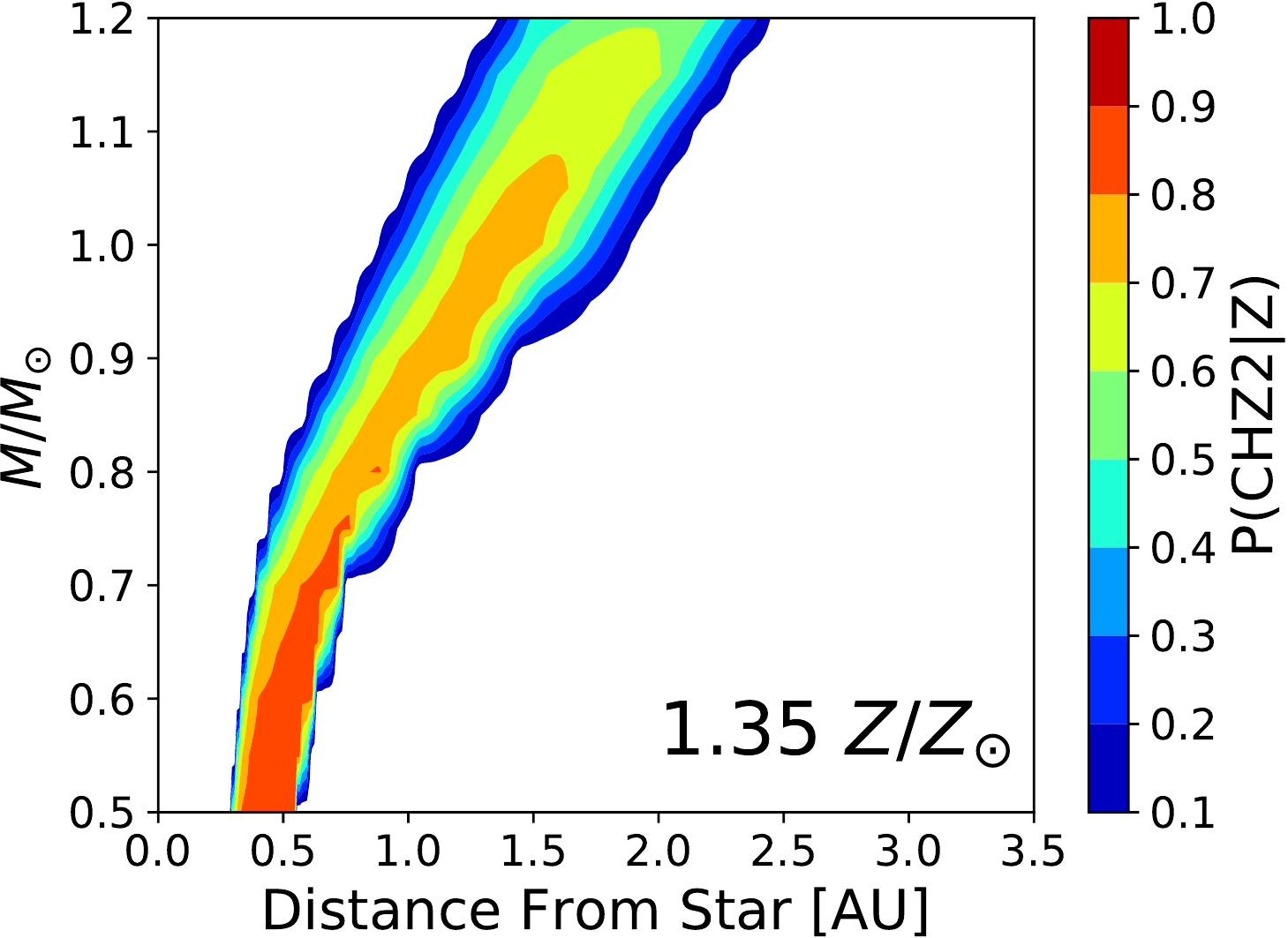}
\includegraphics[height=2.3cm]{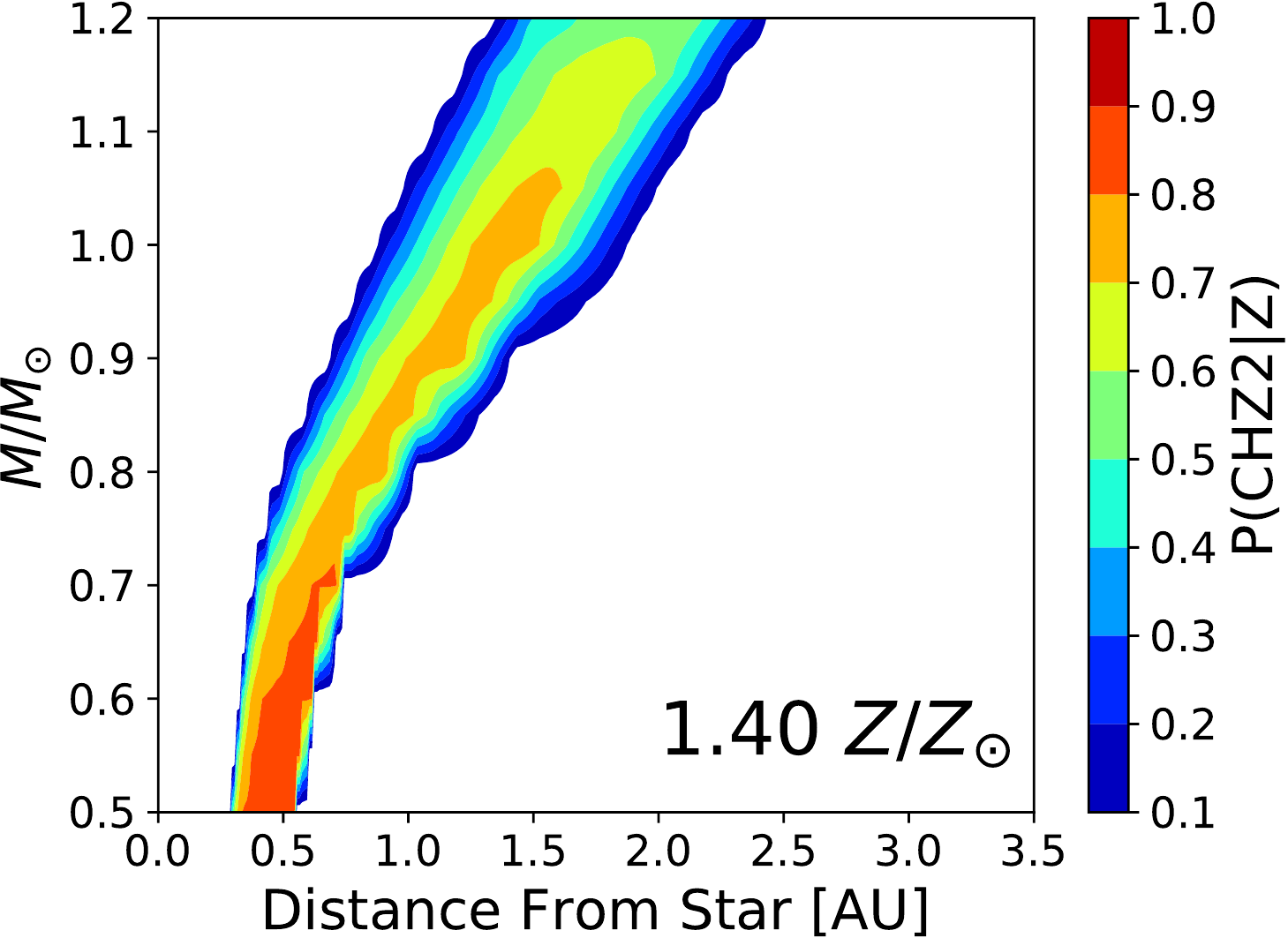}
\includegraphics[height=2.3cm]{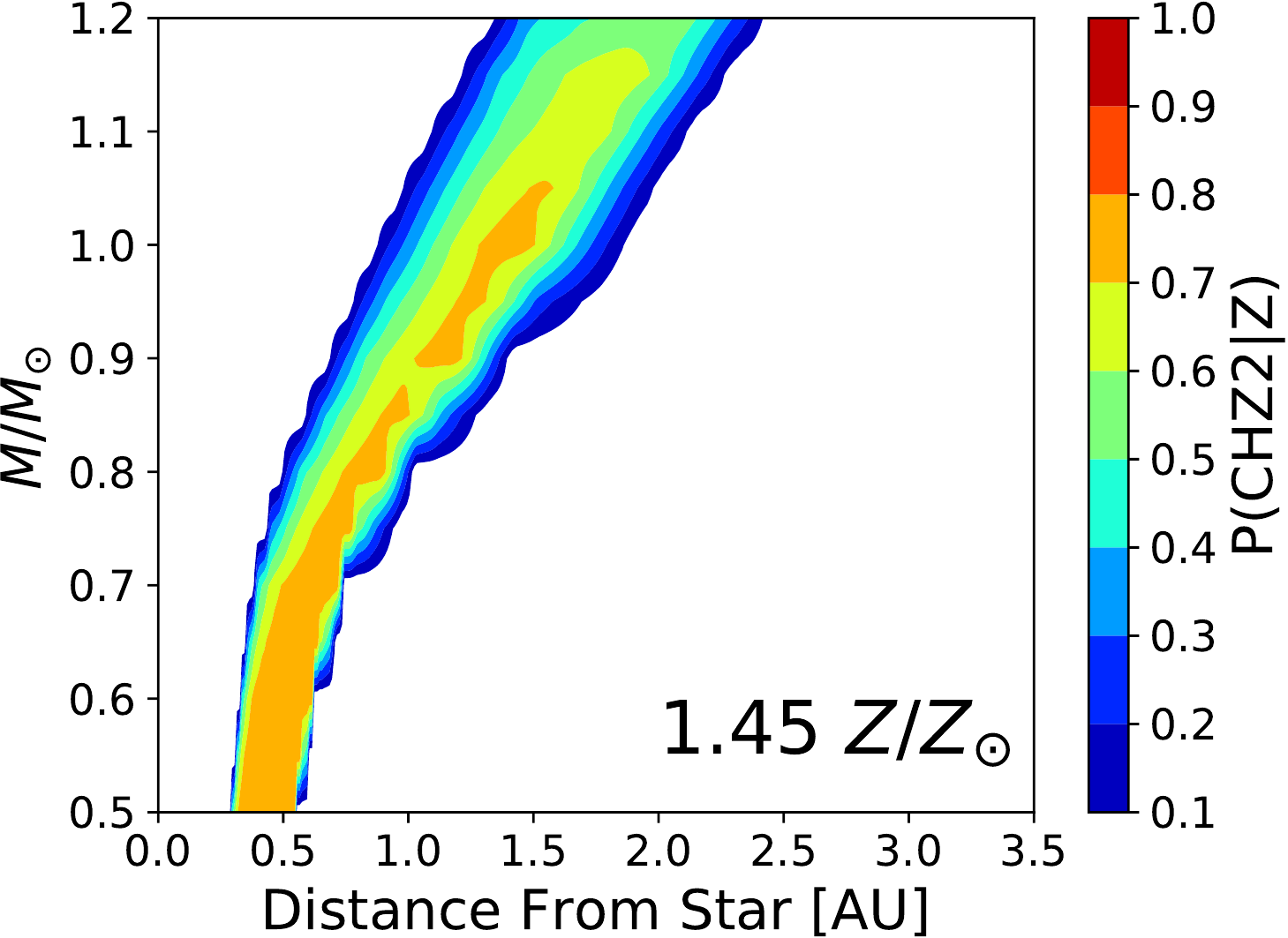}
\includegraphics[height=2.3cm]{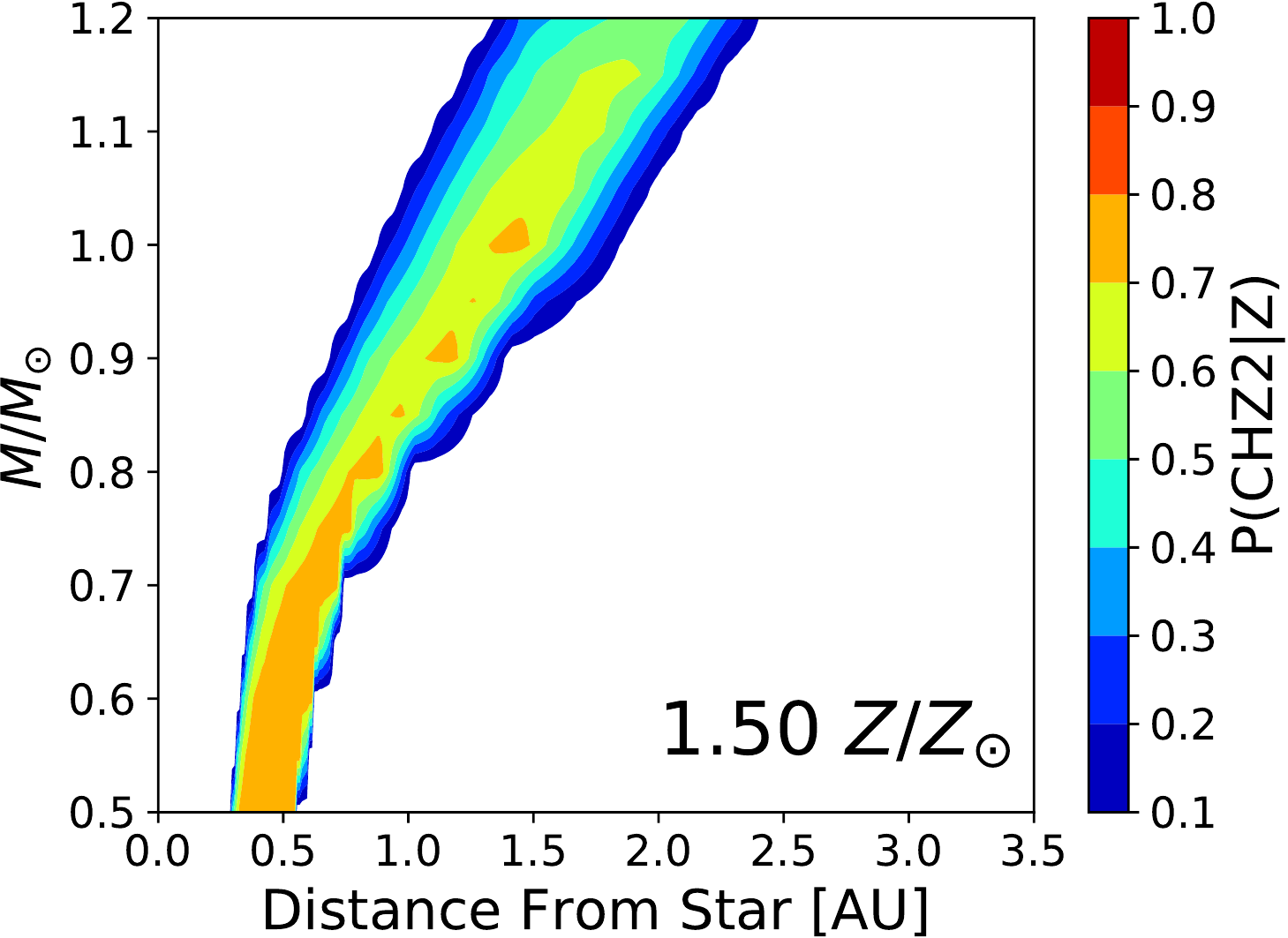}}
\caption{The posterior probability distribution contour plots for each $Z$-value of interest from {\it Hypatia} (end limits of 0.1 Z$_\odot$ and 1.5 Z$_\odot$ are dictated by the grid of Tycho models). Each panel represents a different metallicity value (top left-most panel is for 0.1 $Z\sol$, proceeding to the bottom right-most panel for 1.5 $Z\sol$), the distance from the star is shown on the x-axis, the stellar mass range is indicated on the y-axis, and the contour indicates posterior probability values, color bar at the top of each panel. Note that the stepped edge of the outer boundary of the countours is due to the finite mass resolution of the models plotted.}
\label{fig:29models}
\end{figure}

\begin{figure}[t]
\centerline{\includegraphics[height=6cm,width=8cm]{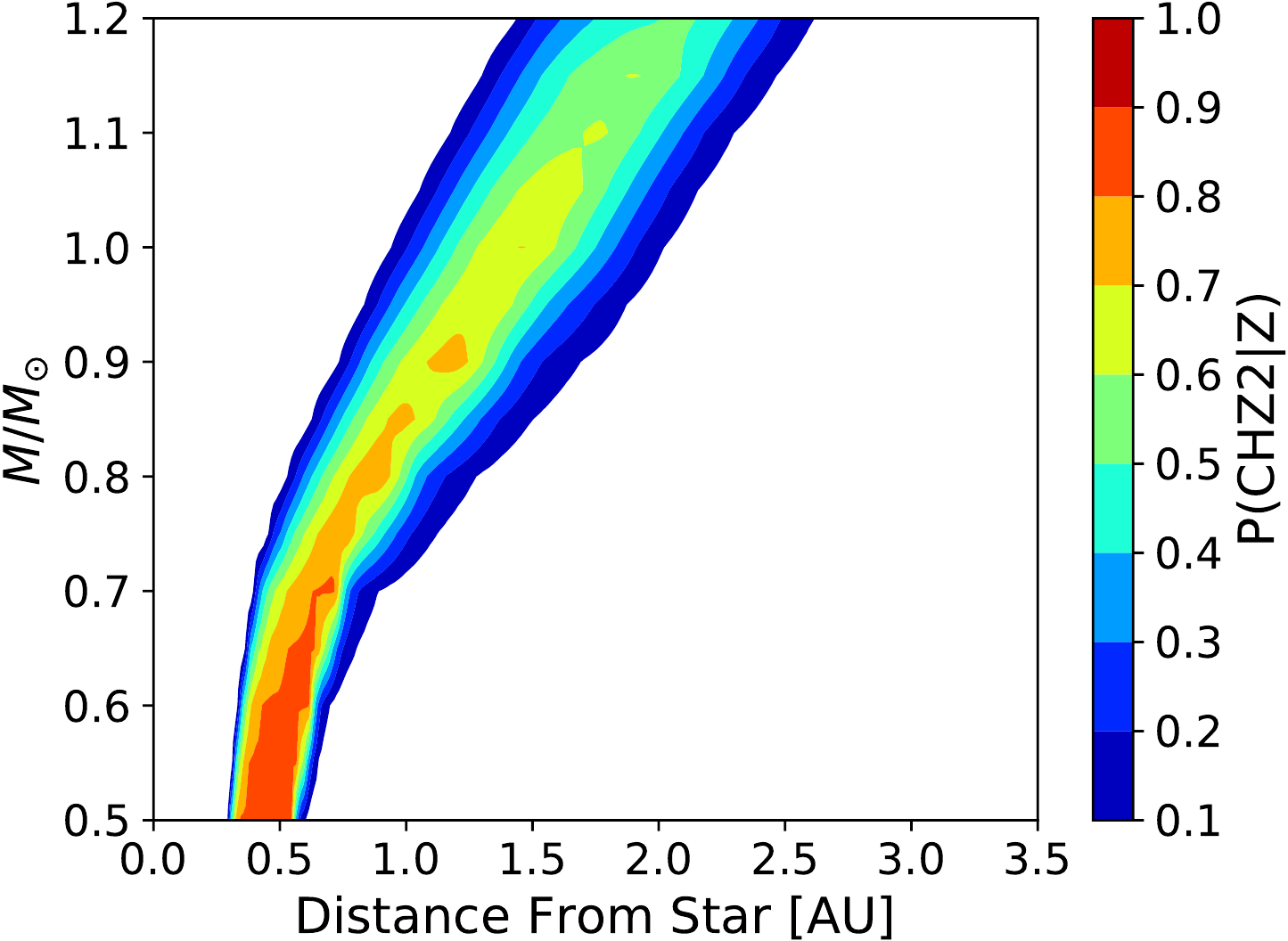}
\includegraphics[height=6cm,width=8cm]{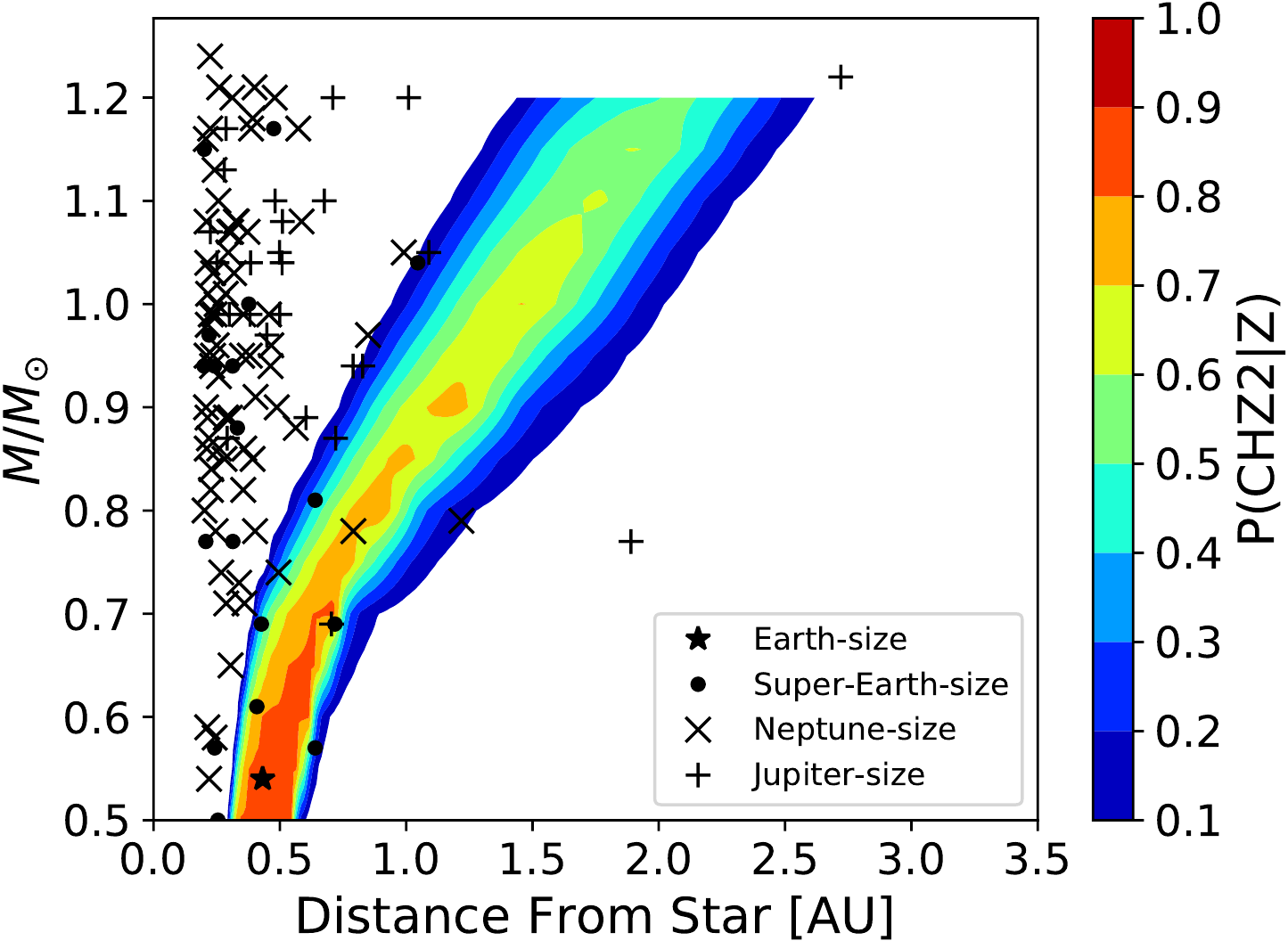}}
\caption{$P(\textrm{CHZ}_2|Z)$ contours (using the {\it Hypatia} catalog distribution) marginalized over all $Z$-values for each mass in the Tycho catalog (i.e. the 15 models in Figure~\ref{fig:29models}), as a function of radius (left), with known exoplanets from the {\it Kepler} database plotted on top (right). Note that there is one promising Earth-like exoplanet (x) and five promising super-Earth-like exoplanets (+) in the 2 Gy continuously habitable zone.}
\label{fig:Pcontour}
\end{figure}

\begin{figure}[t]
\centerline{\includegraphics[height=6cm,width=8cm]{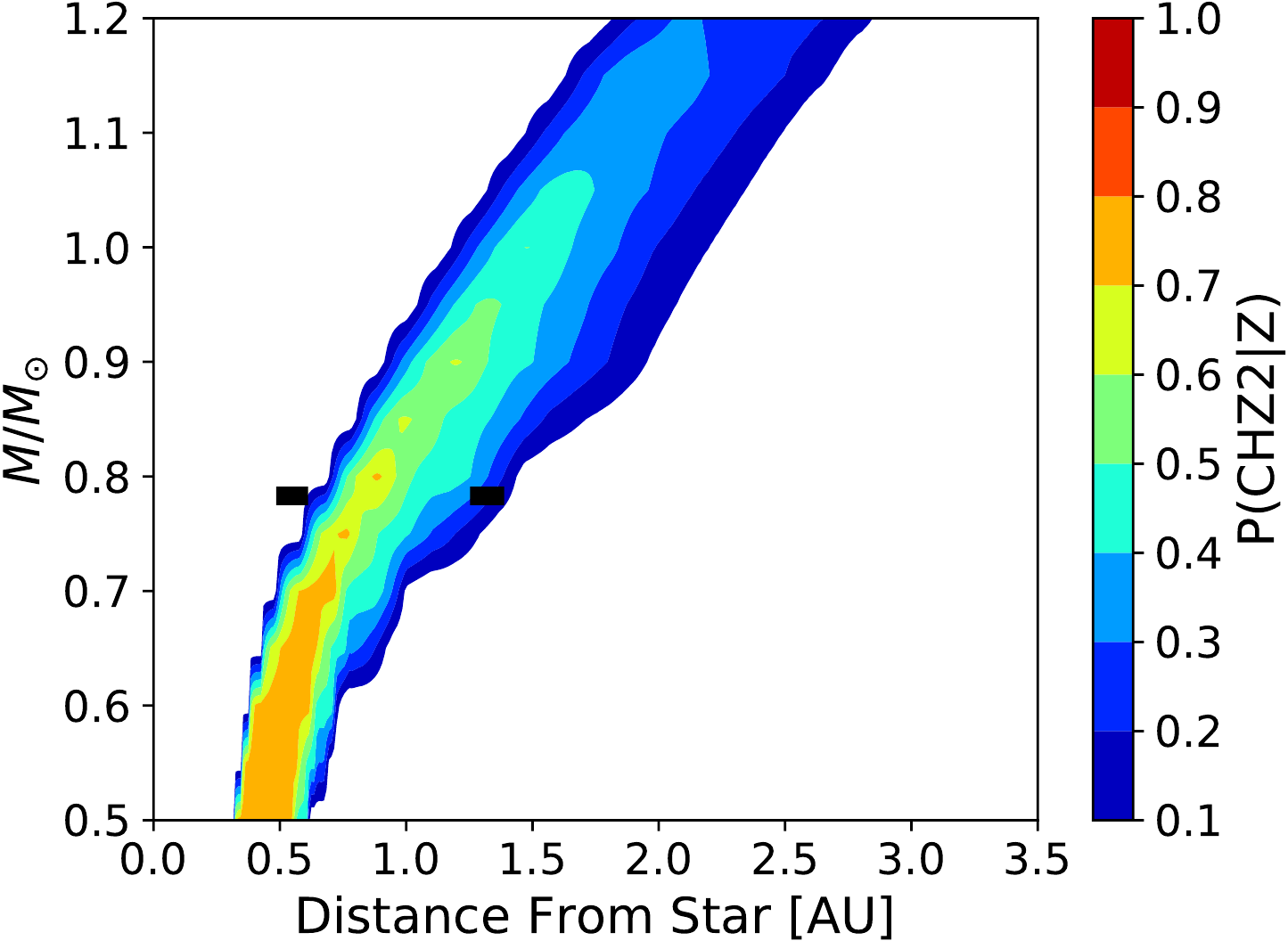}
\includegraphics[height=6cm,width=8cm]{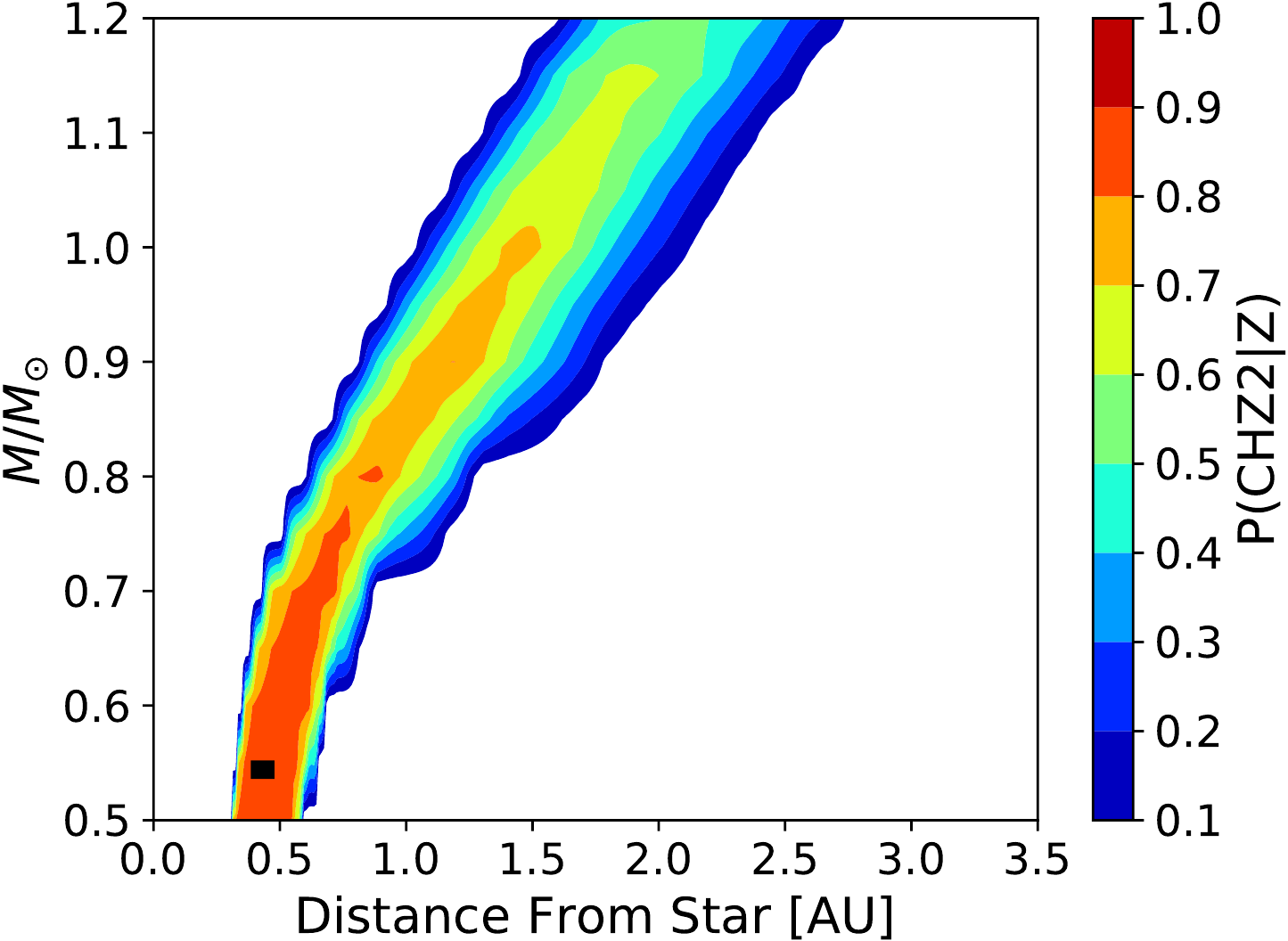}}
\centerline{\includegraphics[height=6cm,width=8cm]{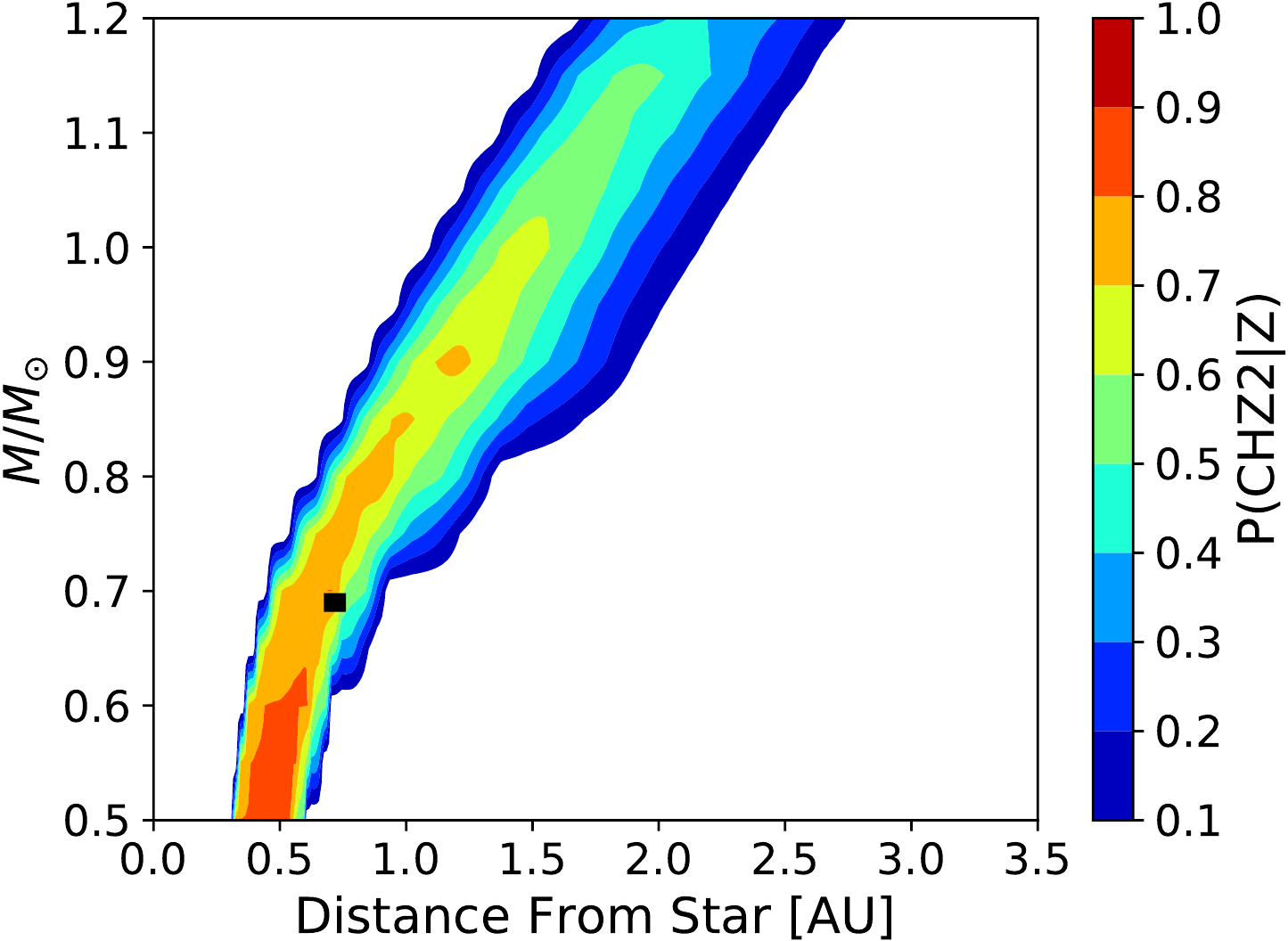}
\includegraphics[height=6cm,width=8cm]{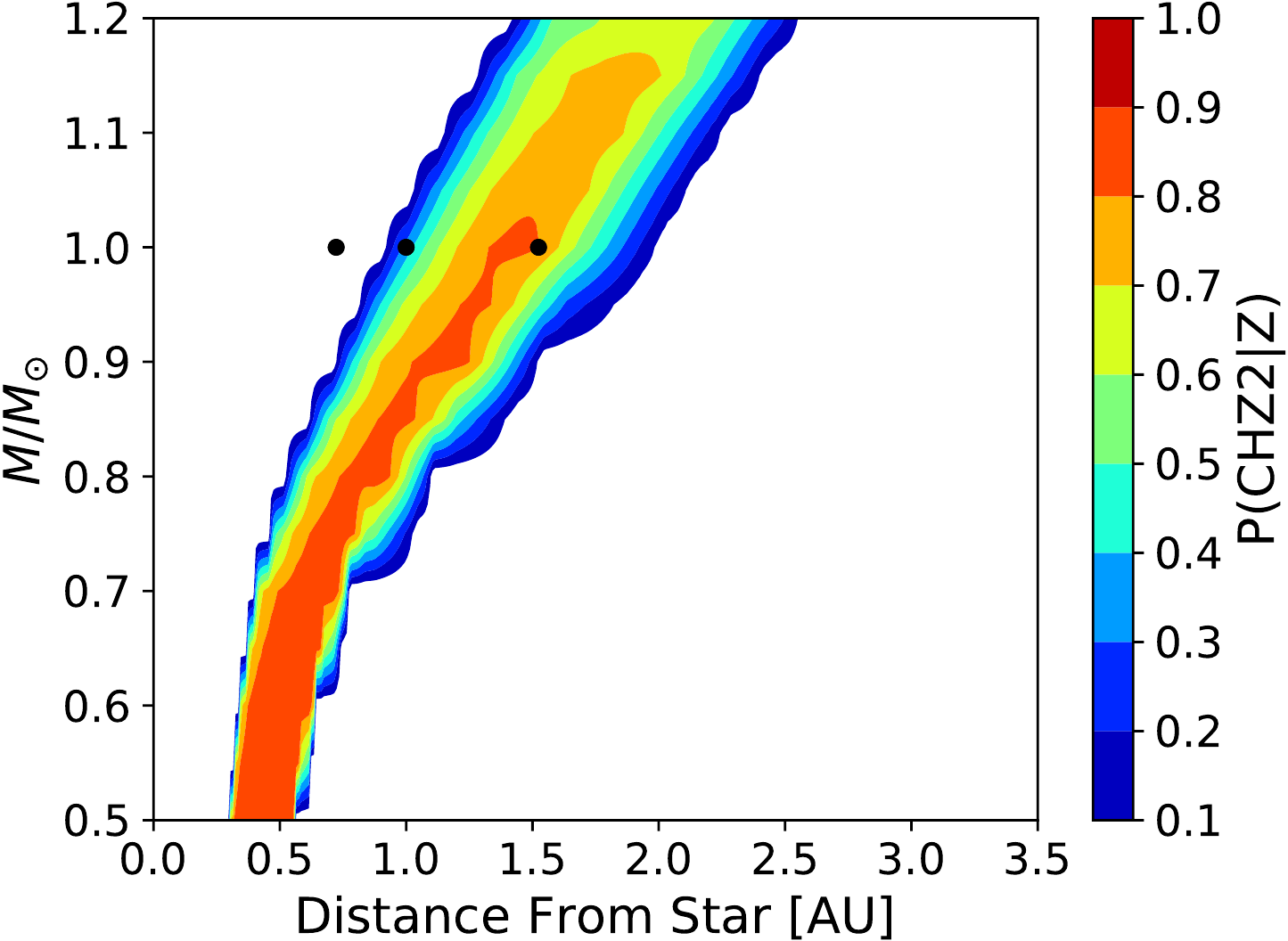}}
\caption{This is representative of the results for the planets indicated in Figure~\ref{fig:ZPplanets}, over-plotted on the corresponding mass and metallicity contour plot (as demonstrated in Figure~\ref{fig:29models}). The black spots indicating the location of each of the planets on this plot are representative of uncertainty in the measurements, i.e. the +/- 1$\sigma$ range.}
\label{fig:planets-contour}
\end{figure}

To begin, we discuss the Hypatia-weighted model probabilities for $P(\textrm{CHZ}_2)$. This corresponds to Case 1, which assumes we only know the stellar mass. This provides a quick estimate for the habitability potential of any given system without followup observations of composition, as is frequently the case with more distant {\it Kepler} candidates that do not have high resolution spectroscopy. The left panel of Figure~\ref{fig:M1_Zall} shows the posterior probability of a given orbital distance having been continuously habitable for at least 2 Gy for metallicity in steps of 0.05 $Z/Z\sol$ from 0.1 to 1.5 $Z/Z\sol$, assuming a fixed mass of $M$ = 1 $M_\odot$ and an unconstrained age. These distributions are mapped out in an associated contour plot (right panel), where P(CHZ$_2|Z)$ is indicated by the color bar at the top. The inner and outer habitable zone limits at the Zero-Age Main Sequence (ZAMS) and the Terminal-Age Main Sequence (TAMS) are also included, where the inner boundary is defined by the conservative Runaway Greenhouse case, and the outer boundary is defined by the Maximum Greenhouse case, discussed in detail in \citet{kopp13}. For the high $Z$-cases, the habitable zone distance is closer-in and narrower, while the highest likelihood of a 2 Gy continuously habitable orbit at some point during the MS evolution occurs between 0.8 and 1.2 $Z/Z\sol$, where the Hypatia distribution peaks. Based on observational data, it appears that while low-$Z$ stars have larger habitable zones, the smaller lifetimes and observational likelihoods give higher-$Z$ stars larger probabilities. This trend is consistent for all masses in Figure~\ref{fig:29models}.

Figure~\ref{fig:29models} shows the $P(\textrm{CHZ}_2)$ contour plots as a function of mass for each $Z$-value. For each panel in the figure, the distance from the star is shown on the x-axis, the stellar mass is given on the y-axis, and the color contour illustrates the associated calculated $P$-value as a visual representation of the habitability potential. As expected for the shorter lifetimes of higher mass stars, the regions of high $P(\textrm{CHZ}_2)$ decrease as mass increases, while the total width of the range of habitable orbits increases due to the faster evolution of the stars.

Figure~\ref{fig:Pcontour} (left panel) shows the $P(\textrm{CHZ}_2|Z)$ contour plot for all masses in our model grid ($0.5-1.2$ M\sol) marginalized over all $Z$-values ($0.1-1.5$ Z\sol) referenced in Figure~\ref{fig:29models}. In the right panel, we over-plot {\it Kepler} exoplanets whose host stars fall within the boundaries of our parameter space. Since the known {\it Kepler} exoplanet host star metallicity distribution is consistent with the {\it Hypatia} distribution (Figure~\ref{fig:kephyp}), this is an appropriate prior.

\begin{figure}[t]
\centerline{\includegraphics[height=6cm,width=8cm]{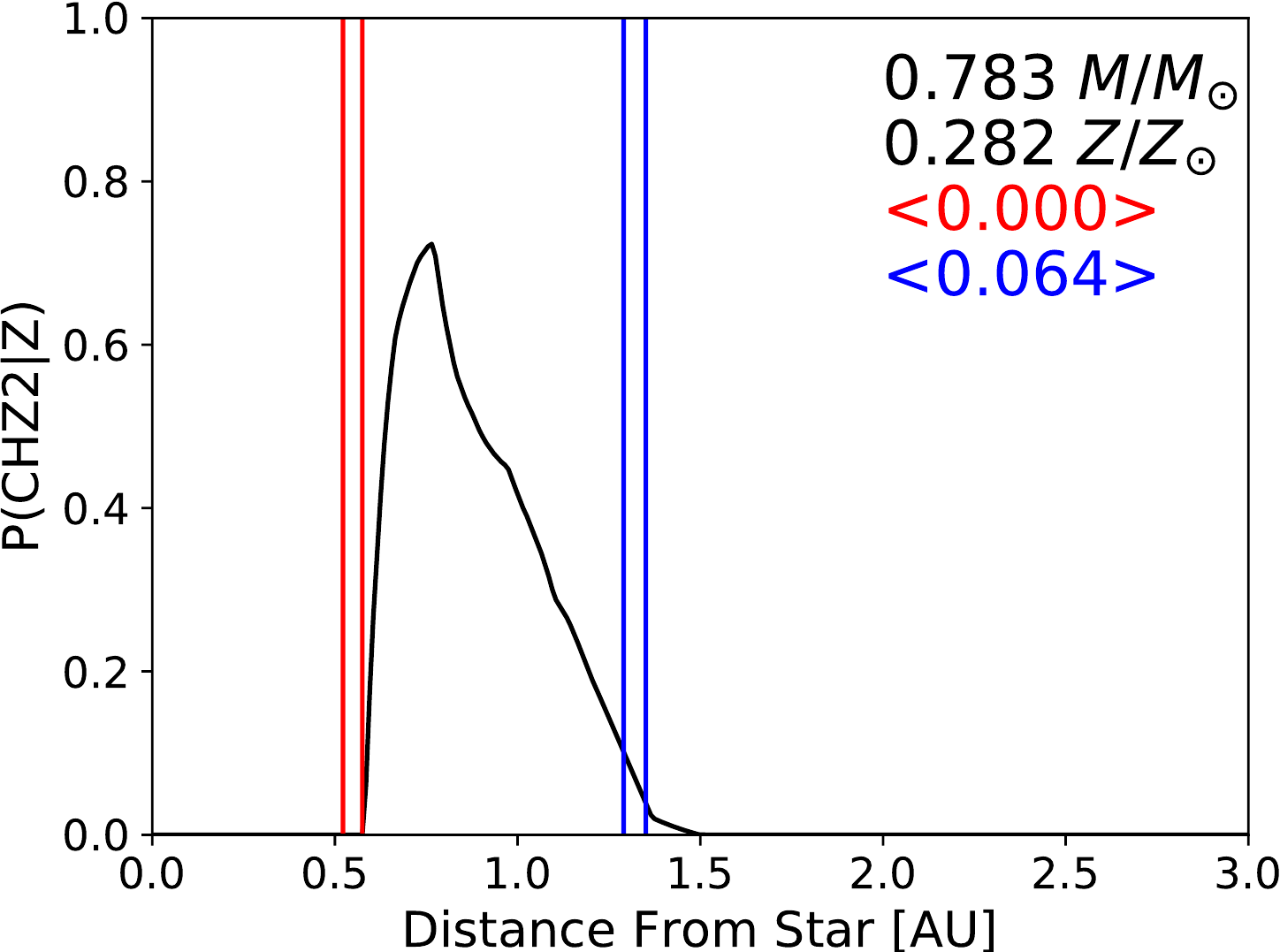}
\includegraphics[height=6cm,width=8cm]{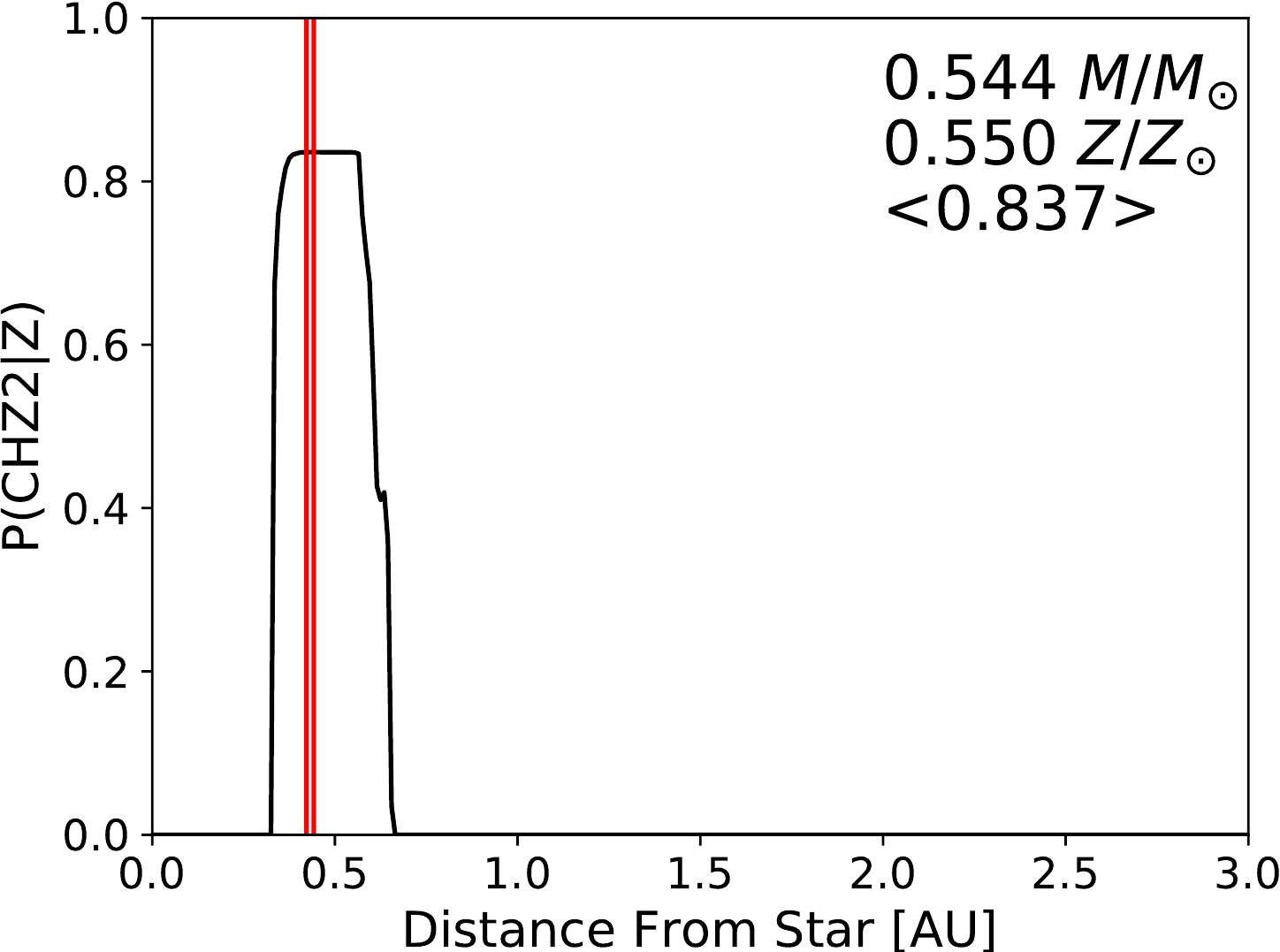}}
\centerline{\includegraphics[height=6cm,width=8cm]{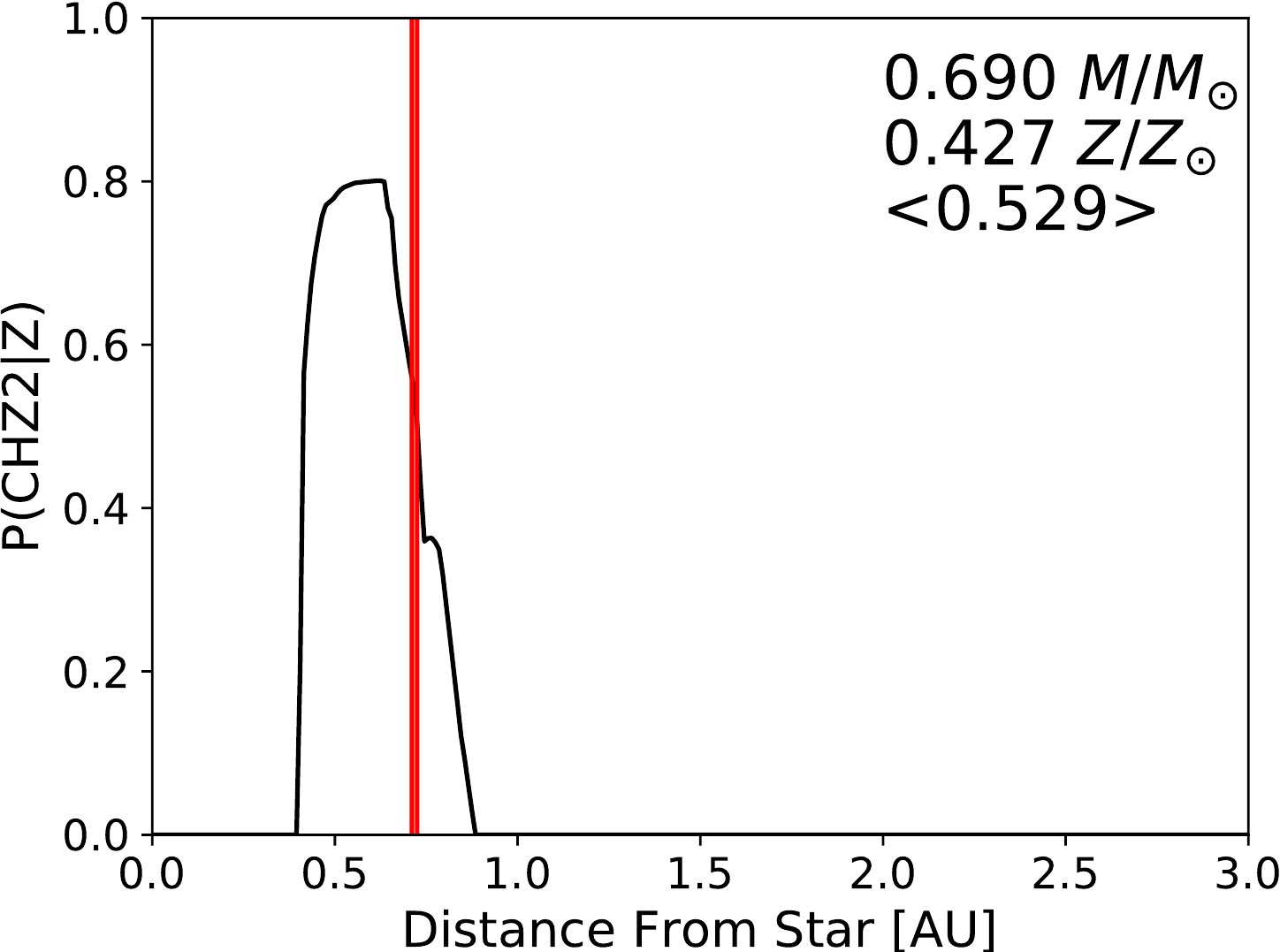}
\includegraphics[height=6cm,width=8cm]{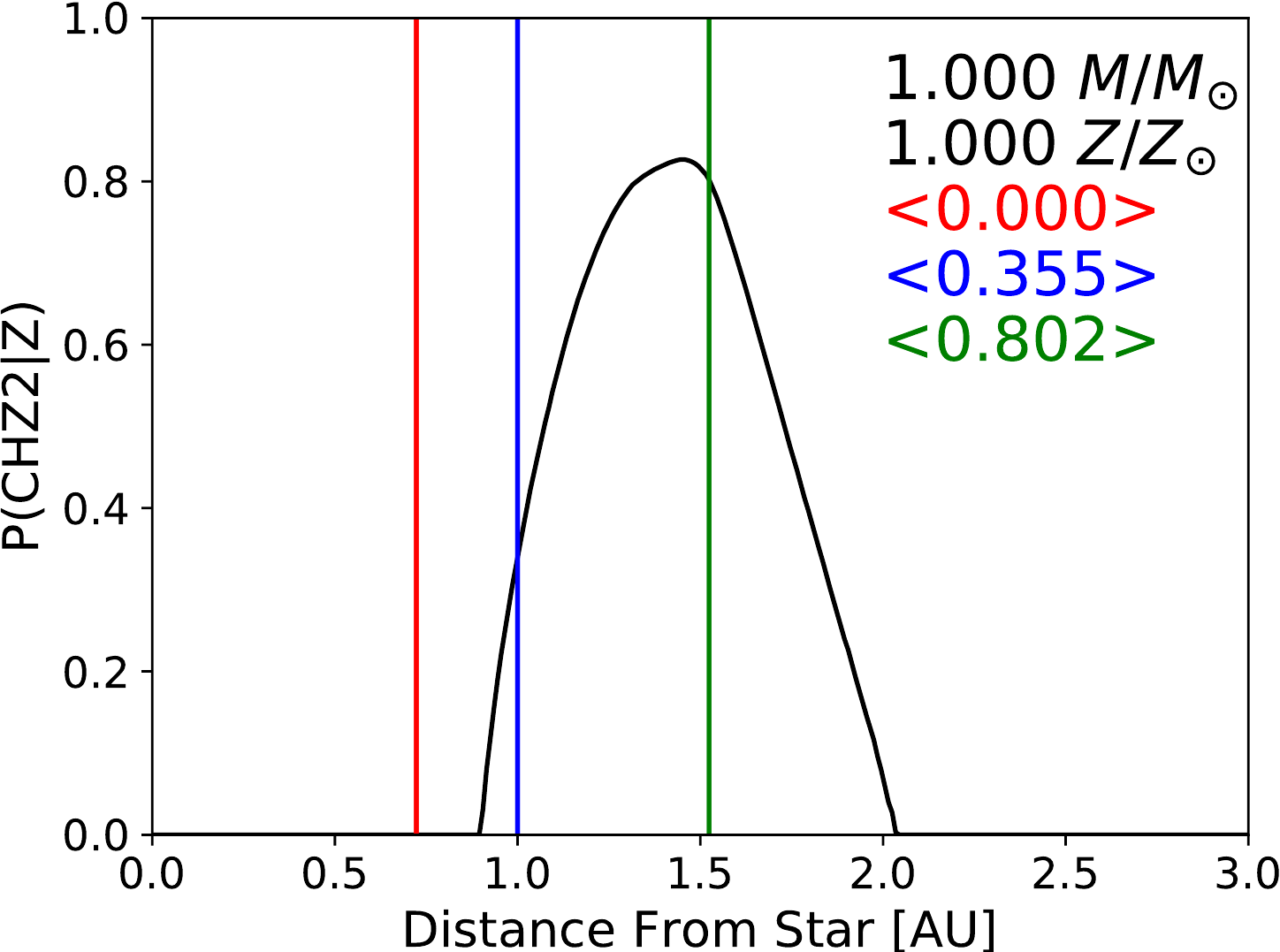}}
\caption{Case 1: $P(\textrm{CHZ}_2|Z)$ (using {\it Hypatia}) for Tau Ceti e, f (upper left; red and blue, respectively), Kepler-186f (upper right), Kepler-62f (lower left), and the Sun with Venus, Earth, and Mars (lower right; red, blue, and green, respectively). The data in the upper right corner of each panel is stellar mass, metallicity (given relative to $Z\sol$), and the $P$-value marginalized over the entire orbital range.}
\label{fig:ZPplanets}
\end{figure}

\begin{figure}[t]
\centerline{\includegraphics[height=6cm,width=8cm]{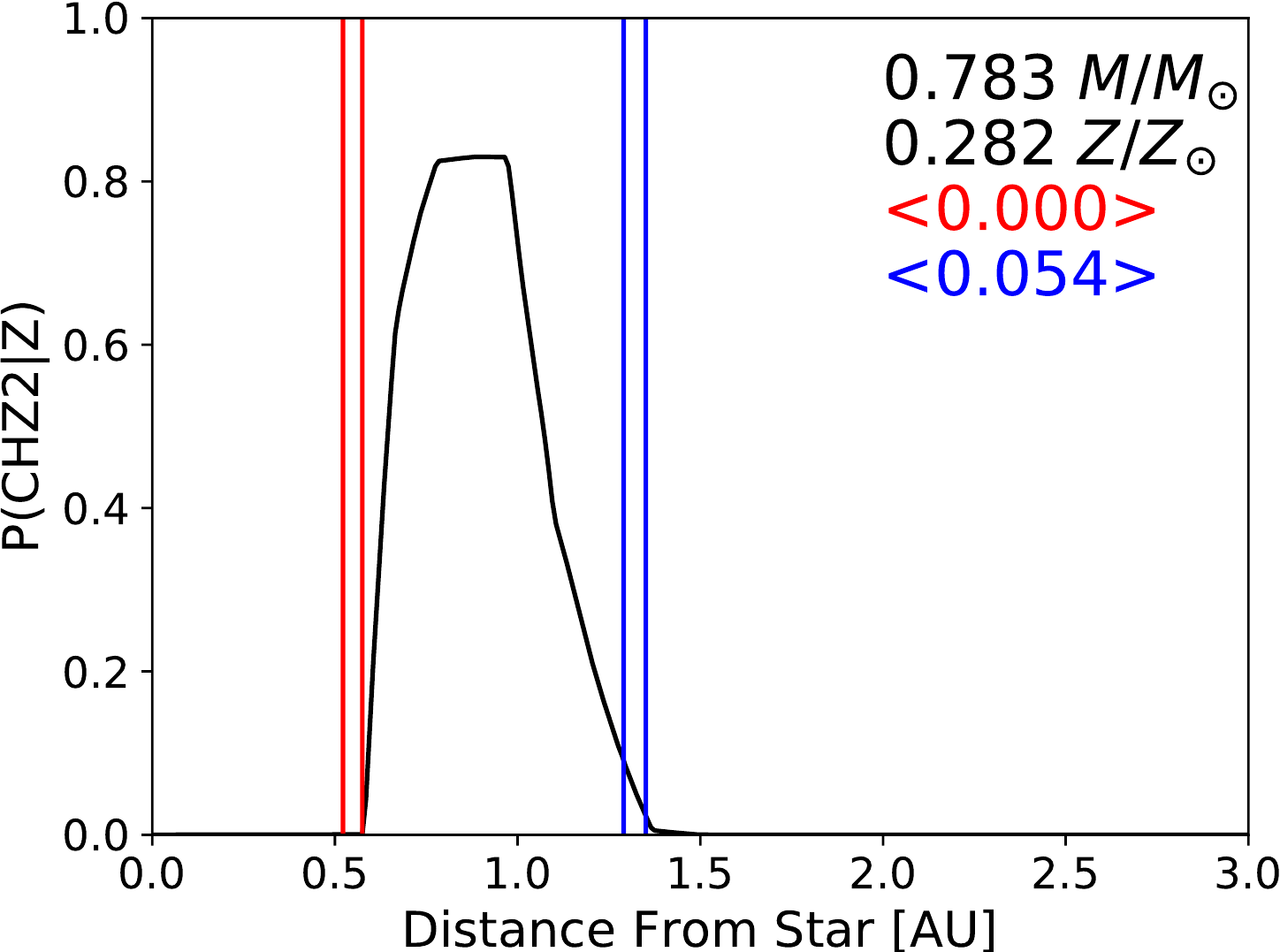}
\includegraphics[height=6cm,width=8cm]{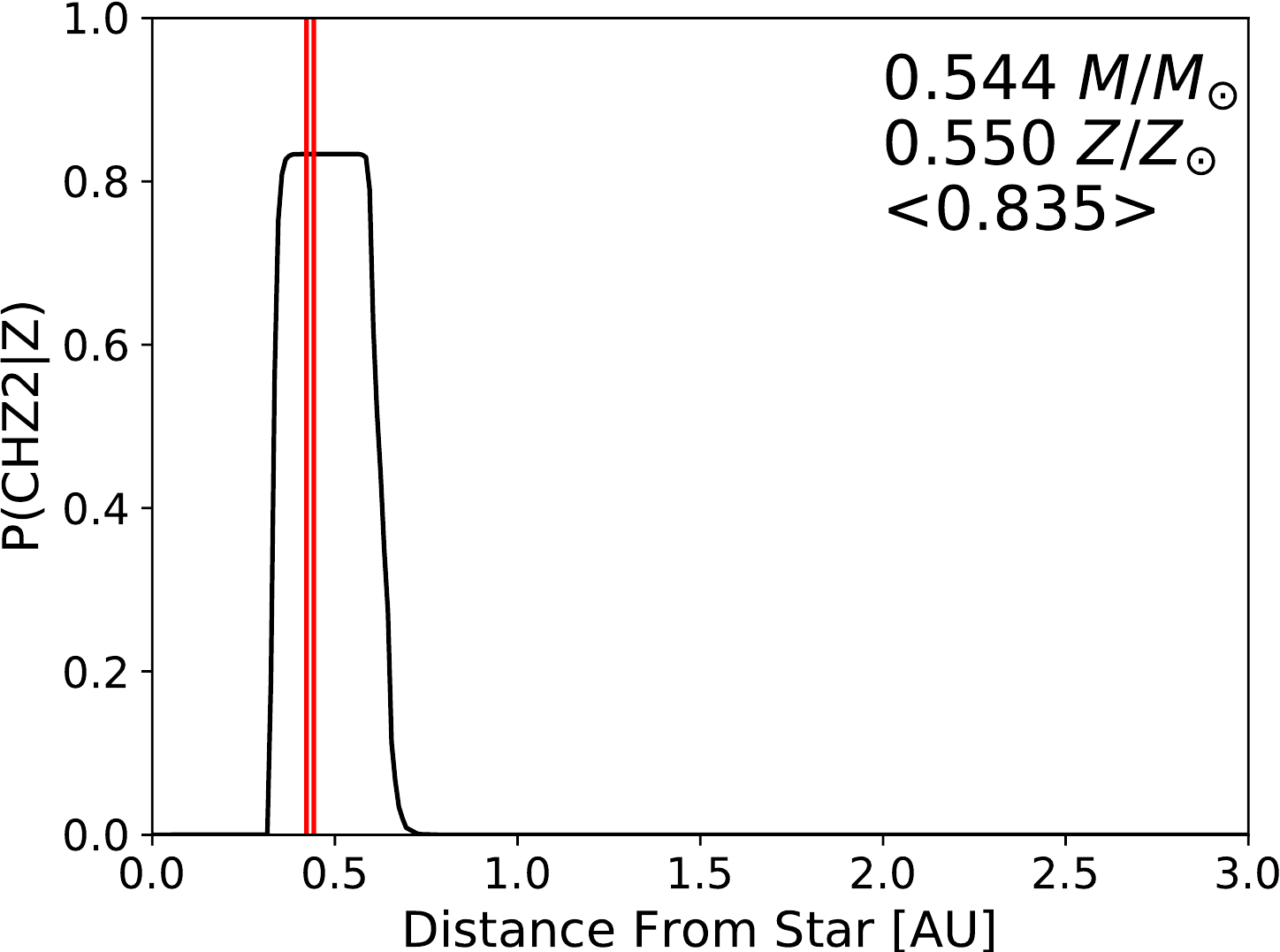}}
\centerline{\includegraphics[height=6cm,width=8cm]{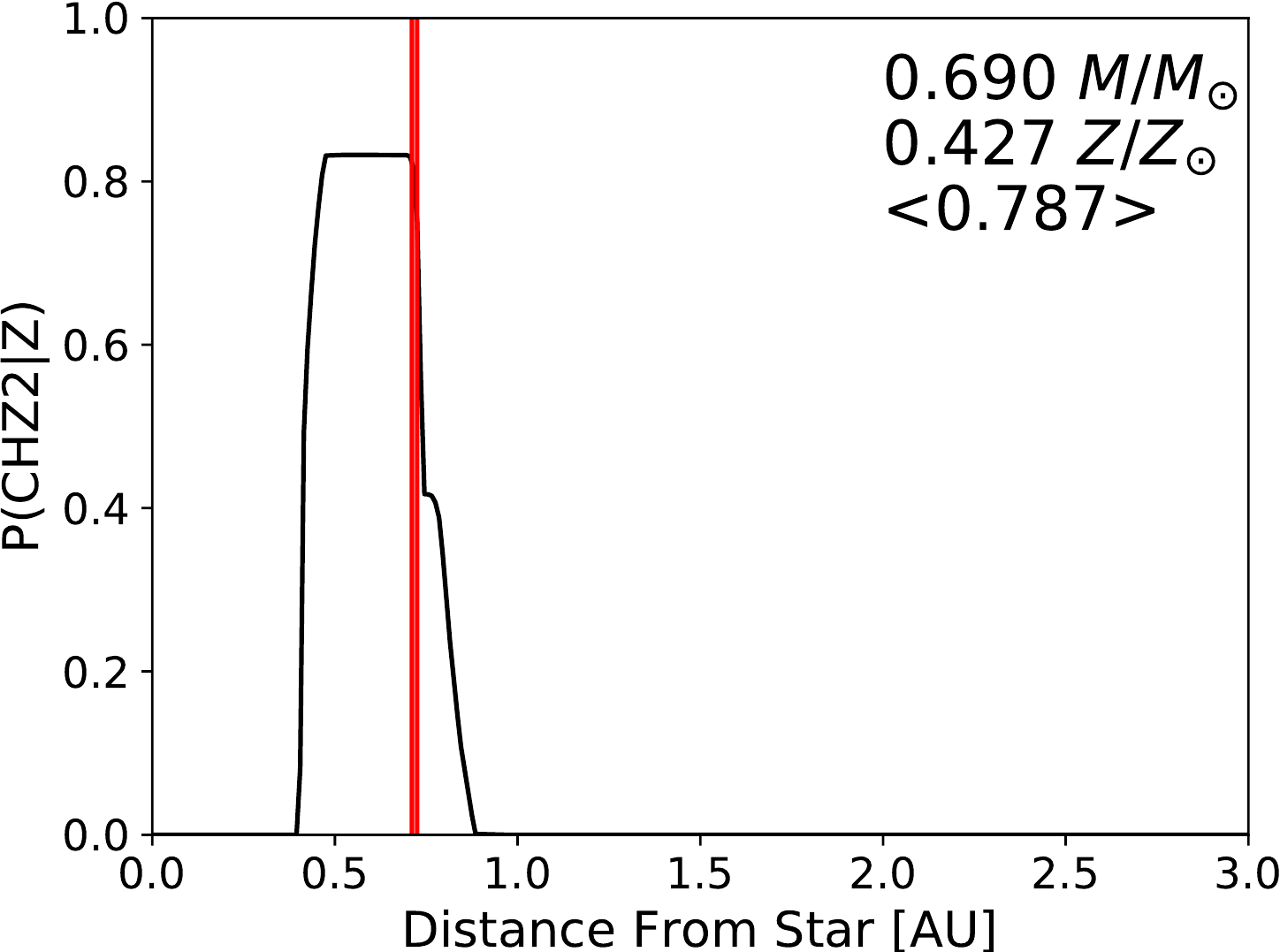}
\includegraphics[height=6cm,width=8cm]{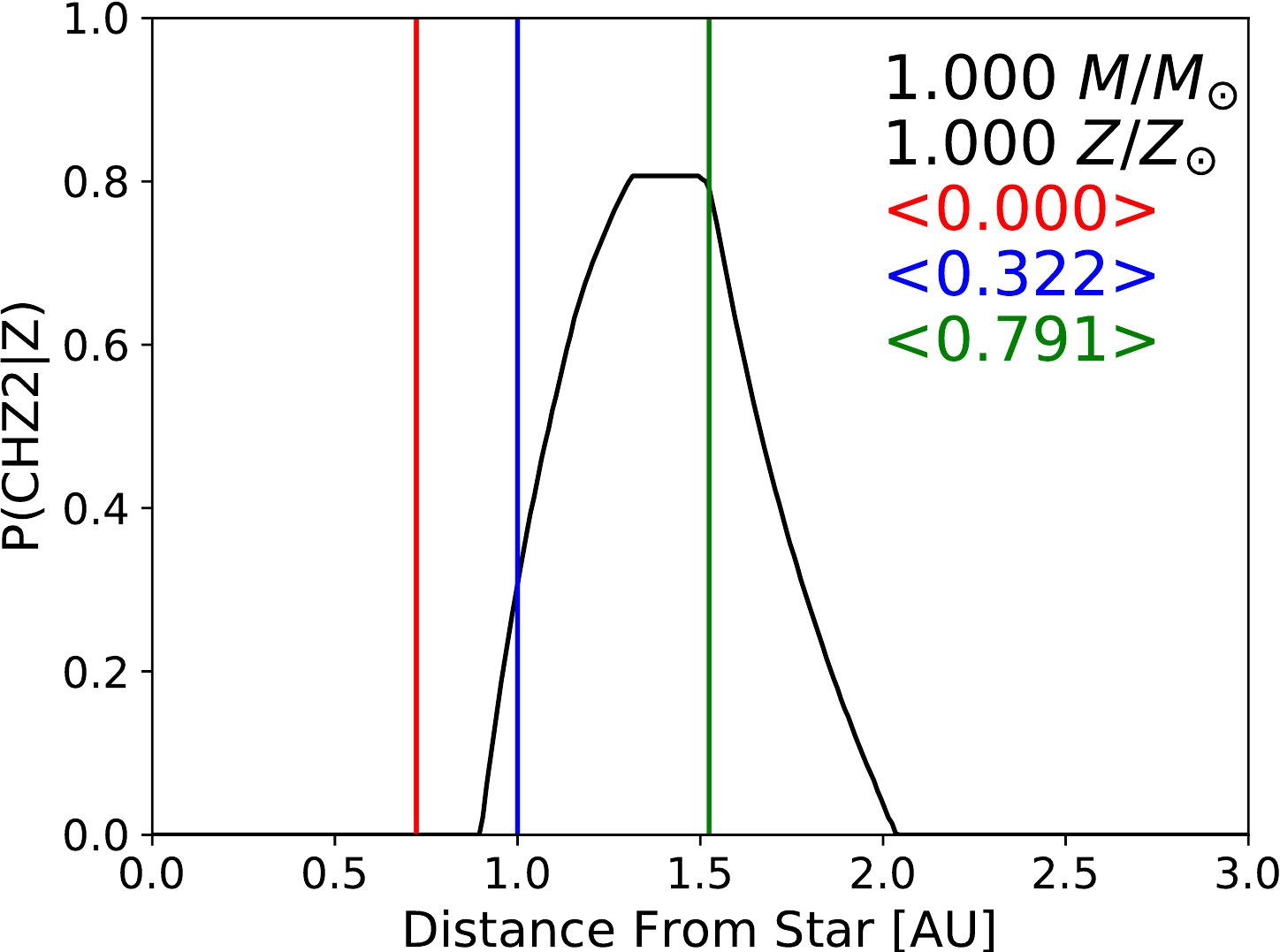}}
\caption{Case 2: $P(\textrm{CHZ}_2|Z)$ (using $Z$-value measurements) for Tau Ceti e, f (upper left; red and blue, respectively), Kepler-186f (upper right), Kepler-62f (lower left), and the Sun with Venus, Earth, and Mars (lower right; red, blue, and green, respectively). The data in the upper right corner of each panel is stellar mass, metallicity (given relative to $Z\sol$), and the $P$-value marginalized over the entire orbital range.}
\label{fig:ZPplanetsMeasure}
\end{figure}

\begin{figure}[t]
\centerline{\includegraphics[height=6cm,width=8cm]{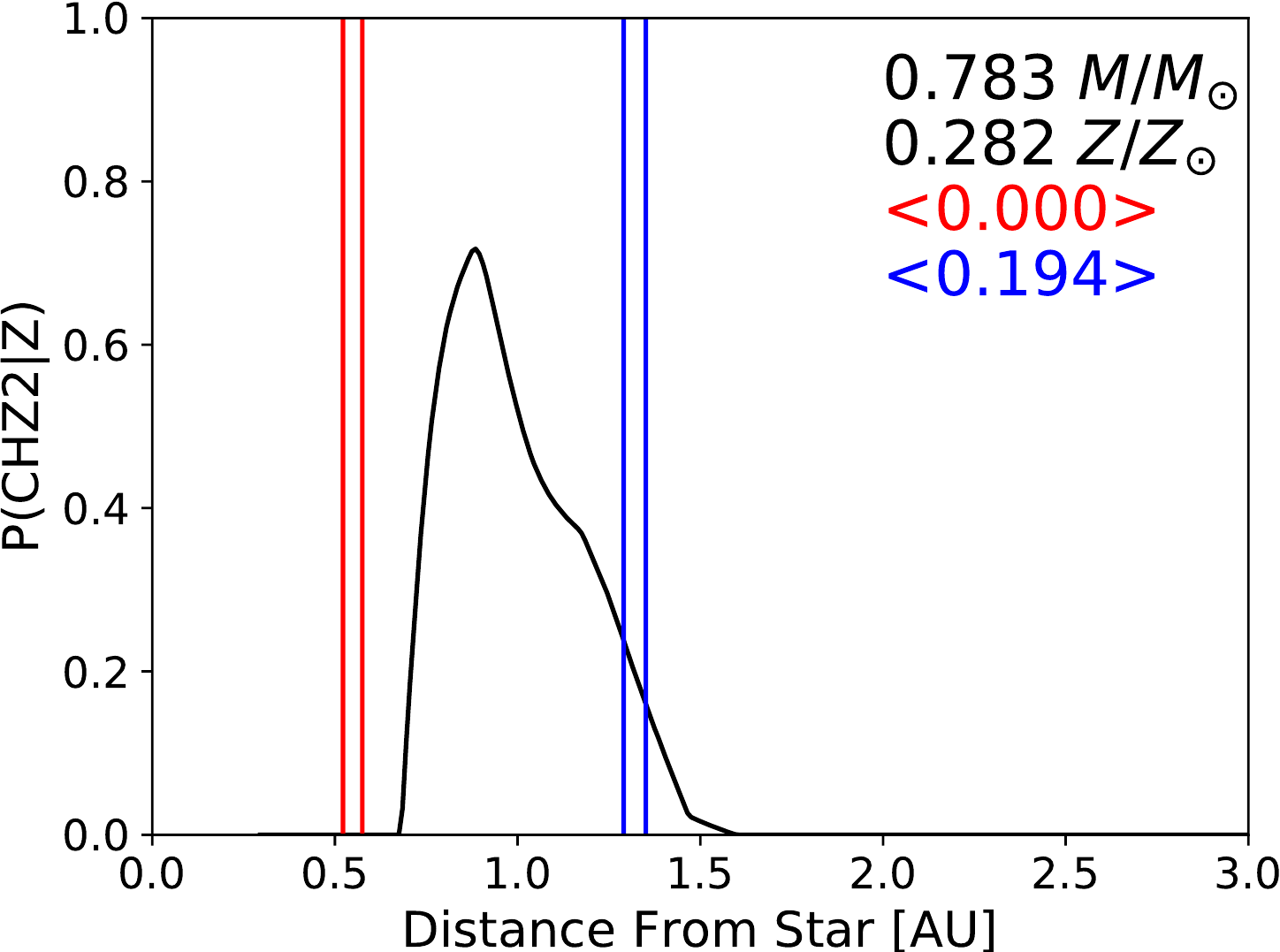}
\includegraphics[height=6cm,width=8cm]{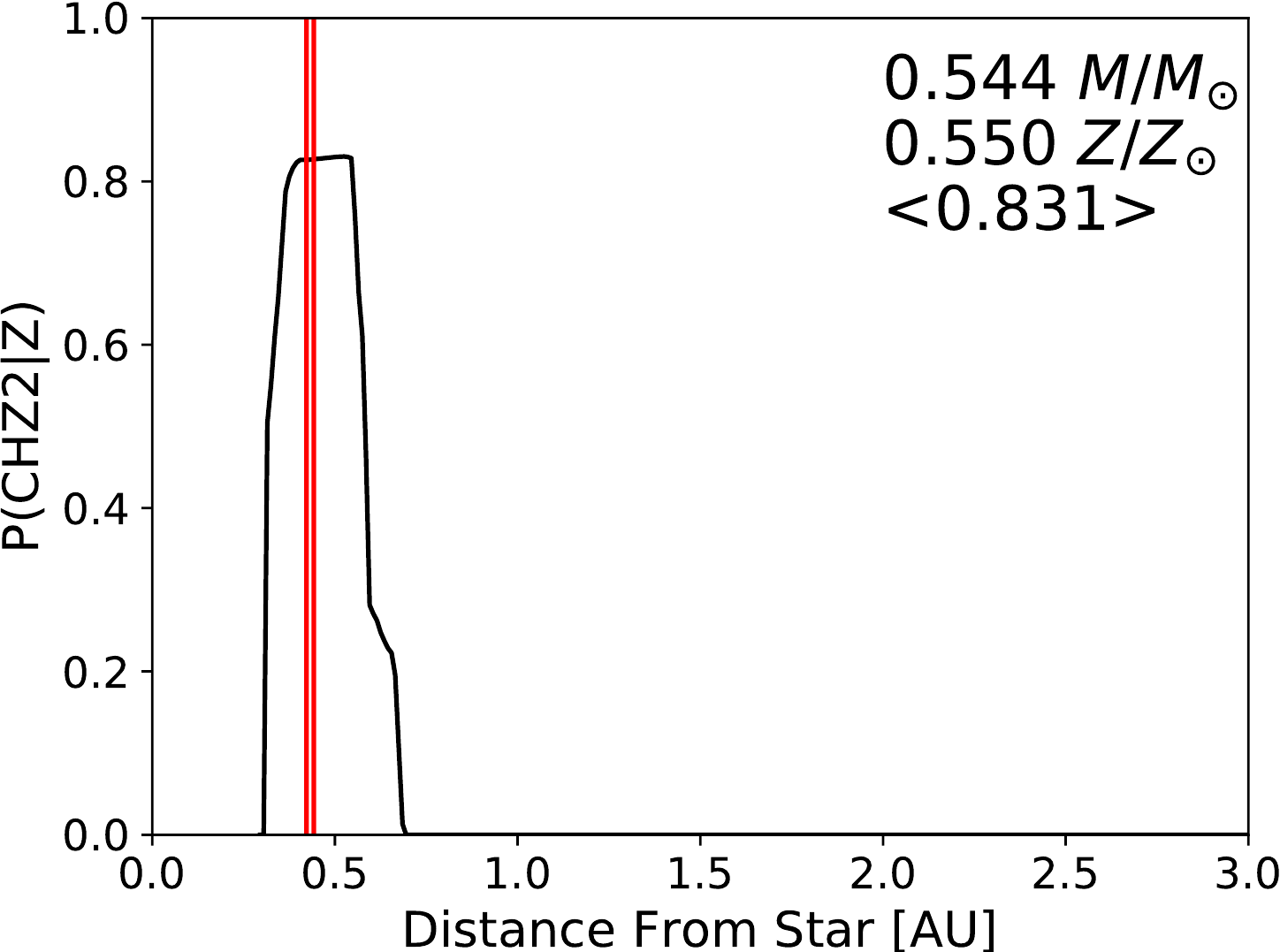}}
\centerline{\includegraphics[height=6cm,width=8cm]{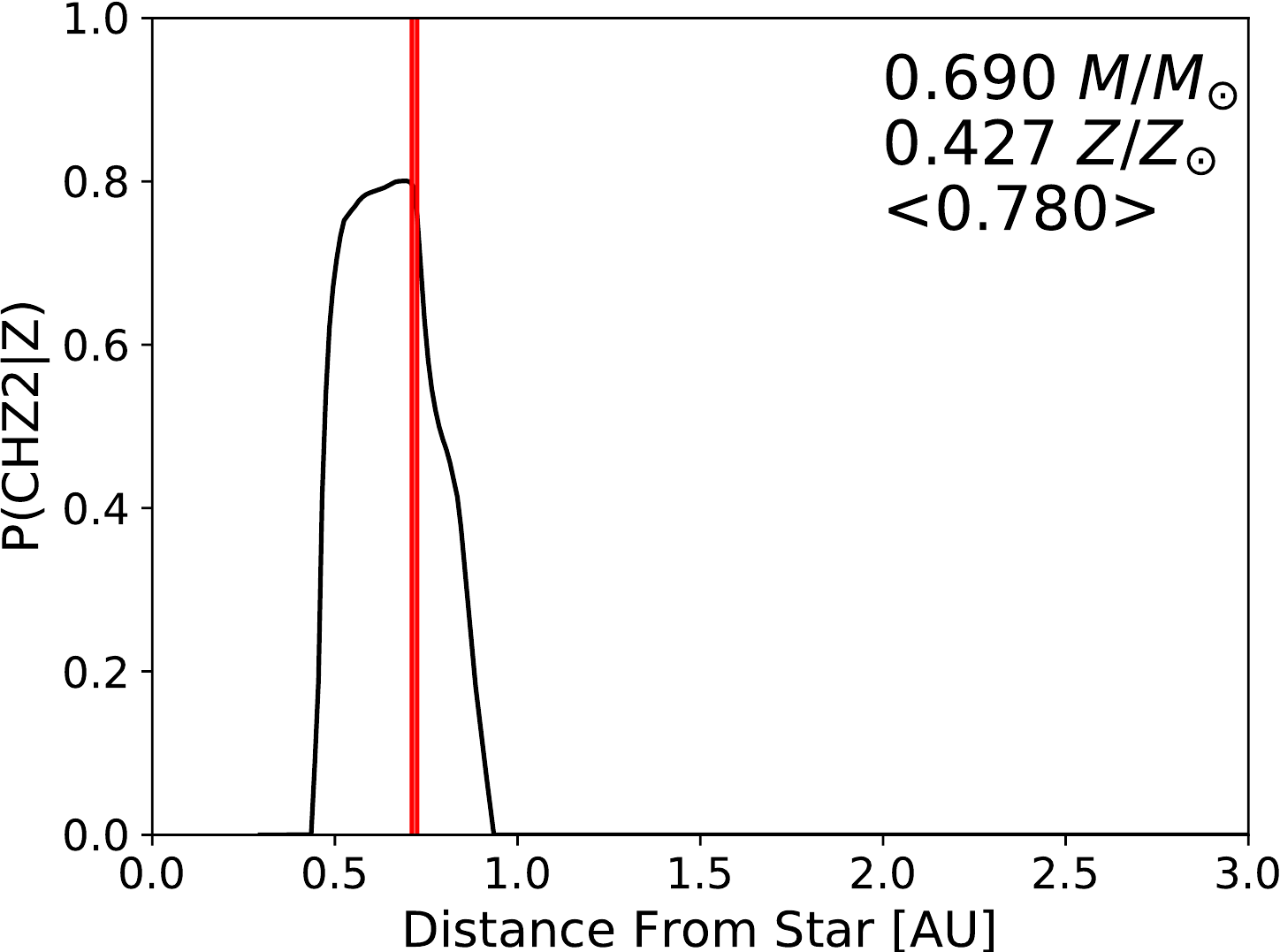}
\includegraphics[height=6cm,width=8cm]{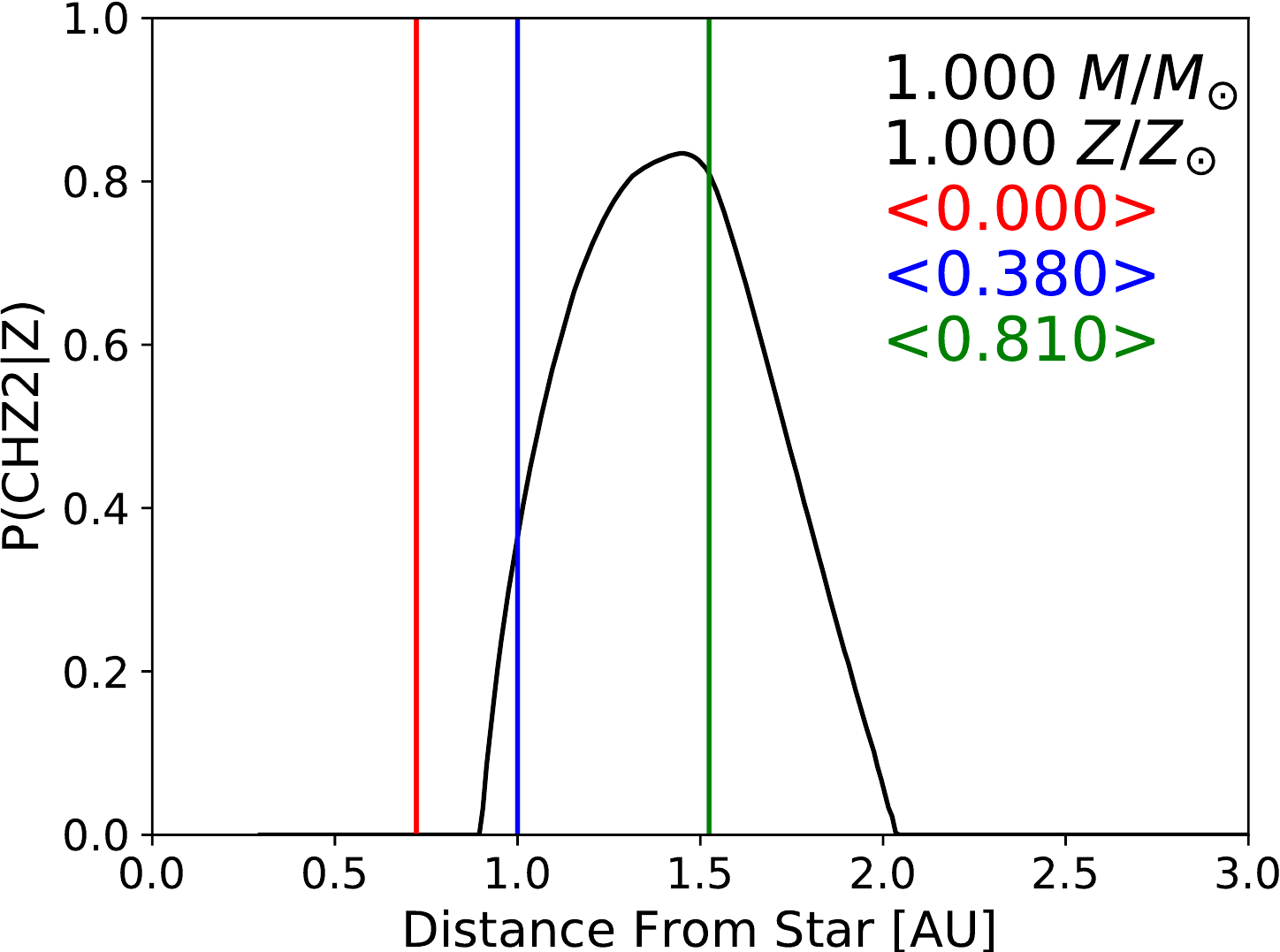}}
\caption{Case 3: $P(\textrm{CHZ}_2|Z,M)$ (using {\it Hypatia}) for Tau Ceti e, f (upper left; red and blue, respectively), Kepler-186f (upper right), Kepler-62f (lower left), and the Sun with Venus, Earth, and Mars (lower right; red, blue, and green, respectively). The data in the upper right corner of each panel is stellar mass, metallicity (given relative to $Z\sol$), and the $P$-value marginalized over the entire orbital range. Note that the Sun's panel is identical to Case 2 (Figure~\ref{fig:ZPplanetsMeasure}) because our $\sigma_M$ is so small, we are effectively using the M = 1 M\sol case and we would indeed expect them to look the same.}
\label{fig:ZMPplanetsHypatia}
\end{figure}

\begin{figure}[t]
\centerline{\includegraphics[height=6cm,width=8cm]{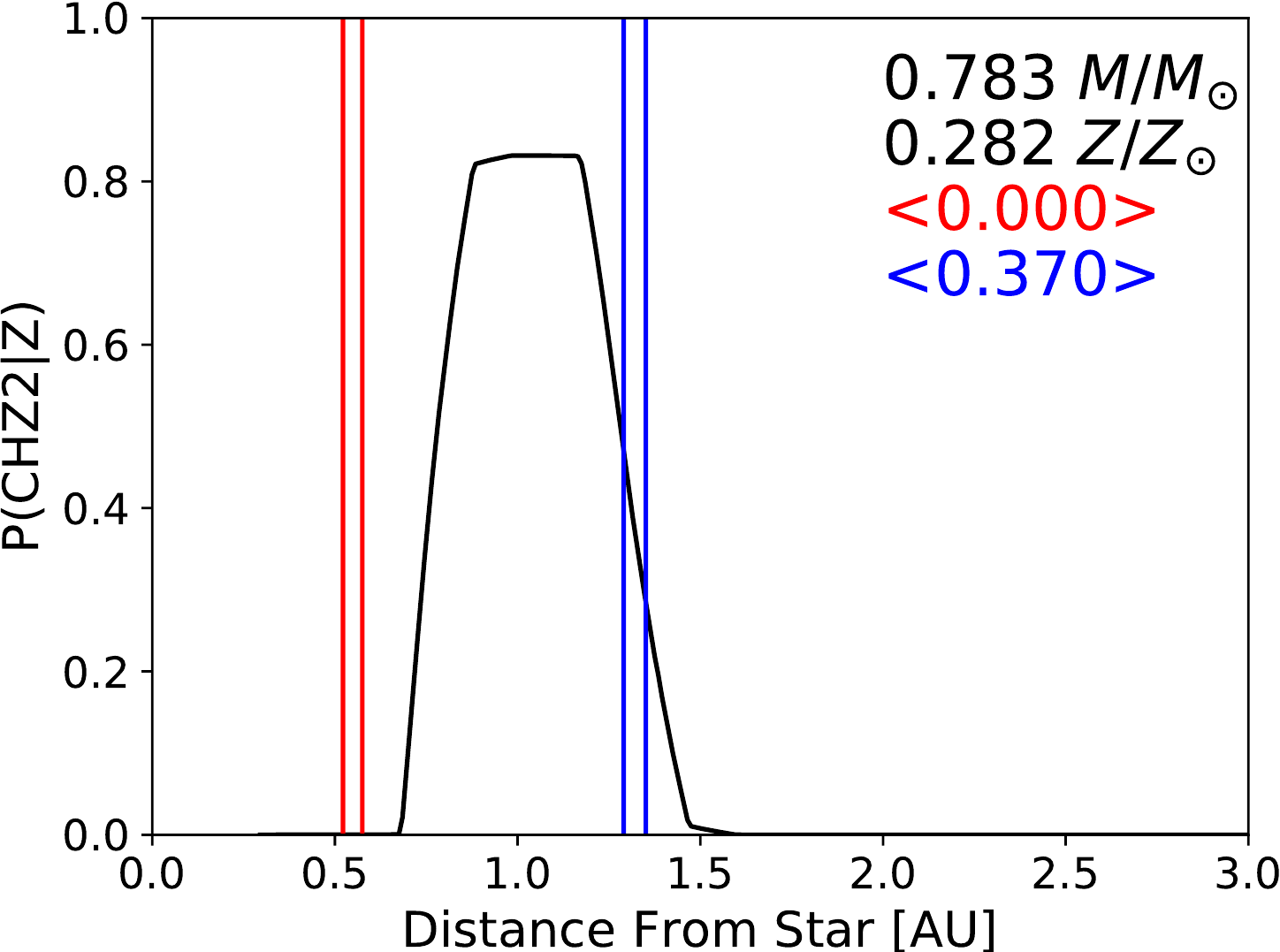}
\includegraphics[height=6cm,width=8cm]{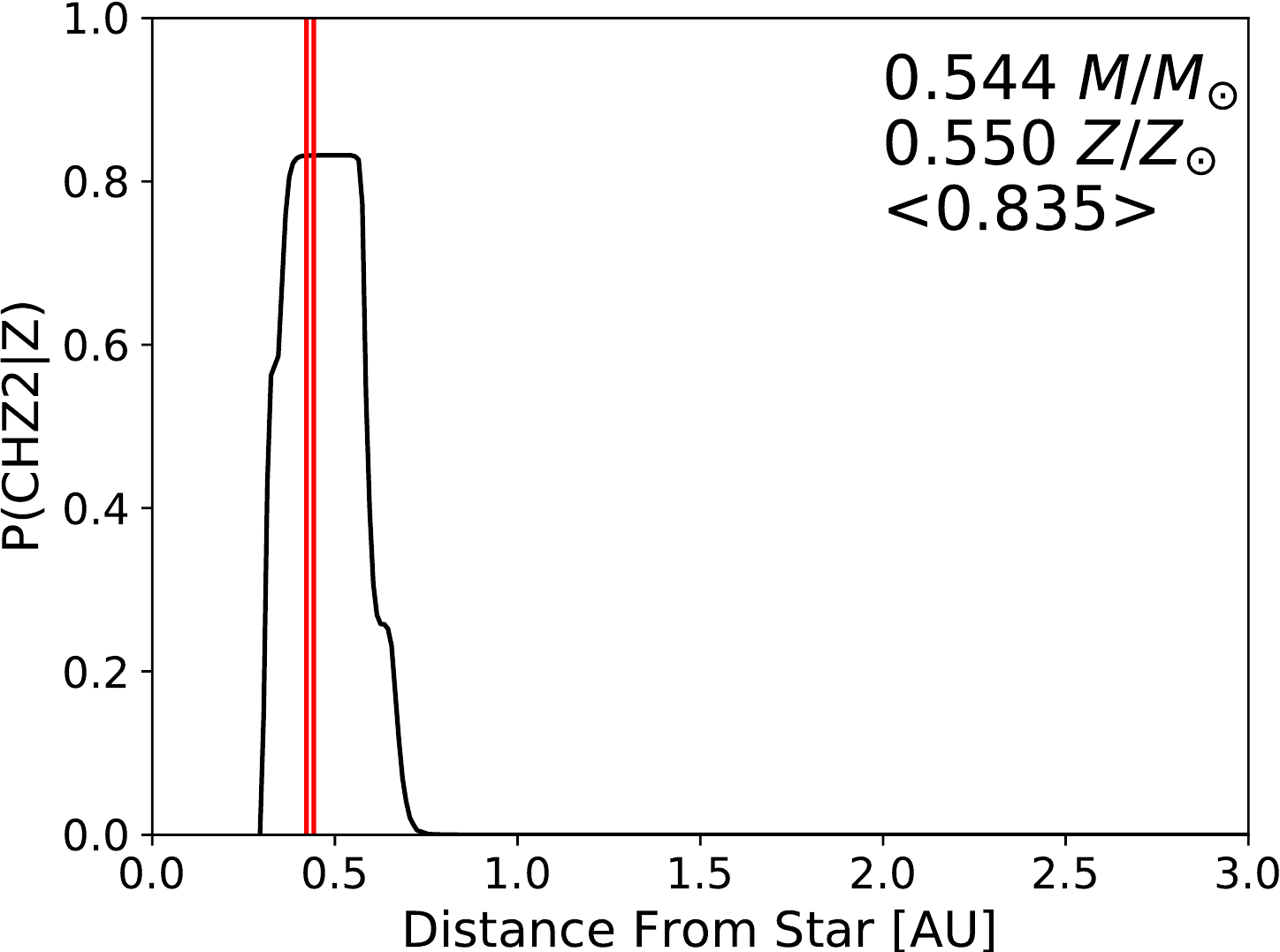}}
\centerline{\includegraphics[height=6cm,width=8cm]{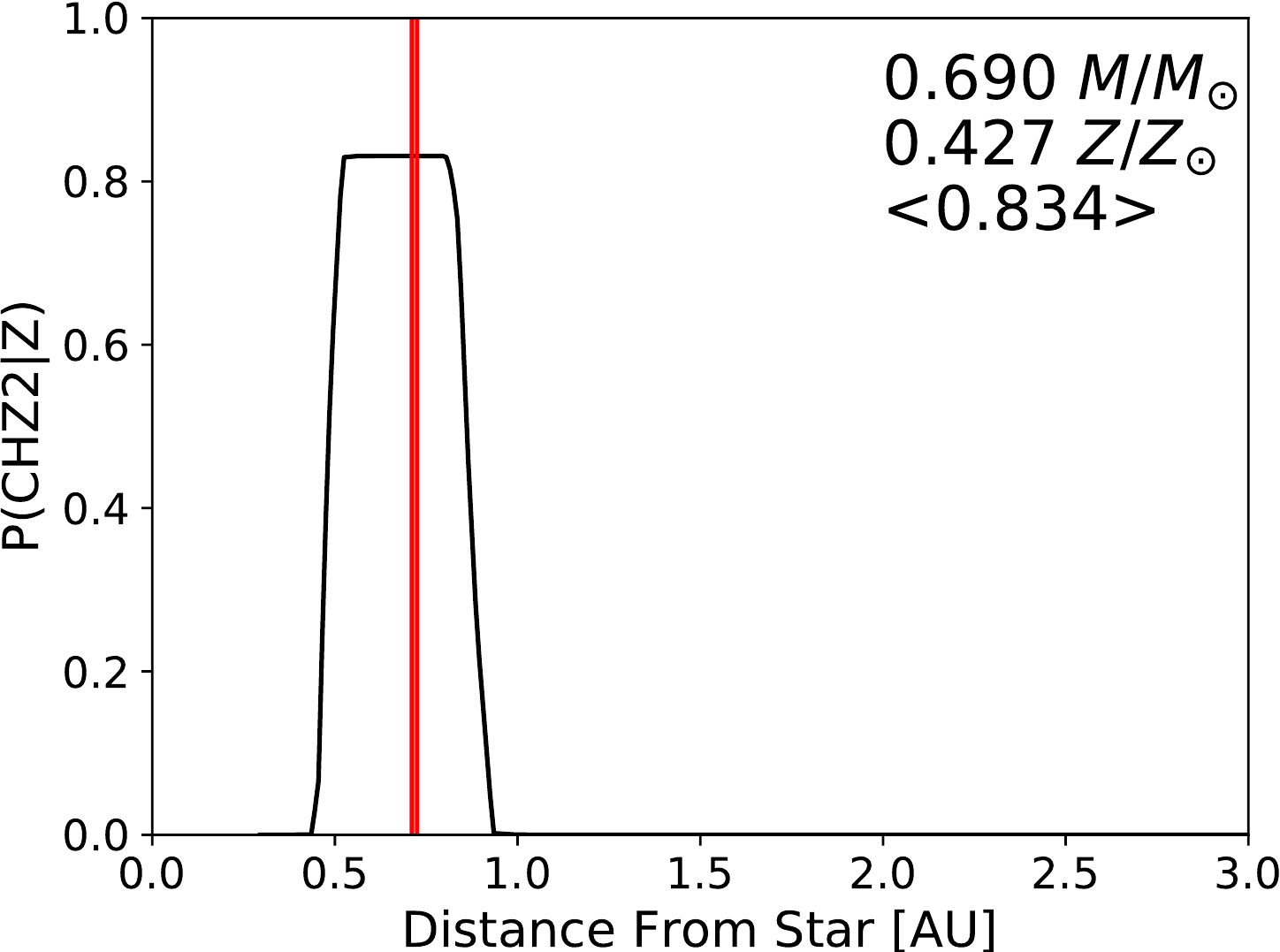}
\includegraphics[height=6cm,width=8cm]{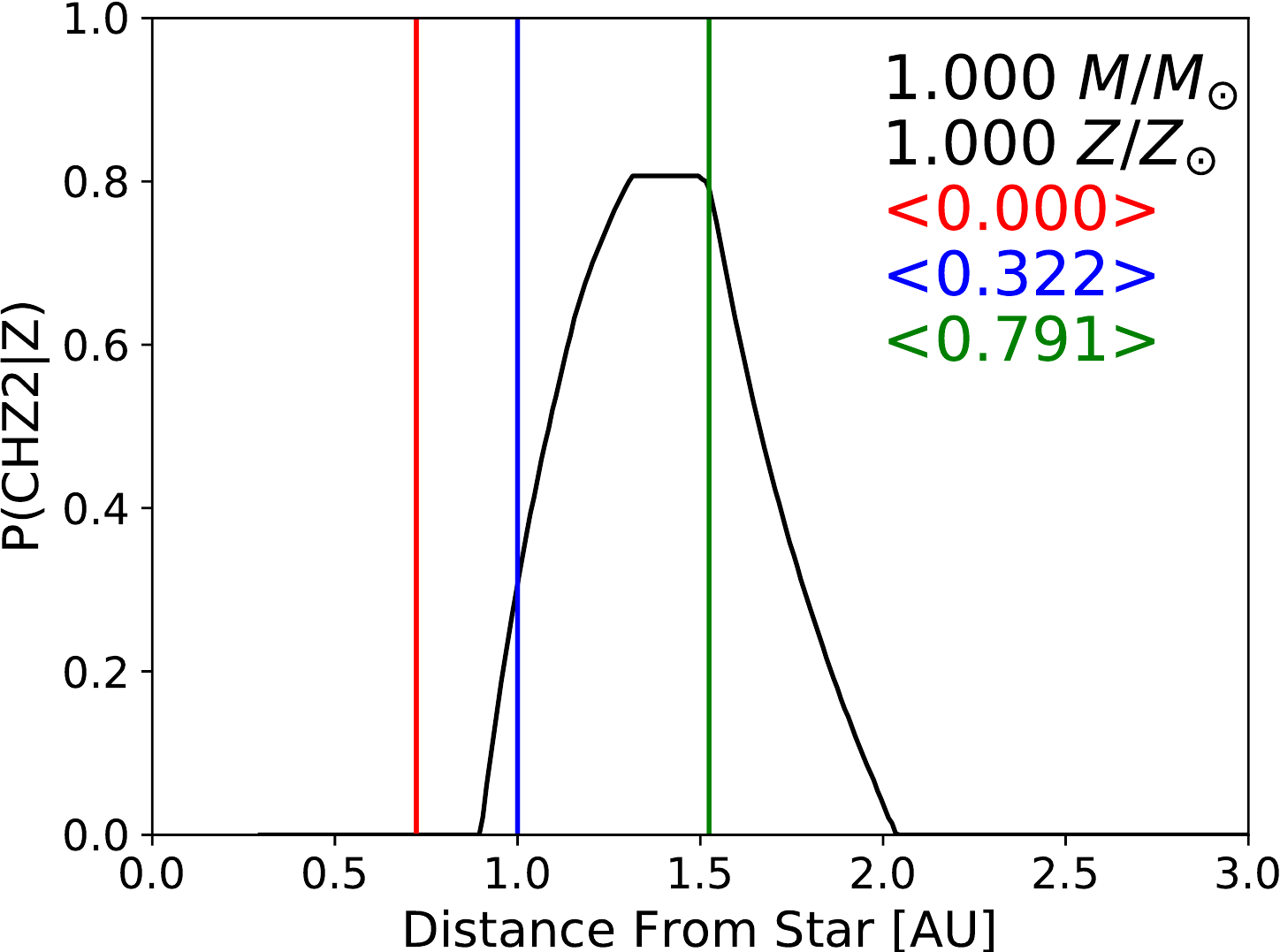}}
\caption{Case 4: $P(\textrm{CHZ}_2|Z,M)$ (using $Z$-value measurements) for Tau Ceti e, f (upper left; red and blue, respectively), Kepler-186f (upper right), Kepler-62f (lower left), and the Sun with Venus, Earth, and Mars (lower right; red, blue, and green, respectively). The data in the upper right corner of each panel is stellar mass, metallicity (given relative to $Z\sol$), and $P$-value marginalized over the entire orbital range.}
\label{fig:ZMPplanetsMeasure}
\end{figure}

Figure~\ref{fig:planets-contour} expands on the results from Figures~\ref{fig:M1_Zall}-\ref{fig:Pcontour}. It is representative of the data for the specific exoplanets of interest indicated in Table~\ref{tab:planettab}, now over-plotted on the corresponding mass and metallicity contour plot (as demonstrated in Figure~\ref{fig:29models}), interpolated to the measured $Z$-value of the host star. The black dots that indicate the location of the planet on this grid indicate the uncertainty in the orbital distance measurements for the planets. Figures~\ref{fig:ZPplanets}-\ref{fig:ZMPplanetsMeasure} (Cases 1-4, respectively) utilize example exoplanetary systems to demonstrate the use of the Bayesian statistical framework for assessing the detectability potential of a given system; we show the Bayesian posterior probability distributions of the habitability potential for several well-known exoplanets of interest: Tau Ceti (e) and (f) \citep[upper left;][]{tu13}, Kepler-186f \citep[upper right;][]{torr15}, Kepler-62f \citep[lower left;][]{bor13}, and the Sun, for comparison. The results for the Sun (shown in the bottom right panels of each of these figures) have particularly interesting implications, in that if we are using our own Solar system as a standard for looking at other potentially habitable systems, we might choose to focus on different planets based on the measured values available.

Some of the sub-panel plots in Figures~\ref{fig:ZPplanets}-\ref{fig:ZMPplanetsMeasure} seem to reach a maximum value ``plateau'' around a value of 0.833, which corresponds to a fractional value of 10/12. This maximum value is inherently defined by our method, due to our inclusion of the 12 Gy MS lifetime cut-off. Because we have limited the MS lifetime of each stellar model in our grid to 12 Gy total, and we are looking specifically at the viability of a 2 Gy continuously habitable zone, the longest possible time that any planet could be in the CHZ$_2$ is 10 Gy. Thus, the highest possible posterior probability is roughly at the value of 10/12. So, a $P$-value of 0.835 indicates that the orbital radius of that particular planet would be considered to be continuously habitable for the entire range of the MS lifetime considered (up to the maximum value of 12 Gy).

Figure~\ref{fig:ZPplanets} is for Case 1, where we calculate $P(\textrm{CHZ}_2|Z)$ using {\it Hypatia} and a fixed mass; Figure~\ref{fig:ZPplanetsMeasure} is for Case 2, where we calculate $P(\textrm{CHZ}_2|Z)$ using $Z$ measurements and a fixed mass; Figure~\ref{fig:ZMPplanetsHypatia} is for Case 3, where we calculate $P(\textrm{CHZ}_2|Z,M)$ using {\it Hypatia} and a mass measurement; and Figure~\ref{fig:ZMPplanetsMeasure} is for Case 4, where we calculate $P(\textrm{CHZ}_2|Z,M)$ using $Z$ measurements and a mass measurement. It will be extremely important to obtain the best possible measurements for stellar mass and metallicity to calculate the most accurate $P$-value for habitability potential. For purposes of comparison in Cases 2-4 we assign a Gaussian uncertainty of $\sigma_M$ and $\sigma_Z = 10^{-3}$ for both solar mass and metallicity. The data in the upper right corner of each panel are the associated $P$-values for the planets and exoplanets of interest.

For the $M$ measurement and Hypatia Bayesian distribution (Case 3, Figure~\ref{fig:ZMPplanetsHypatia}), we find the posterior probability of Venus to be in the CHZ$_2$ is $P$=0 (this is expected, as Venus would not be considered habitable by most definitions), Earth is $P$=0.380, and Mars is the highest with a $P$=0.810. This by itself is due to the fact that in these habitable zone models, Mars would be continuously habitable over a longer fraction of the Sun's overall MS lifetime. When we impose measurements for both $Z$ and $M$ (Case 4, Figure~\ref{fig:ZMPplanetsMeasure}) we again calculate a $P$-value of $P$=0 for Venus, and $P$=0.322 for Earth, $P$=0.791 for Mars. This is a useful illustration of how the posterior probability distribution will change as constraints are added. With a wide range of possible compositions, there are more evolutionary histories that allow a given orbit to be in the CHZ$_2$. As a result, the distribution is wider, with a higher posterior probability that a given orbit has been habitable. When the value of $Z$ is more constrained, the distribution is narrower.

The {\it Kepler} exoplanets included here have a high likelihood to be terrestrial, based on the Earth Similarity Index (ESI) discussed in \citet{sm11}. All have a radius between 0.5-1.5 R$_\Earth$ and masses between 0.1-5 M$_\Earth$; they also fall within the limits of the conservative habitable zones around the associated host star. $P(\textrm{CHZ}_2|Z)$ and $P(\textrm{CHZ}_2|Z,M)$ values are calculated for each exoplanet of interest. The data in the upper right corner of each panel are the stellar mass, the metallicity value ($Z/Z\sol$) of the host star, and the $P$-value marginalized over the entire orbital range.

The Tau Ceti exoplanets were chosen specifically as a follow-up investigation to the work done in \citet{pagano15}. In Figure~\ref{fig:ZMPplanetsMeasure}, $P(\textrm{CHZ}_2|Z,M)$ for planet (e) of the Tau Ceti system (blue) is $P$=0, since the planet is quite close-in to the star. We consider planet (f) as more solidly dwelling in the habitable zone of Tau Ceti (red) with $P$=0.370, but it still has a relatively low posterior probability for long-term habitability (in terms of the CHZ$_2$) due to the fact that (f) has likely only recently entered the habitable zone. These calculations confirm the result of \citet{pagano15}, who noted that planet (e) would be considered to be in the habitable zone only with very optimistic assumptions. Planet (f) would initially appear to be a somewhat more robust candidate as a potentially habitable exoplanet, but the associated stellar evolution track shows that due to the increased luminosity of Tau Ceti over its MS lifetime, it is likely the planet has only recently moved into the star's conservative habitable zone. We find that Kepler-186f has the highest posterior probability, with $P$=0.835 and would likely be the best candidate for in-depth follow-up observations with JWST or other applicable missions. Kepler-62f has only marginally less habitability potential, with $P$=0.834, and would likely also make a good candidate for follow-up characterization. These posterior probability values are consistent with what we would expect for the overall habitability potential for these planets, in terms of the ESI of {\it Kepler} exoplanets, as well as the assessment outlined in \citet{pagano15}.

\section{Conclusions \label{conclusions}}

The intent of this paper has been to demonstrate that we can use our catalog of Tycho stellar evolution tracks, along with observational data, to estimate the likelihood that an observed exoplanet is in its host star's 2 Gy continuously habitable zone (CHZ$_2$) at the time of observation, even if certain stellar properties are poorly measured or unknown. We have shown that the {\it Hypatia} distribution of iron abundance measurements in the included stars is a reasonable representation for the ``metallicities'' catalogued in the {\it Kepler} exoplanet host star database. By constructing a Bayesian statistical framework wherein we can utilize model constraints and observations of individual systems or populations as priors, we can draw meaningful conclusions about the potential for long-term planetary habitability. We have shown that we have the capability to do this for specific well-studied exoplanets of interest. The instantaneous present day location of the habitable zone is an insufficient criterion for assessing the detectability of life on a planet. We conservatively assess the 2 Gy continuously habitable zone, which is based on the timescale for the concurrent evolution of life and conceivably detectable gases in Earth's atmosphere.

A statistical approach toward the characterization of detectable, potentially habitable exoplanetary systems is an important aspect to be considered in the ongoing search for life elsewhere. In this proof of concept, we confirm that stellar mass and composition (specifically, ``metallicity'' [Fe/H], or $Z/Z\sol$ values) play an important role in the evolution of a star's habitable zone, as well as influencing an orbiting exoplanet's long-term habitability potential. We have shown that by implementing a Bayesian statistical approach, we can use Tycho stellar evolution models to inform what kind of stars and what specific orbital radii (habitable distances) are potentially the best to look at with follow-up observations with upcoming space telescope missions. Since our catalog of Tycho models is a reasonable representation of the intrinsic spread in values observed for the mass and metallicity of real, nearby host star candidates, we are now better able to leverage the database of stellar evolution tracks to assess the long-term habitability potential of a given planetary system of interest.

The results of this work can also be complemented by deeper, more specific analysis and observations to better understand the potential habitability of a given planet. For example, it is understood that a planet's habitability can vary significantly within the HZ (e.g. \citet{vanlaer2014}). When looking at a subset of promising targets, additional planetary and stellar characteristics can be taken into account when trying to better understand the potential for habitability, especially for candidates near the extremes of the HZ.

This kind of statistical assessment gives us more analytical power, since we do not need to know every single stellar or planetary parameter in order to calculate a Bayesian \textbf{posterior} probability and ultimately gauge the habitability potential of a particular stellar system. Conversely, it is relatively straightforward to add additional constraints that increase the robustness of this kind of likelihood assessment. These can range from observational parameters, such as stellar age or O/Fe abundance ratios, to choices of theoretical models. For example, $P(CHZ_2)$ can be calculated from models with more sophisticated HZ prescriptions. $P(CHZ_2)$ itself can be used as a prior for calculating the likelihood of a higher level of detectability, perhaps combining it with the probability of geological activity given a measured age and stellar radionuclide abundances. For a subset of potentially habitable exoplanets, we have provided a quick and efficient method for observers to directly compare the estimated detectability potential for exoplanet host stars. This systematic statistical analysis will help to further narrow down the potentially habitable exoplanet candidates that could be targeted for in-depth follow-up characterization with upcoming space telescopes like JWST and future, more sensitive missions.

{}

\end{document}